\documentclass[aps,twocolumn,superscriptaddress]{revtex4}
\usepackage{graphicx}
\begin{document}
\bibliographystyle{apsrev}


\title{Evolution of the resonance and incommensurate spin fluctuations in 
superconducting YBa$_2$Cu$_3$O$_{6+x}$\\}



\author{Pengcheng Dai}
\email{daip@ornl.gov}
\affiliation{Solid State Division, Oak Ridge National Laboratory, Oak Ridge, Tennessee 37831-6393}
\author{H. A. Mook}
\affiliation{Solid State Division, Oak Ridge National Laboratory, Oak Ridge, Tennessee 37831-6393}

\author{R. D. Hunt}
\affiliation{Chemical Technology Division, 
Oak Ridge National Laboratory, Oak Ridge, 
Tennessee 37831-6221
}

\author{F. Do$\rm\breve{g}$an}
\affiliation{Department of Materials Science and Engineering, 
University of Washington, Seattle, Washington 98195
}

\date{\today}

\begin{abstract}
Polarized and unpolarized neutron triple-axis spectrometry was used to 
study the dynamical magnetic susceptibility 
$\chi^{\prime\prime}({\bf q},\omega)$ as a function of energy ($\hbar\omega$) 
and wave vector (${\bf q}$) in a wide temperature range  
for the bilayer superconductor YBa$_2$Cu$_3$O$_{6+x}$ with oxygen concentrations, $x$, 
of 0.45, 0.5, 0.6, 0.7, 0.8, 0.93, and 
0.95. 
The most prominent features in the magnetic spectra 
include a spin gap in the superconducting state, a pseudogap in the normal
state, the much-discussed resonance, and
incommensurate spin fluctuations below the resonance. 
We establish the doping dependence of the spin gap 
in the superconducting state, the resonance energy, 
and the incommensurability of the spin fluctuations.
The magnitude of the spin gap ($E_{sg}$) up to the optimal doping 
is proportional to the superconducting transition temperature $T_c$ with 
$E_{sg}/k_BT_c=3.8$. The resonance, which exists exclusively below $T_c$ for highly doped 
YBa$_2$Cu$_3$O$_{6+x}$ with $x=0.93$ and 0.95, appears above $T_c$ for underdoped compounds 
with $x\leq 0.8$. 
The resonance energy ($E_r$) also scales with
$k_BT_c$, but saturates at $E_r\approx 40$ meV for $x$ 
close to 0.93. The incommensurate spin fluctuations 
at energies below the resonance 
have structures similar to that of the single-layer superconducting La$_{2-x}$Sr$_x$CuO$_4$.
However, there are also important differences.
While the incommensurability ($\delta$)
of the spin fluctuations in La$_{2-x}$Sr$_x$CuO$_4$ is proportional to $T_c$ for  
the entire hole-doping range up to the optimal value, the  
 incommensurability in YBa$_2$Cu$_3$O$_{6+x}$ increases with $T_c$ for low oxygen doping 
and saturates to $\delta=0.1$ for $x\geq0.6$. In addition,
the incommensurability decreases with increasing energy close to the resonance. 
Finally, the incommensurate spin fluctuations appear above $T_c$ in underdoped
compounds with $x\leq 0.6$ but for highly doped materials they are only observed below 
$T_c$. We discuss
 in detail the procedure used for separating the magnetic scattering from phonon and 
other spurious effects. 
In the comparison of our experimental results with various microscopic 
theoretical models, particular emphasis was made to address the similarities and differences
in the spin fluctuations of the two most studied superconductors. Finally, we briefly mention 
recent magnetic field dependent studies of the spin fluctuations and discuss their relevance in understanding
the microscopic origin of the resonance. 
\end{abstract}

\maketitle

\section{Introduction}

The parent compounds of the high-transition-temperature (high-$T_c$) copper-oxide 
superconductors are antiferromagnetic (AF) insulators characterized by a simple 
doubling of the crystallographic unit cell in the CuO$_2$ planes \cite{vaknin,tranquada1}. 
When holes are doped into these planes, the long-range AF-ordered phase is destroyed and the 
lamellar copper-oxide materials become metallic and superconducting. This raises 
the fundamental question of the role of the magnetism in the superconductivity 
of these materials.
Over the past decade, experimental probes such as nuclear magnetic resonance (NMR), muon spin 
resonance ($\mu$SR), and neutron scattering have been used to address this fundamental 
question. In particular, neutron scattering experiments have unambiguously shown the presence
of short-range AF spin correlations (fluctuations)
in cuprate superconductors such as La$_{2-x}$Sr$_x$CuO$_4$ 
\cite{yoshizawa,birgeneau,cheong,mason,thurston,yamada,yamadaprl,lake,kastner} and 
YBa$_2$Cu$_3$O$_{6+x}$ 
\cite{mignod,tranquada,mook93,sternlieb,fong95,bourges96,fong96,dai96,fong97,bourges97,dai98,mook98,mookn98,arai,hayden98,dai99}
at all doping levels, $x$. 
Even so, the role of such fluctuations in the pairing mechanism and superconductivity
is still a subject of controversy \cite{conference}.

If spin fluctuations are important for the mechanism of high-$T_c$ superconductivity, 
they should be universal for all copper oxide systems.  
For the La$_2$CuO$_4$ family of materials, which   
has a single CuO$_2$ layer per unit cell,
superconductivity can be induced by the substitution of 
Sr, Ba, or Ca for La or by intercalation of excess oxygen. 
The low-energy spin fluctuations have been  
observed at a quartet of incommensurate wave vectors (Fig. 1)
away from the AF Bragg positions, often referred to as $q_0=(\pi,\pi)$, in the  $x=0$ compounds 
 \cite{kastner}.
These modulated low-frequency spin fluctuations persist in both the normal and superconducting states  
with a suppression of their intensity below $T_c$ \cite{yamadaprl,lake}. As a function of the doping,
the incommensurability $\delta$ increases linearly with $x$ until saturating at 
$\delta \simeq 1/8$ for $x\geq 1/8$ \cite{cheong,yamada,noteincommb}. Furthermore, the
incommensurability depends only on the doping level and the unbound charge introduced by the
dopants,  and not on the method for introducing the charge \cite{kastner}.
As a function of increasing 
frequency, the incommensurability does not change \cite{mason}, and 
no particularly sharp features have been identified \cite{kastner}. 
For YBa$_2$Cu$_3$O$_{6+x}$, which has two CuO$_2$ planes per unit cell (bilayer), 
 the situation
is more complex.  Early measurements \cite{mignod,mook93,fong95,bourges96,fong96}
have shown that the most prominent feature in the spin fluctuations spectra 
in the highly doped YBa$_2$Cu$_3$O$_{6+x}$ ($x\geq 0.9$) is a sharp peak which 
appears below $T_c$ at an energy of $\sim$41 meV. When scanned at fixed energy as a function of 
wave vector, the sharp peak is centered at $(\pi,\pi)$ and generally referred to 
as a resonance. For underdoped YBa$_2$Cu$_3$O$_{6+x}$, the resonance also peaks at 
$(\pi,\pi)$ but with a reduced energy \cite{dai96,fong97,bourges97}. 
In YBa$_2$Cu$_3$O$_{6.6}$, it occurs at 34 meV and is superposed on a continuum 
which is gapped at low energies \cite{dai99}.
Below the resonance frequency, the continuum is actually strongest at a quartet of
incommensurate positions consistent with those in La$_{2-x}$Sr$_x$CuO$_4$ of the same hole 
doping (Fig. 1) \cite{dai98,mook98,mookn98}. This discovery along with similar observations
in YBa$_2$Cu$_3$O$_{6.7}$ \cite{arai} suggests that incommensurate 
spin fluctuations may be universal to all high-$T_c$ superconductors. 

\begin{figure}
\includegraphics[width = 3 in]{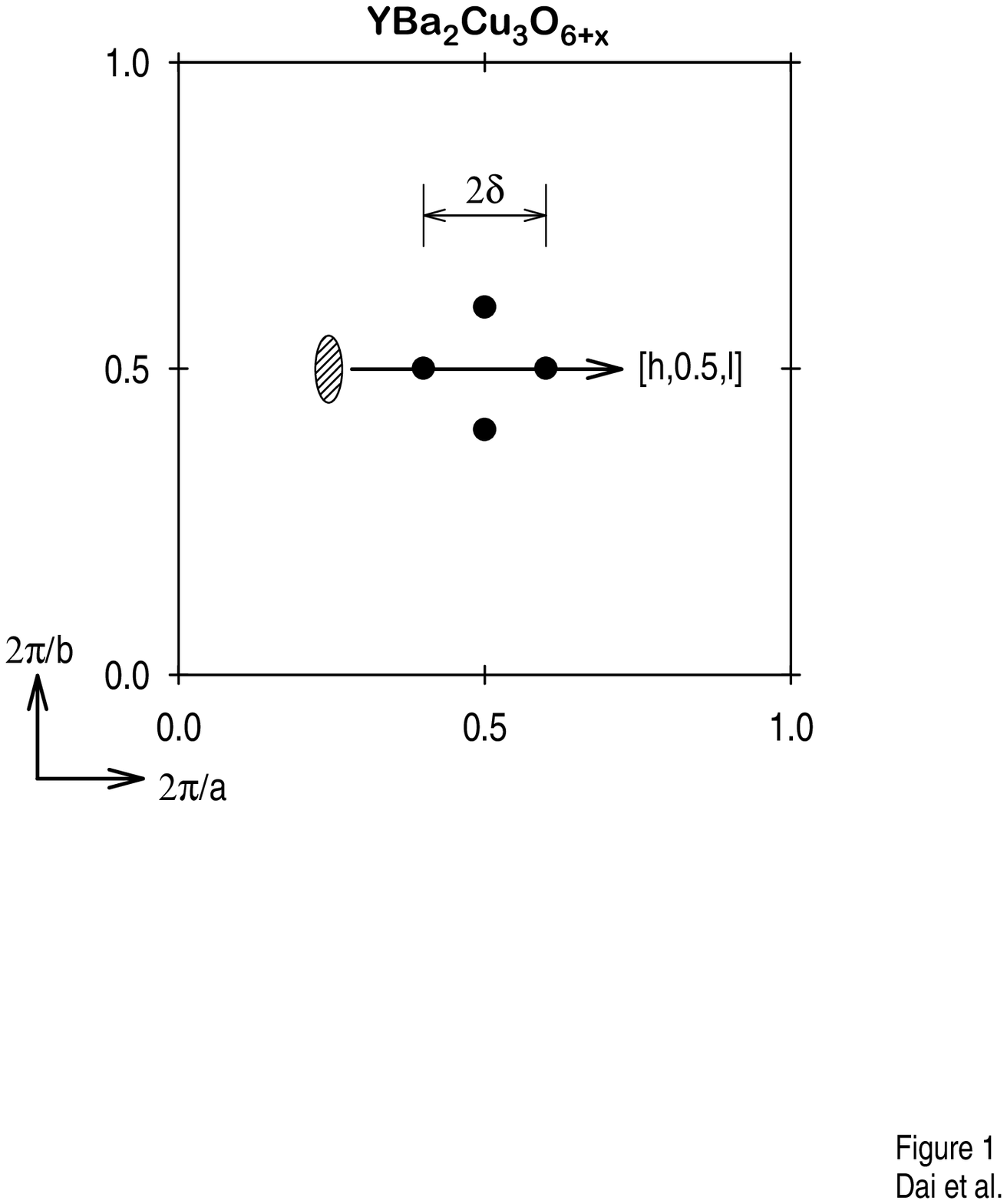}
\caption{
The reciprocal space diagram and scattering geometry used in the experiments.
Note that the scan direction for most of the measurements is along $[h,0.5,l]$. The structure 
of the incommensurate spin fluctuations is very similar to that of La$_{2-x}$Sr$_x$CuO$_4$ as shown 
by Mook {\it et al.} [24]. In this article, 
we define the separation between the incommensurate peaks as $2\delta$ as shown in the figure.
The shaded ellipse shows that the narrow part of the resolution volume is 
along the scan direction.
}
\end{figure}

In this article, we report systematic unpolarized and polarized inelastic neutron scattering 
studies of spin fluctuations 
in YBa$_2$Cu$_3$O$_{6+x}$. By carefully separating the magnetic signal from phonons and
other spurious effects, we determine the wave vector and energy dependence of 
the spin dynamical susceptibility,  
$\chi^{\prime\prime}({\bf q},\omega)$, as a function of temperature and doping concentration. 
For YBa$_2$Cu$_3$O$_{6+x}$ with oxygen concentrations at 
all doping levels addressed in this article, we observe a spin gap in the superconducting state,
a commensurate resonance, and incommensurate spin fluctuations at energies 
below the resonance. The magnitude of the spin gap in the superconducting state ($E_{sg}$) is
proportional to $k_BT_c$ with $E_{sg}/k_BT_c=3.8$. The resonance appears above $T_c$ for 
underdoped compounds but occurs exclusively below $T_c$ for optimal and overdoped materials. 
The incommensurability of the excitations below the resonance decreases with increasing
energy until merging into the resonance. For YBa$_2$Cu$_3$O$_{6+x}$ with 
$x\leq 0.6$, the incommensurability is approximately proportional to $T_c$. For oxygen concentrations
$x$ above 0.6, the incommensurability is independent of hole-doping and 
saturates at $\delta=1/10$.
In the case of YBa$_2$Cu$_3$O$_{6.8}$, the commensurate 
low-frequency normal state magnetic excitations centered around $(\pi,\pi)$
are suppressed with the opening of
a large spin gap in the superconducting state. 
For YBa$_2$Cu$_3$O$_{6.93}$ and YBa$_2$Cu$_3$O$_{6.95}$, 
the normal state spin fluctuations are below the detection limit of current 
triple-axis spectroscopy, and the magnetic fluctuations in the superconducting state
are dominated by the well-discussed resonance at $\sim40$ meV 
\cite{mignod,mook93,fong95,bourges96,fong96} and incommensurate spin fluctuations below the resonance.

The rest of the paper is organized as follows. The next section describes the 
experimental procedures, including details of sample characterization and spectrometer 
setup. In Sec. IIIA, polarized and unpolarized neutron scattering data for 
YBa$_2$Cu$_3$O$_{6+x}$ with oxygen concentrations, $x$, of $\sim 0.45$, 0.5, 0.6, 0.7, 0.8, 0.93, and 0.95
 are presented with a detailed description of the procedures used to extract magnetic scattering.
In Sec. IIIB, a  
comparison of the results with those for La$_{2-x}$Sr$_x$CuO$_4$ is presented. 
The issues of the doping dependence of the incommensurability, the 
spin gap in the superconducting state,
the resonance in the normal state of underdoped materials, 
and resonance energy are addressed. Finally,  
the results presented in this paper and recent magnetic field dependence
measurements of the spin fluctuations
are discussed with the predictions and consequences of various theoretical models. 
Section IV gives a brief summary of the 
major conclusions of this work.

\section{Experimental details}
\subsection{Sample characterization}
The experiments were performed 
on large twined crystals of YBa$_2$Cu$_3$O$_{6+x}$ that were
prepared by the melt-grown technique at the University of 
Washington. As emphasized in previous publications \cite{mook93,dai96},
these samples contain considerable amounts of  
Y$_2$BaCuO$_5$ \cite{raveau} as 
an impurity. The structural and superconducting properties of
these melt-grown YBa$_2$Cu$_3$O$_{6+x}$ samples with different 
Y$_2$BaCuO$_5$ concentrations have been studied in detail by 
Gopalan {\it et al.} \cite{gopalan}. These authors found that the
YBa$_2$Cu$_3$O$_{6+x}$ platelet width and crack width decrease with increasing
Y$_2$BaCuO$_5$ content. The reduced average YBa$_2$Cu$_3$O$_{6+x}$ grain width 
 may facilitate the formation of equilibrium crystal structures at
various oxygen concentrations. In a recent
study of oxygen equilibration and chemical diffusion in massive 
YBa$_2$Cu$_3$O$_{6+x}$ crystals, Lindemer \cite{lindemer} found that 
melt-grown samples containing a significant amount of Y$_2$BaCuO$_5$ 
 reached their equilibrium values of oxygen concentration in a dynamic flow of oxygen 
considerably faster than samples without this impurity. These results are
consistent with the notion that oxygen invariant Y$_2$BaCuO$_5$, micro-cracks, and 
the boundaries between twin grains are important in providing 
oxygen pathways through these large samples of YBa$_2$Cu$_3$O$_{6+x}$.

For nonstoichiometric pure phase YBa$_2$Cu$_3$O$_{6+x}$,
Lindemer and coworkers \cite{lindemer1}
have established the thermodynamic equilibrium phase diagram of 
oxygen concentration versus 
temperature and oxygen pressure ($T$-$p$).
We oxygenate YBa$_2$Cu$_3$O$_{6+x}$ in a vacuum thermogravimetric apparatus (TGA) by
annealing to the equilibrium weight ($\pm0.1$ mg), 
equivalent to a
sensitivity in $6+x$ of 0.0003 per mol, at $T$-$p$[O$_2$] conditions for
the desired oxygen doping following previously established models \cite{lindemer1}. 
Instead of quenching the sample 
after annealing, the partial pressure of oxygen in the TGA was adjusted during
the cooling process  to maintain a constant sample mass, thus ensuring uniformity in
oxygen content and an equilibrium crystal structure. Seven single 
crystals of YBa$_2$Cu$_3$O$_{6+x}$ were oxygenated in this way
to obtain $x$ of 0.45, 0.5, 0.6, 0.7, 0.8, 0.93, and 0.95.

The superconducting properties of these YBa$_2$Cu$_3$O$_{6+x}$ crystals 
were characterized by a SQUID magnetometer. 
Susceptibility measurements 
made on small pieces cut from these samples show 
superconducting transition temperatures of $\sim48\pm 6$ K, $52\pm3$ K, 
$62.7\pm 2.7$ K, $74\pm 2.7$ K, $82\pm1.25$ K, and $92.5\pm 0.5$ K, 
for YBa$_2$Cu$_3$O$_{6.45}$, YBa$_2$Cu$_3$O$_{6.5}$, 
YBa$_2$Cu$_3$O$_{6.6}$, YBa$_2$Cu$_3$O$_{6.7}$, YBa$_2$Cu$_3$O$_{6.8}$, and YBa$_2$Cu$_3$O$_{6.93}$, 
 respectively. The transition temperature for YBa$_2$Cu$_3$O$_{6.95}$
 of 92 K is
estimated from the temperature dependence of the resonance \cite{mook93,dai99}.
Although the widths of superconducting transition
temperatures for YBa$_2$Cu$_3$O$_{6+x}$ with $x\geq 0.6$ are usually sharp ($< 3$ K), the
transition widths for lower doping samples are considerably broader. This
appears to be the intrinsic property of the underdoped YBa$_2$Cu$_3$O$_{6+x}$ \cite{lindemer1}.

To quantitatively determine 
the effectiveness of this oxygenation technique and 
the composition of the melt-grown crystal, 
two pieces were cut from a 96-g sample of YBa$_2$Cu$_3$O$_{6.93}$.
Sample $A$ weighed approximately 15-g and was crushed into fine
powder. Sample $B$ weighed 14.98-g. 
Powder diffraction measurements were performed 
on sample $A$ at room temperature using the
 HB-4 high-resolution powder diffractometer with an incident 
beam wavelength of 1.0314 \AA\  at the
High-Flux Isotope Reactor (HFIR) 
at Oak Ridge National Laboratory \cite{dai1}.  Rietveld analysis
of the powder data
($R_{wp}=6.55\%$, $\chi^2=1.447$) revealed that the sample 
contains 83.6 mol\% YBa$_2$Cu$_3$O$_{6+x}$ with the partial 
oxygen occupancies [O(1)] at $0.96\pm 3$ 
and 16.4 mol\% Y$_2$BaCuO$_5$. The lattice parameters 
for the YBa$_2$Cu$_3$O$_{6.93}$ [$a=3.8188(2)$ \AA, 
$b=3.8856(2)$ \AA, and $c=11.6920(5)$ \AA] 
are slightly different from  
that of the YBa$_2$Cu$_3$O$_{6.93}$ sample reported by 
Jorgensen {\it et al.} \cite{jorgensen}, but they agree
well with those of the YBa$_2$Cu$_3$O$_{6+x}$ ($0.93\leq x<1$)
crystals of Altendorf {\it et al.} \cite{altendorf}. 

To further verify the partial 
oxygen content, sample $B$ was
subjected to independent thermogravimetric 
analysis. The 14.98-g specimen ($B$) was subjected to the de-oxygenation study  
in the TGA. 
 The weight
change of $B$ relative to the room-temperature weight was
measured after equilibration at 573 K and at
successively higher temperatures.
The results were converted to $6+x$ values
and compared to the expected 
behavior \cite{lindemer1}. This analysis 
demonstrated that the initial oxygen 
content of $B$ was $6.93\pm0.005$, thus confirming the
effectiveness of our oxygenation technique. 

In an unpublished work, Chakoumakos and Lindemer \cite{chakoumakos} have 
systematically studied the lattice parameters of the powder YBa$_2$Cu$_3$O$_{6+x}$ 
oxygenated using this technique. They performed Rietveld analysis of the powder
neutron diffraction data obtained on HB-4 for a series YBa$_2$Cu$_3$O$_{6+x}$ samples
and established the relationship between oxygen concentration and lattice parameters (Fig. 2). 
High resolution measurements on the single crystals oxygenated in our study indicated that 
the lattice parameters have virtually the same value as powders. As a consequence, we used 
the lattice parameters established in their powder work for our single-crystal inelastic 
neutron scattering measurements. 

\begin{figure}
\includegraphics[width = 3 in]{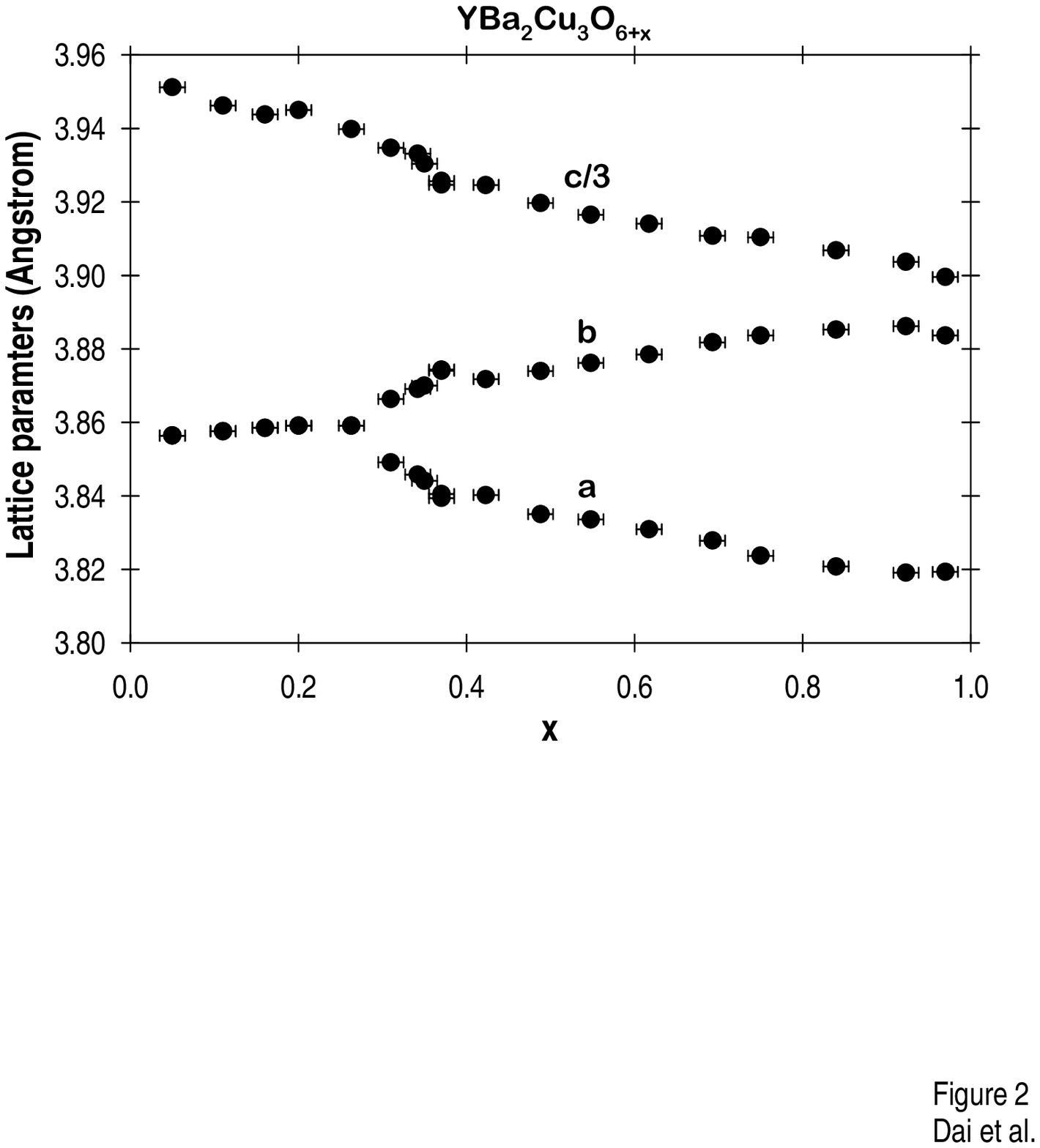}
\caption{
Lattice parameters of YBa$_2$Cu$_3$O$_{6+x}$ as a function of oxygen concentration 
$x$ from Chakoumakos and Lindemer [37]. The results are obtained using a high-resolution 
powder diffractometer on powder samples oxygenated in the same way as the 
single crystals in this work. The vertical error bars are smaller than the size of the symbols.
}
\end{figure}

\subsection{Neutron scattering}

Inelastic neutron scattering 
measurements were made at HFIR  
using the HB-1 and HB-3 triple-axis spectrometers. 
The momentum transfer $(q_x,q_y,q_z)$ is measured 
in units of \AA$^{-1}$ and reciprocal space positions are specified in reciprocal lattice units (rlu)
$(h,k,l)=(q_xa/2\pi,q_yb/2\pi,q_zc/2\pi)$, where $a$, $b$ ($\approx a$), and $c$ are the
lattice parameters of the orthorhombic unit cell of YBa$_2$Cu$_3$O$_{6+x}$ (see Figs. 1 and 2).
The magnetic neutron scattering directly measures the imaginary
part of the generalized spin susceptibility $\chi^{\prime\prime}({\bf q},\omega)$ 
for momentum transfer ${\bf q}={\bf k}_i-{\bf k}_f$ and energy transfer $\hbar\omega$,
where ${\bf k}_i$ and ${\bf k}_f$ are the incident and final neutron wave vectors, respectively.
The scattering cross-section for an isotropic system \cite{lovesey} is 
\begin{equation}
{{\rm d^2}\sigma\over{\rm d}\Omega{\rm d}E}=
{2\over\pi g^2}{k_f\over k_i}r_0^2
|F({\bf q})|^2{1\over 1-exp(-\hbar\omega/k_BT)}\chi^{\prime\prime}({\bf q},\omega)
/ \mu_{\rm B}^2
\label{eq:two}
\end{equation}
where $g$ is the Lande factor ($\approx 2$), $r_0$ is $5.4\times 10^{-13}$ cm,  
$|F({\bf q})|$ is the magnetic form factor, $[n(\omega)+1]=1/[1-exp(-\hbar\omega/k_BT)]$
is the Bose population factor, and $\mu_B$ is the Bohr magneton.  
The major difficulty in studying spin fluctuations in 
high-$T_c$ materials is to separate the magnetic scattering from 
(single- and multi-) phonon scattering and other spurious processes.
While many spurious events such as accidental Bragg scattering can be identified 
by performing the desired inelastic scan in the two-axis mode \cite{tranquada},
two approaches are used to separate magnetic from phonon scattering. 

We first describe the use of neutron polarization analysis \cite{moon} which, in principle,
can unambiguously separate magnetic from non magnetic processes. 
For the polarized experimental setup,  
a crystal of $^{57}$Fe(Si) [$d(110)=2.0202$ \AA] 
and crystals of Heusler alloy Cu$_2$MnAl [$d(111)=3.4641$ \AA] were
used as a monochromator and analyzer, respectively. 
The polarization state of the neutron beam can be changed by a
spin-flipping device (Mezei coil)
 that is placed in the incident beam {\it before} the sample.
The combination of a $^{57}$Fe(Si) (110) monochromator and 
a Heusler (111) analyzer permits spin-flip (SF) measurements 
without using the Mezei coil because the polarization
of the neutrons scattered by the $^{57}$Fe(Si) is along
the applied field direction while it is opposite for the Heusler 
crystal. For the measurements, 
the final neutron energy was fixed at 30.5 meV, and 
a pyrolytic graphite (PG) filter was used before the analyzer. 
The polarization is usually determined 
by measuring the flipping ratio $R$, defined
as the ratio of neutrons reflected by the analyzer with the flipper (Mezei coil)
``on'' [which corresponds to neutron nonspin flip (NSF), or $++$] and ``off'' [neutron spin-flip (SF) or $+-$]. 
Assuming the flipping ratios for guide field parallel 
(horizontal field or HF) and perpendicular (vertical 
field or VF) to the wave vector $\bf {q}$ are $R_H$ and $R_V$, respectively, the observed neutron
cross sections for a paramagnetic system are given by \cite{moon,note}:
\begin{eqnarray}
{\rm HF}:   & \sigma^{+-}_{ob}  = & {R_H\over R_H +1} (M+{2\over 3}NSI+BG)+ \nonumber \\
            &   & {1\over R_H+1}(N+{1\over 3}NSI+BG) \nonumber \\
  & \sigma^{++}_{ob} = &{R_H\over R_H +1} (N+{1\over 3}NSI+BG)+ \nonumber \\
  &                    &{1\over R_H+1}(M+{2\over 3}NSI+BG) \nonumber \\
{\rm VF}: & \sigma^{+-}_{ob} = & {R_V\over R_V +1}({1\over 2}M+{2\over 3}NSI+BG)+ \nonumber \\
  &                    &{1\over R_V+1}(N+{1\over 2}M+{1\over 3}NSI+BG) \nonumber \\
 & \sigma^{++}_{ob} = &{R_V\over R_V +1} ({1\over 2}M+N+{1\over 3}NSI+BG)+ \nonumber \\
 &                    &{1\over R_V+1}
          ({1\over 2}M+{2\over 3}NSI+BG) \label{eq:one}
\end{eqnarray}
where $M$ is the magnetic, $N$ the nuclear, $NSI$ the nuclear spin incoherent neutron
cross section, and $BG$ is the background. Inspection of Eq. [2.2] 
reveals that the most effective way to detect a magnetic 
signal is to measure SF scattering with HF. Although 
such method has 
been successfully used to identify the 
magnetic origin of resonance peaks
in YBa$_2$Cu$_3$O$_{6+x}$ \cite{mook93,fong96,dai96,fong97}, the advantage of 
the technique comes at a considerable cost in intensity which makes it impractical 
for observing small magnetic signals. 

As a consequence, most of our experiments were performed with unpolarized neutrons with 
Be (002) or PG (002) as monochromators as specified in the figure captions. 
In these experiments, we also used 
PG(002) as analyzers. Fast neutrons were removed from the incident beam by 
an in-pile sapphire filter, and PG filters were used to 
remove high-order neutron contaminations where appropriate.
The horizontal collimations were controlled with Soller slits and are 
specified in the figures containing the experimental data.  
To separate the magnetic from phonon scattering, we  
utilize the differences in their temperature and wave vector dependence of
the cross sections. While the phonon scattering gains intensity on warming due to 
the thermal population factor, the magnetic signal usually becomes weaker because it
spreads throughout the energy and momentum space at high temperatures. 
In addition, magnetic intensity should decrease at large 
wave vectors due to the reduction in the magnetic form factor  
while phonon and multi-phonon scattering are expected to increase with increasing 
wave vector. Therefore,  
in an unpolarized neutron scattering measurement the net 
intensity gain above the multi-phonon background on cooling at appropriate wave vectors 
is likely to be magnetic in origin. Previous experiments have utilized this method to discover  
incommensurate spin fluctuations in 
YBa$_2$Cu$_3$O$_{6.6}$ \cite{dai98}.

Figure 1 depicts the reciprocal lattice of the YBa$_2$Cu$_3$O$_{6+x}$ ${\bf a^\ast}$ ($=2\pi/a$), 
${\bf b^\ast}$ ($=2\pi/b$) 
directions shown in square lattice notation. Our crystals are all highly twined so 
we cannot distinguish ${\bf a^\ast}$ from ${\bf b^\ast}$. Because the CuO$_2$ planes in 
YBa$_2$Cu$_3$O$_{6+x}$ actually appear in coupled bilayers that are separated by a much
larger distance between the bilayer, spin fluctuations
that are in-phase (acoustic or $\chi^{\prime\prime}_{ac}$) or out-of-phase (optical or 
$\chi^{\prime\prime}_{op}$) with respect to the neighboring plane will have
different spectra \cite{mignod,tranquada,mook93}.
It is therefore convenient to separate the magnetic response of YBa$_2$Cu$_3$O$_{6+x}$ 
into two parts 
$\chi^{\prime\prime}(q_x,q_y,q_z,\omega)=\chi^{\prime\prime}_{ac}(q_x,q_y,\omega)\sin^2(q_zd/2)+
\chi^{\prime\prime}_{op}(q_x,q_y,\omega)\cos^2(q_zd/2)$, where $d$ (3.342 \AA) is the spacing
between the nearest-neighbor CuO$_2$ planes along $c$. For the AF insulating 
parent compound YBa$_2$Cu$_3$O$_6$, $\chi^{\prime\prime}_{ac}$ and $\chi^{\prime\prime}_{op}$ 
correspond to the acoustic and optical spin-wave excitations, respectively \cite{stephen96,reznik}.
Because the interesting features 
such as the resonance and incommensurate fluctuations 
\cite{mignod,tranquada,mook93,sternlieb,fong95,bourges96,fong96,dai96,fong97,bourges97,dai98,mook98,mookn98,arai,hayden98,dai99}
 are only observed in the 
acoustic channel of spin susceptibility 
$\chi^{\prime\prime}_{ac}(q_x,q_y,\omega)$ and the optical fluctuations 
occur at higher energies than the acoustic excitations, we focused on the temperature,
frequency, and composition dependence of spin excitations in the acoustic mode which 
has a modulation $\sin^2(q_zd/2)$ along the $c$-axis. 

For this purpose, we used several 
crystal orientations and scattering geometries.  
Previous work by different groups 
\cite{mignod,tranquada,mook93,sternlieb,fong95,bourges96,fong96,dai96,fong97,bourges97,dai98,mook98,mookn98,arai,hayden98,dai99}
were focused on 
the behavior of magnetic excitations 
for ${\bf q}$ 
in the $(h,h,l)$ and/or $(h,3h,l)$ zone.
 In view of the fact that the low frequency 
spin fluctuations in YBa$_2$Cu$_3$O$_{6+x}$ peak 
at a quartet of incommensurate positions  
$\delta$ rlu away from $(1/2,1/2)$, $(\pi,\pi)$, as depicted in Fig. 1 \cite{mookn98}, we 
scanned along the arrow direction in Fig. 1  
from $(0,1/2)$ to $(1,1/2)$ with fixed $l$. 
To perform these scans, we first aligned the crystal 
with its [1,0,0] and [0,0,1] axes along the 
rotational axes of the 
lower ($\alpha$-axis) and upper ($\chi$-axis) arcs of the
spectrometer goniometer, respectively. The crystal was then rotated 
$\alpha=\arctan(c/2al)$ degrees such that 
after the rotation the $[1/2,0,0]$ and $[0,1/2,l]$ axes were in the horizontal scattering plane.
To access the maxima of acoustic spin fluctuations, we chose 
$l\approx 2$ or 5 rlu so the 
$\omega$ rotation angle would be approximately 37 or 17 degrees, respectively. 
One advantage of performing scans in this manner
is that the scan direction is along the direction of
incommensurate peaks and the narrow part of the resolution
ellipse. As shown in Fig. 1, the narrow part of the spectrometer resolution 
matches the separation of the incommensurate peaks. 
Therefore, this is a favorable geometry to
study the evolution of the incommensurate magnetic excitations.
In addition to the new experimental geometry, we also used the 
conventional $(h,3h,l)$ zone as specified in the figure captions.

\section{Results and discussion}
\subsection{Wave vector and energy dependence of the acoustic spin fluctuations}
We begin this section by noting that inelastic neutron
scattering experiments over the last decade 
\cite{mignod,tranquada,mook93,sternlieb,fong95,bourges96,fong96,dai96,fong97,bourges97,dai98,mook98,mookn98,arai,hayden98,dai99}
have provided valuable information about the evolution of the 
spin dynamics in YBa$_2$Cu$_3$O$_{6+x}$. The key features of the spin excitations
spectra include a gap in the superconducting state, a pseudogap in the normal state,
the resonance peak, and the incommensurate fluctuations below the resonance. 
The observation of remarkable similarities in spin incommensurate structure for the bilayer 
YBa$_2$Cu$_3$O$_{6.6}$ \cite{mookn98} and the single-layer 
La$_{2-x}$Sr$_x$CuO$_4$ of the same hole doping density \cite{yamada}
suggests that incommensurate magnetic fluctuations may be a universal feature of the
cuprate superconductors. For the La$_{2-x}$Sr$_x$CuO$_4$ system, 
Yamada and coworkers established the Sr-doping dependence of the incommensurability $\delta$ 
of the spin fluctuations and found a remarkable linear relationship between 
$T_c$ and $\delta$ \cite{yamada}. In the case of YBa$_2$Cu$_3$O$_{6+x}$, 
Balatsky and Bourges \cite{balatsky}
suggested that a similar linear relationship exists 
between $T_c$ and the 
width in momentum space ($q$) of the ``normal'' spin fluctuations, i.e., fluctuations at frequencies 
away from the resonance.
The authors argued that spin excitations of YBa$_2$Cu$_3$O$_{6+x}$ have two 
components and the low-energy incommensurate fluctuations are different than 
the commensurate resonance. The conclusions
of \cite{balatsky} were reached based mostly on data where explicit   
incommensurability in the low-frequency spin fluctuations was not observed. Therefore, it is still not clear
whether the linear relationship between $T_c$ and $\delta$ is a universal feature of
the cuprates. Below, 
we present our systematic investigation of spin excitations in YBa$_2$Cu$_3$O$_{6+x}$.
\subsubsection{YBa$_2$Cu$_3$O$_{6.45}$}
As stressed in Sec. IIB, it is not a trivial matter to separate  
the magnetic scattering from phonon and other spurious
processes in metallic crystals of YBa$_2$Cu$_3$O$_{6+x}$. 
Although the use of polarized neutrons can provide an unambiguous measure of
the magnetic and phonon scattering \cite{moon}, the technique usually   
suffers from the limited flux available with polarized neutrons. Nevertheless,
polarization analysis has distinctive advantages in separating the magnetic signal from 
phonon and other spurious processes. 
As an example, we show in Figure 3 the polarization analysis of the 
spin excitations for YBa$_2$Cu$_3$O$_{6.45}$ at temperatures above and below 
$T_c$. 
  
\begin{figure}
\includegraphics[width = 3 in]{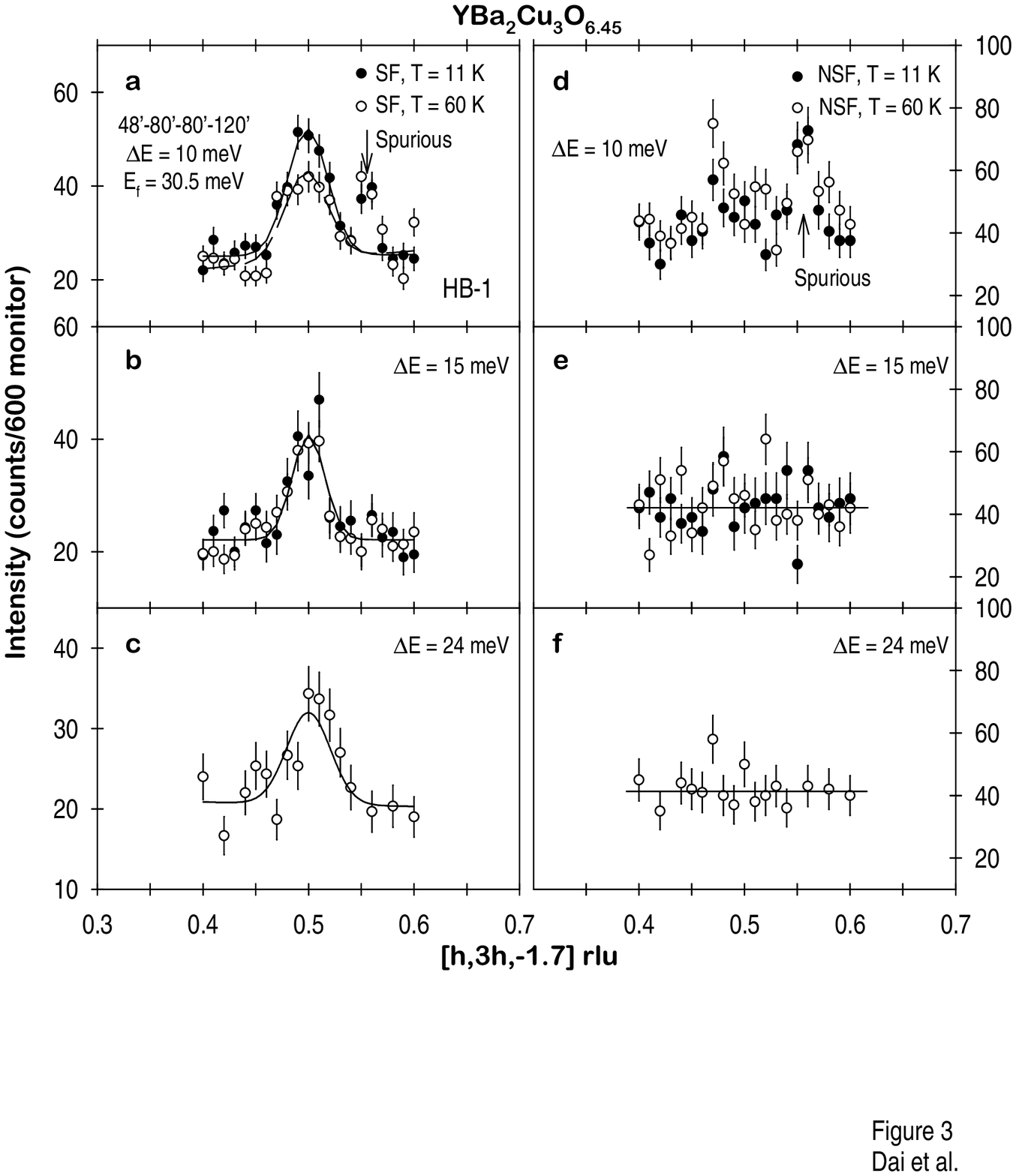}
\caption{
The spin-flip (SF) scattering of YBa$_2$Cu$_3$O$_{6.45}$ 
in the polarized scans along $(H,3H,-1.7)$ direction with $\hbar\omega=10$ meV (a), 
15 meV (b), and 24 meV (c) at 11 K and 60 K.
The nonspin-flip (NSF) scattering of the same scans at 
$\hbar\omega=10$ meV, 
15 meV, and 24 meV are shown in (d), (e), and (f), respectively. 
The solid and dashed lines in (a-c) are
Gaussian fits to the data on linear backgrounds. 
The solid lines in (e) and (f) are guides to the eye.
}
\end{figure}

Because magnetic fluctuations in YBa$_2$Cu$_3$O$_{6+x}$ are centered around $(\pi,\pi)$ 
with a 
sinusoidal modulation along the $(0,0,l)$ direction, we oriented the crystal in 
the $(h,3h,l)$ zone and searched for the magnetic signal 
  with scans along $(h,3h,-1.7)$ which 
corresponds to the maximum intensity of the acoustic modulation. 
Figures 3(a-c) show the raw 
data of SF scattering for $\hbar\omega=10$, 15, and 24 meV at 60 and 11 K.
The NSF scattering of the same scans are shown in Figs. 3(d-f). 
While SF scattering clearly peaks around $(\pi,\pi)$, there are no similar 
features in 
the NSF scattering at the same position. Therefore, it is clear that 
the scattering around $(\pi,\pi)$ in the SF scattering are magnetic in origin. 
To quantitatively parameterize 
the SF scattering data, 
we used a Gaussian line shape on a linear background
to fit the data as shown in solid lines in Figs. 3(a-c) \cite{mignod,tranquada,mook93,sternlieb,fong95,bourges96,fong96,dai96,fong97,bourges97}.
The scattering function $S({\bf q},\omega)$ for such line shape is related
to the in-plane dynamical spin susceptibility $\chi^{\prime\prime}_{ac}(q_x,q_y,\omega)$ by 
\begin{eqnarray}
S({\bf q},\omega)& = & I(\omega)\sin^2(q_zd/2)\chi^{\prime\prime}_{ac}(q_x,q_y,\omega) \nonumber \\
                 & = & I(\omega)\sin^2(q_zd/2)e^{\left[-{\Delta q^2/ 2\sigma^2}\right]},
\label{eq:three}
\end{eqnarray}
where $\Delta {\bf q}=2\pi(h-1/2,3h-3/2,0)/a$, $I(\omega)$ is the peak susceptibility and
$\sigma$ is related to the half width at half maximum (HWHM) of the excitations via 
${\rm HWHM}=\sigma\sqrt{2ln2}$.

\begin{figure}
\includegraphics[width = 3 in]{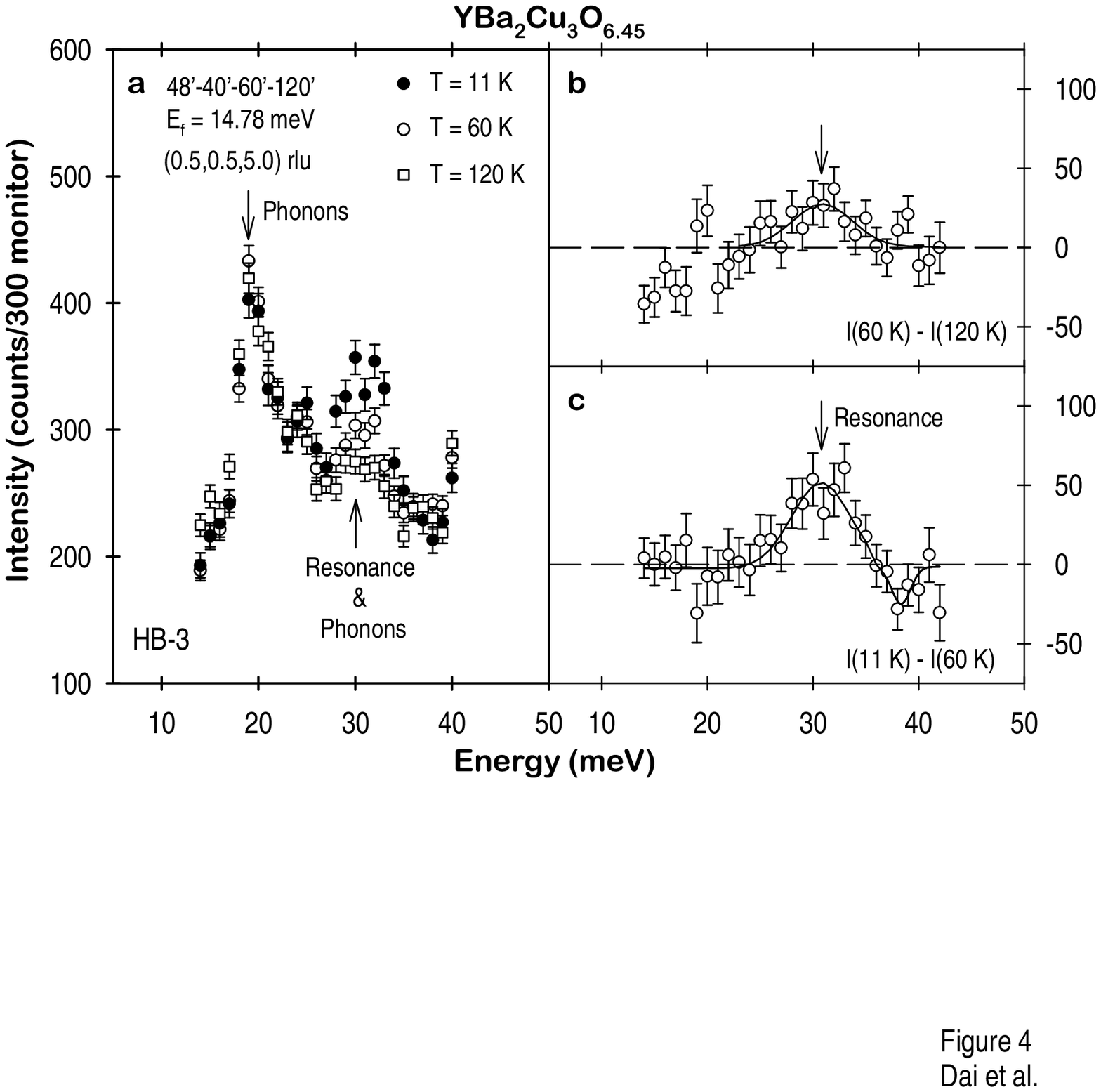}
\caption{
The unpolarized constant-${\bf q}$ scans at (0.5,0.5,5) rlu 
for YBa$_2$Cu$_3$O$_{6.45}$. Data were taken with a PG(002) monochromator
and PG analyzer with 40$^{\prime}$-40$^{\prime}$-60$^{\prime}$-120$^{\prime}$ 
collimation and $E_f=14.78$ meV. (a) Raw scattering at three different 
temperatures with phonons marked by arrows.  
(b) The temperature difference between 60 K ($T_c+12$ K) and 120 K. The positive scattering around 30 meV
is consistent with the precursor of the resonance in the normal state. (c) 
The temperature difference 
between 11 K ($T_c-37$ K) and 60 K. The solid lines are Gaussian fits to
the  data. The resonance intensity gain at 11 K is partially compensated by the 
loss of the spectral weight at energies above it.
}
\end{figure}

Although polarization analysis is excellent in identifying the magnetic nature of 
Gaussian peaks around $(\pi,\pi)$, the limited flux available with the technique
means that detailed wave vector and energy dependence of the scattering need to 
be obtained with unpolarized neutrons. For this purpose, we realigned the sample in the 
$[1/2,0,0]$ and $[0,1/2,l]$ zone. Figure 4 shows the constant-$q$ scans at various 
temperatures at $(\pi,\pi)$. The spectra consist of two major peaks at $\sim19$ 
and 31 meV. While the intensity of the $\sim19$ meV mode follows the Bose 
population factor $[n(\omega)+1]$ consistent
with phonons, the scattering at $\sim30$ meV decreases with warming which is indicative of its
magnetic origin. As we shall demonstrate below, phonons are also present at 
$\sim31$ meV. However, phonon scattering in this energy range 
is expected to change negligibly with temperature for $T\leq 120$ K. Therefore,  
the difference spectra in Fig. 4 can be regarded as changes in 
the dynamical susceptibility. Clearly, there are spectral weight enhancements around 30 meV 
with decreasing temperature above and below $T_c$. 
If we define the susceptibility gain on cooling above $T_c$ at the resonance energy as
the precursor of the resonance, we find that the spectral weight of the resonance 
increases slightly from the normal to the superconducting state [Fig. 4(b) and (c)] and  
is only partially 
compensated by the loss at energies above it [Fig. 4(c)].

\begin{figure}
\includegraphics[width = 3 in]{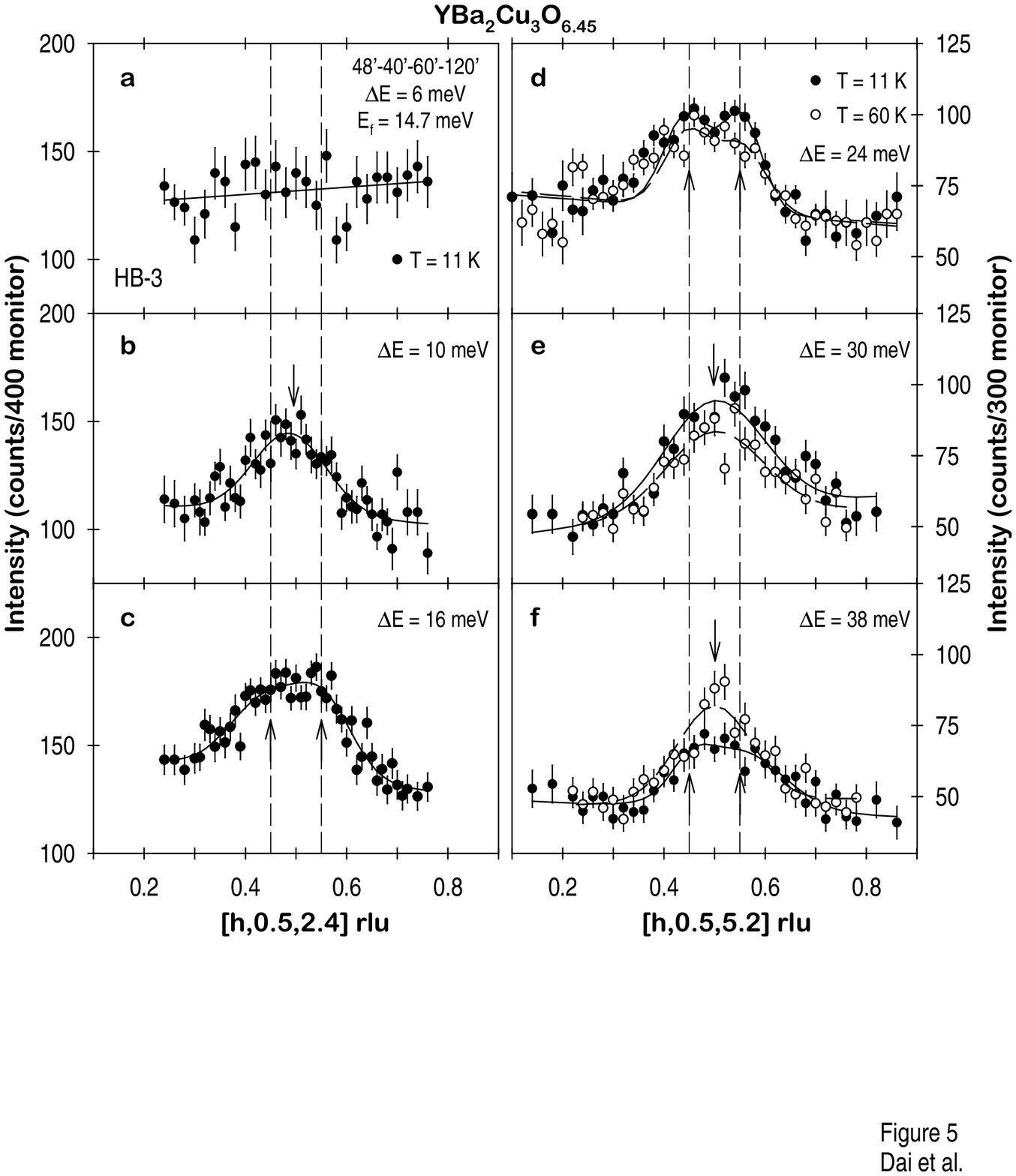}
\caption{
Constant-energy scans along the $[h,0.5,2.4]$ (a-c) 
and $[h,0.5,5.2]$ (d-f) directions. The energy transfers are $\hbar\omega=6$ meV (a),
10 meV (b), 16 meV (c), 24 meV (d), 30 meV (e), and 38 meV (f).
Incommensurate spin fluctuations are clearly observed at $\hbar\omega=24$ meV 
both above and below $T_c$. In the superconducting state, the flattish top profiles 
are observed at energies above and below the resonance energy. 
Solid lines are Gaussian fits to the data in (b-f).
}
\end{figure}

Figure 5 shows the momentum dependence of the response at energies above and below the 
magnetic resonance. At $\hbar\omega=6$ meV, the scattering [Fig. 5(a)] is featureless around $(\pi,\pi)$,
suggesting the presence of a spin gap in the superconducting state \cite{pseudogap}. 
On increasing the energy to 10 meV, the scattering shows a broad peak that can be 
well described by a single Gaussian centered at $(\pi,\pi)$. At $\hbar\omega=16$ meV,
the profile around $(\pi,\pi)$ develops a flattish top similar to previous observations 
\cite{tranquada,sternlieb}. Although detailed analysis suggests that the profile is better 
described by a pair of peaks rather than a single Gaussian, well defined incommensurate peaks 
at $\delta=0.054\pm0.004$ rlu 
from $(\pi,\pi)$ are only observed 
 at $\hbar\omega=24$ meV [Fig. 5(d)]. At the resonance energy of 
$\sim30$ meV, commensurate profiles are observed above and below $T_c$ [Fig. 5(e)]. 
For an energy above the resonance (at 38 meV), the peak appears to be commensurate 
in the normal state, but changes to a flat top (incommensurate) profile in the 
low-temperature superconducting state [Fig. 5(f)].

\subsubsection{YBa$_2$Cu$_3$O$_{6.5}$, YBa$_2$Cu$_3$O$_{6.6}$, and YBa$_2$Cu$_3$O$_{6.7}$}

In this section, we discuss results on three samples with superconducting 
transition temperatures of 52 K, 62.7 K, and 74 K.   
The fractional hole concentration
per Cu atom in the CuO$_2$ sheet, $p$, 
for YBa$_2$Cu$_3$O$_{6+x}$ around 
the 60-K plateau phase is close to $\sim0.1$ \cite{tallon}.
For the single-layer superconducting La$_{1.9}$Sr$_{0.1}$CuO$_4$ which also has $p=0.1$, 
the incommensurability of the spin fluctuations is at $\delta = 0.1$ \cite{yamada}.
The discovery of the same incommensurate structure and $\delta$ ($\approx 0.1$) for
the equivalent hole concentration ($p \approx 0.1$) of 
the bilayer YBa$_2$Cu$_3$O$_{6.6}$ \cite{mookn98} has generated 
considerable interests because such observation suggests that incommensurate
spin fluctuations  may be the common feature of these two most studied families of cuprates.

For La$_{2-x}$Sr$_{x}$CuO$_4$, it is generally believed that
the presence of dynamic stripes is 
the microscopic origin of the observed 
incommensurate spin fluctuations \cite{tranquada95}.
Thus, if the incommensurate spin fluctuations in YBa$_2$Cu$_3$O$_{6+x}$ have 
similar behavior as La$_{2-x}$Sr$_{x}$CuO$_4$ at all doping concentrations,
it is likely that dynamic stripes should also be present in the bilayer cuprates. 
Although the explicit incommensurate structure \cite{mookn98} 
and the one-dimensional nature of the 
incommensurate spin fluctuations in 
YBa$_2$Cu$_3$O$_{6.6}$ \cite{mook00} favor the dynamic striped-phase interpretation,
the detailed energy and doping dependence of the incommensurate spin fluctuations 
is still lacking. In the discussions below, we attempt to remedy this situation. 

\begin{figure}
\includegraphics[width = 3 in]{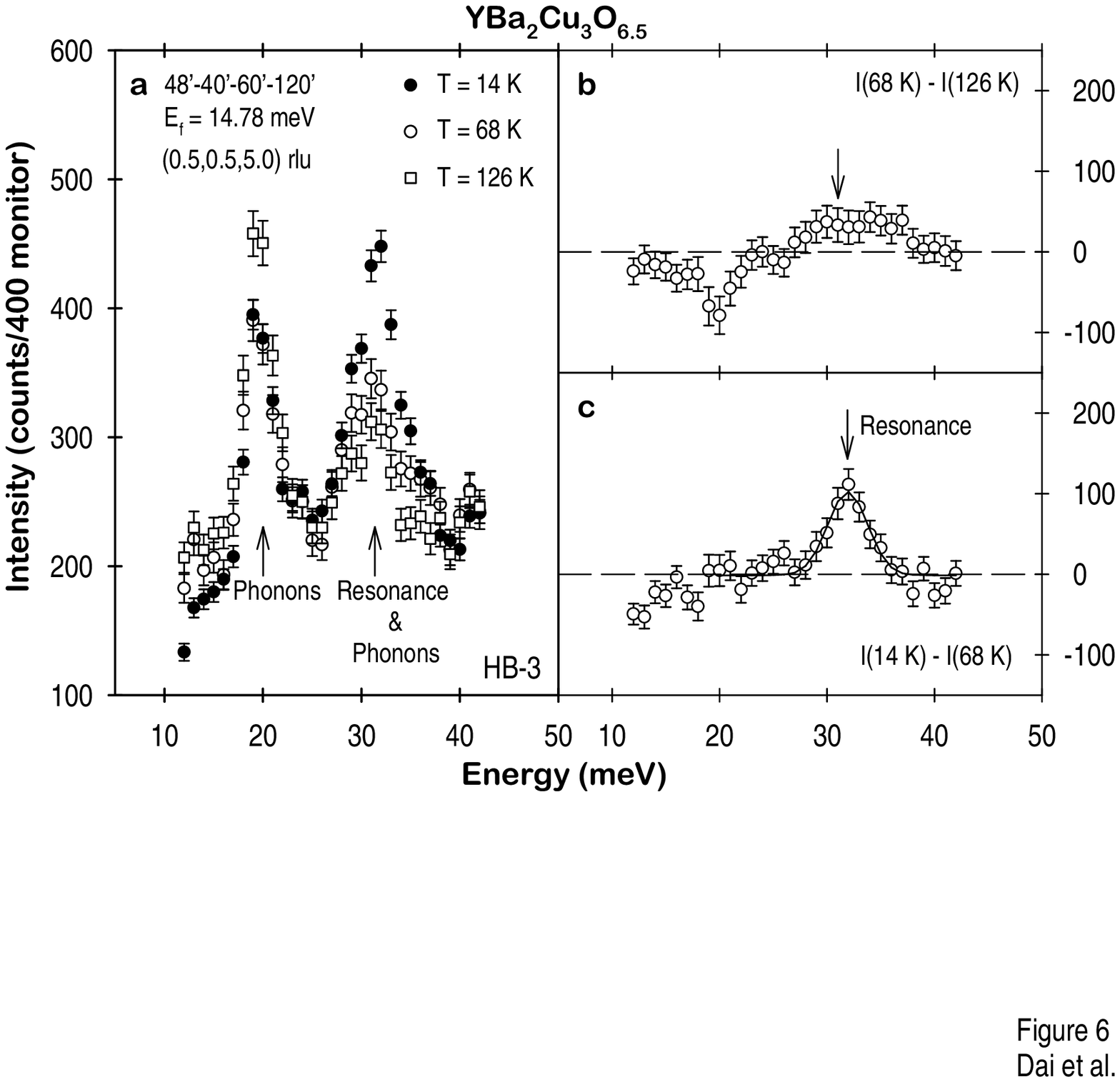}
\caption{
Constant-${\bf q}$ scans at (0.5,0.5,5) rlu  
for YBa$_2$Cu$_3$O$_{6.5}$.  (a) Raw scattering at three different 
temperatures with phonons and resonance marked by arrows. 
(b) The temperature difference between 68 K ($T_c+16$ K) and 126 K. The positive scattering around 32 meV
is consistent with the precursor of the resonance in the normal state. (c) The temperature 
difference between 14 K ($T_c-38$ K) and 68 K. The solid lines are Gaussian fits to the 
data.
}
\end{figure}

We first describe measurements on YBa$_2$Cu$_3$O$_{6.5}$. 
Although the rocking curves of all other crystals of YBa$_2$Cu$_3$O$_{6+x}$  
show single Gaussian peaks with mosaic spreads of $\sim$1--2.5 degrees,
the rocking curve around $(0,0,6)$ Bragg reflection for YBa$_2$Cu$_3$O$_{6.5}$
displays two distinct peaks separated by $\sim$3 degrees. This 
means that the YBa$_2$Cu$_3$O$_{6.5}$ sample is composed of two major grains
$\sim$3 degrees apart.
Figure 6 summarizes constant-$q$ scans at various temperatures at $(\pi,\pi)$ 
for YBa$_2$Cu$_3$O$_{6.5}$. While the $\sim$20 meV phonons change 
negligibly from those for YBa$_2$Cu$_3$O$_{6.45}$ (Fig. 4), there are 
two features in 
the raw spectra of Fig. 6(a) worth noting. First, the 
resonance energy shifted from $\sim$30.5 meV for YBa$_2$Cu$_3$O$_{6.45}$ 
to $\sim$32 meV for YBa$_2$Cu$_3$O$_{6.5}$. 
In addition, the spectral weight of the resonance relative to that of the 20 meV phonons
is clearly enhanced in YBa$_2$Cu$_3$O$_{6.5}$. This observation indicates that the intensity gain 
of the resonance in this material below $T_c$ is considerably bigger than that for
YBa$_2$Cu$_3$O$_{6.45}$. Similar to YBa$_2$Cu$_3$O$_{6.45}$, 
there are magnetic spectral weight enhancements around the resonance energy 
above and below $T_c$. High intensity, low-resolution measurements in the $[h,3h,l]$ zone
confirm this conclusion (Fig. 7). 

\begin{figure}
\includegraphics[width = 3 in]{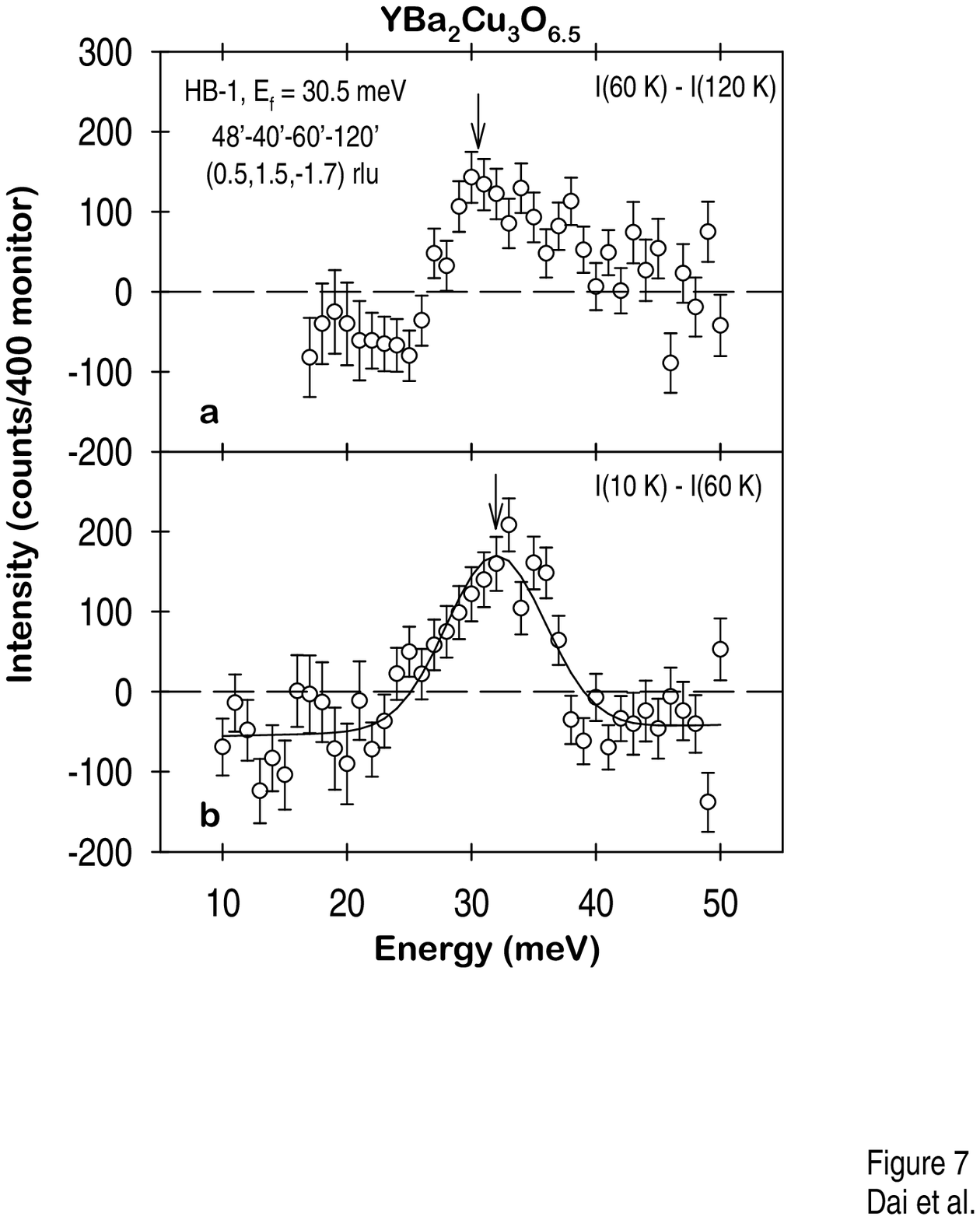}
\caption{
Constant-${\bf q}$ scans at (0.5,1.5,-1.7) rlu above and 
below $T_c$ for YBa$_2$Cu$_3$O$_{6.5}$ with the crystal in the $[h,3h,l]$ zone.  
(a) Difference scattering between 60 K ($T_c+8$ K) and 120 K. 
(b) Difference scattering 
between 10 K ($T_c-42$ K) and 60 K. The solid lines are Gaussian fits to
the  data.
}
\end{figure}

To determine the magnitude of the spin gap and search for incommensurate 
spin fluctuations, we 
performed constant-energy scans at various energies along the $[h,0.5,5]$ direction above 
and below $T_c$ as shown in Fig. 8.
At 10 and 16 meV, the scattering shows weak features around $(\pi,\pi)$. 
This is different from YBa$_2$Cu$_3$O$_{6.45}$ where 
spin fluctuations centered around $(\pi,\pi)$ 
are found at energies above 10 meV (Figs. 3 and 5). At 24 meV [Fig. 8(c)],
the scattering displays a flattish top in both the normal and superconducting
states \cite{notemosaic}. At energies near 
[27 meV, Fig. 8(d)] and at the resonance [31.5 meV, Fig. 8(e)], 
the scattering narrows in width and increases in intensity. For an energy 
above the resonance, flattish top profiles 
indicative of incommensurate fluctuations are observed at 40 meV as shown in Fig. 8(f),
which is similar to Fig. 5(f) of YBa$_2$Cu$_3$O$_{6.45}$. 

\begin{figure}
\includegraphics[width = 3 in]{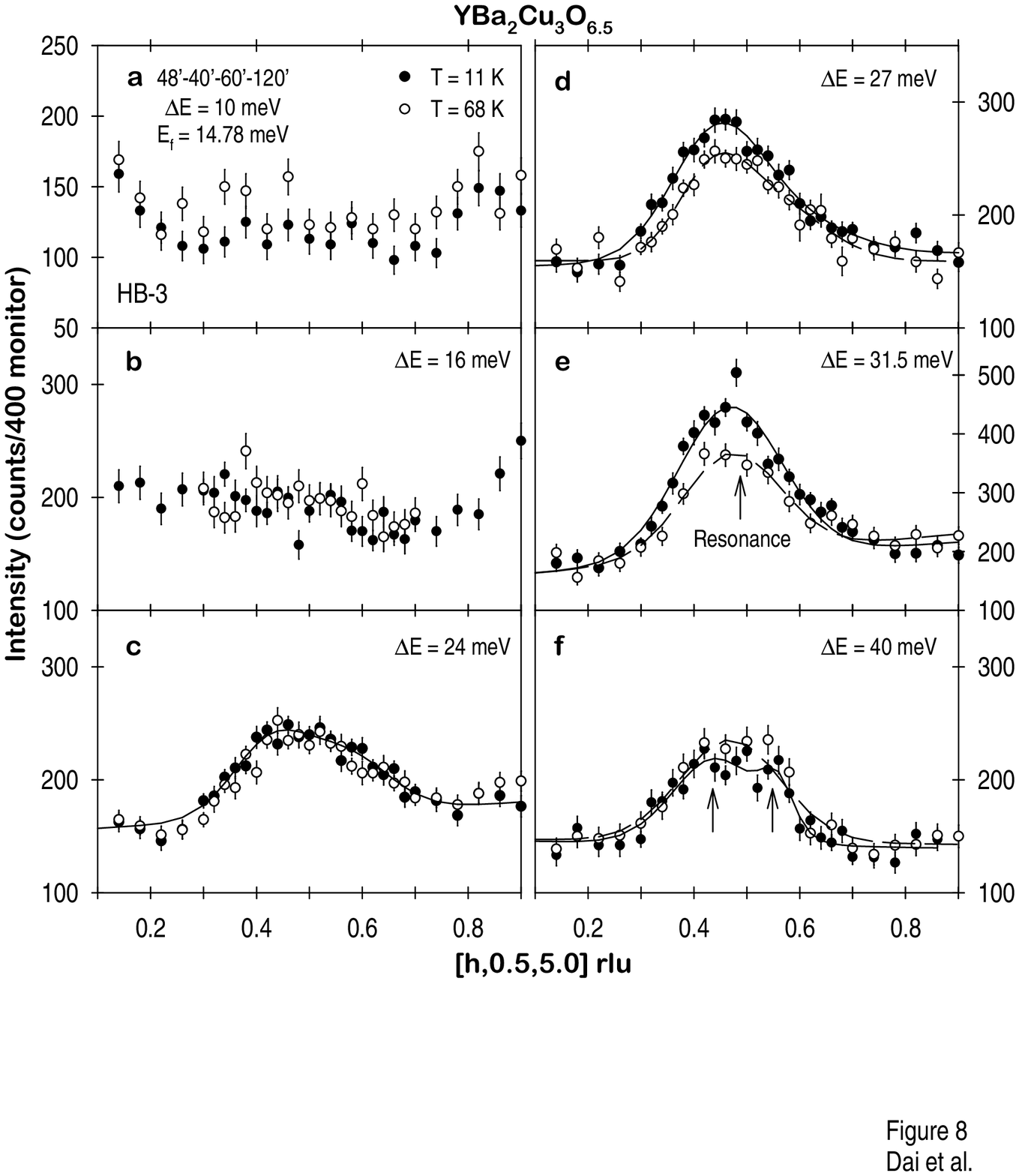}
\caption{
Constant-energy scans along the  
$[h,0.5,5.0]$ direction for YBa$_2$Cu$_3$O$_{6.5}$. 
The energy transfers are $\hbar\omega=10$ meV (a),
16 meV (b), 24 meV (c), 27 meV (d), 31.5 meV (e), and 40 meV (f).
The flattish top profiles indicative of the 
incommensurate spin fluctuations are observed at $\hbar\omega=24$ meV 
both above and below $T_c$. In the superconducting state, flattish top profiles 
are also observed above the resonance $\hbar\omega=40$ meV. 
The solid lines are Gaussian fits to the data, and the 
arrows in (c) and (f) mark the fitted incommensurate positions.
}
\end{figure}

Next we describe measurements on YBa$_2$Cu$_3$O$_{6.6}$ where the incommensurate 
spin fluctuations at 24 meV and the resonance at 34 meV 
have been extensively studied \cite{dai96,dai98,mook98,mookn98,dai99}. For completeness,
we plot in Fig. 9 the difference spectra of constant-$q$ scans showing the 
presence of the 34 meV resonance above and below $T_c$. 
Detailed temperature evolution of the 
resonance and the magnitude of the spin gap has already been reported \cite{dai99}.
To establish the energy dependence of the incommensurate spin fluctuations, we 
carried out high-resolution measurements using Be(002) as the 
monochromator and fixed the final neutron energy at 14.78 meV. Figure 10 shows a 
series of constant-energy scans at energies below and above the 34 meV resonance.
Consistent with previous results \cite{dai98,mook98,mookn98}, constant-energy scans
at 24 meV [Fig. 10(a)] show a pair of well-defined peaks separated by $2\delta=0.20\pm 0.01$
rlu in the low-temperature superconducting state and a flattish top profile just 
above $T_c$. Although the same scans at 27 meV and 30 meV 
also exhibit the same double peak structure below $T_c$ [Figs. 10(b) and (c)],
the separation between the peaks or the incommensurabilities clearly decrease
with increasing energy. In addition, the normal state flat top profile at 24 meV
is replaced by a single Gaussian peak centered around $(\pi,\pi)$ at 30 meV. 
As energy increases, the scattering profile 
continues to sharpen, with the 
narrowest width and highest intensity profile 
occuring at the resonance energy  [Fig. 10(e)].
When moving to energies above the resonance [Figs. 10(f-h)], the profiles increase 
in width and decrease in intensity. At 41 meV, the scattering becomes incommensurate 
in both the normal and superconducting states [Fig. 10(h)]. 

\begin{figure}
\includegraphics[width = 3 in]{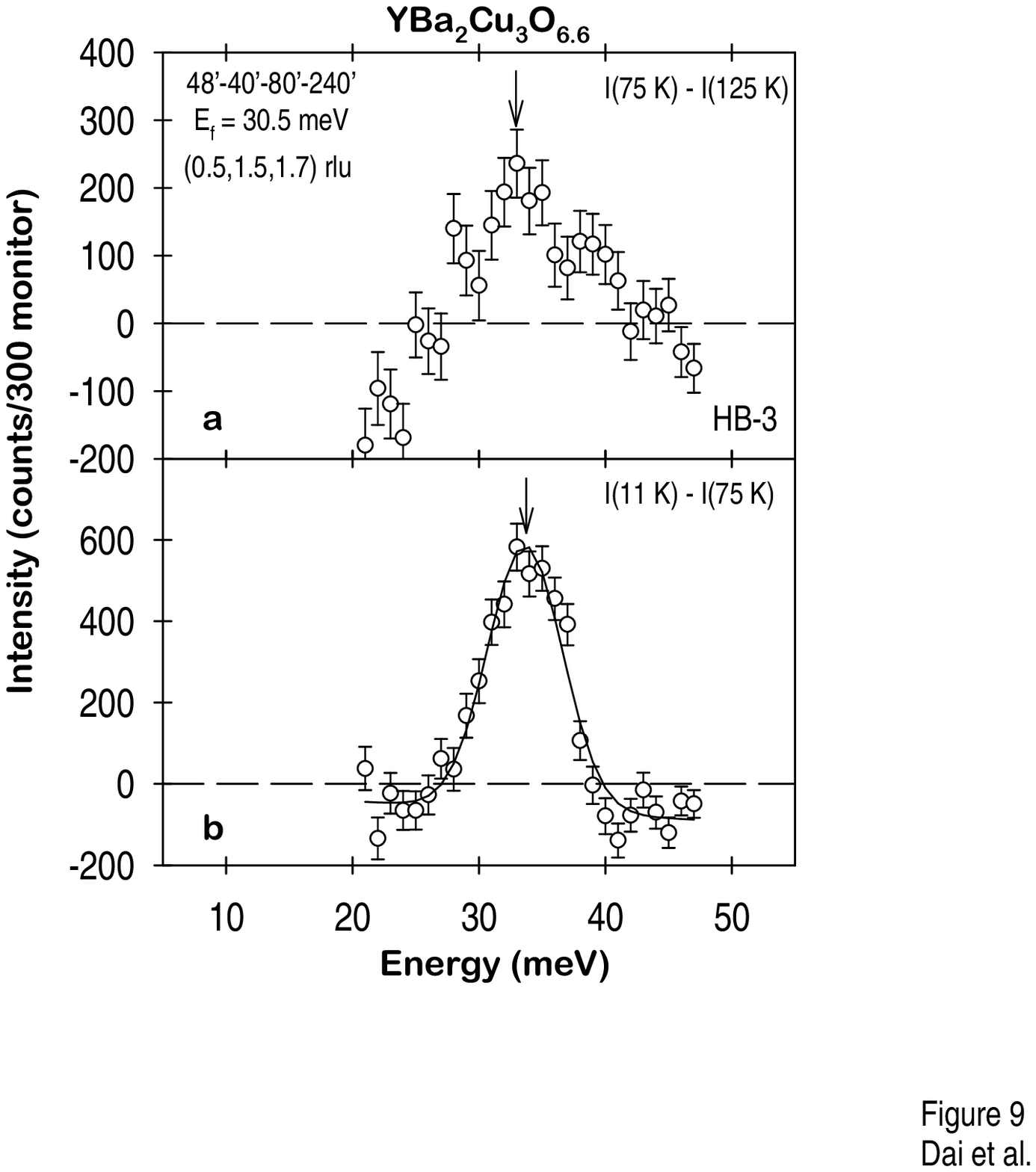}
\caption{
Constant-${\bf q}$ scans at (0.5,1.5,1.7) rlu 
above and below $T_c$ for YBa$_2$Cu$_3$O$_{6.6}$ with the crystal in the $[h,3h,l]$ zone.  
(a) The temperature difference between 75 K ($T_c+12$ K) and 125 K. 
(b) The temperature difference between 11 K ($T_c-52$ K) and 75 K. The solid line is a Gaussian fit
to the  data.
}
\end{figure}

\begin{figure}
\includegraphics[width = 3 in]{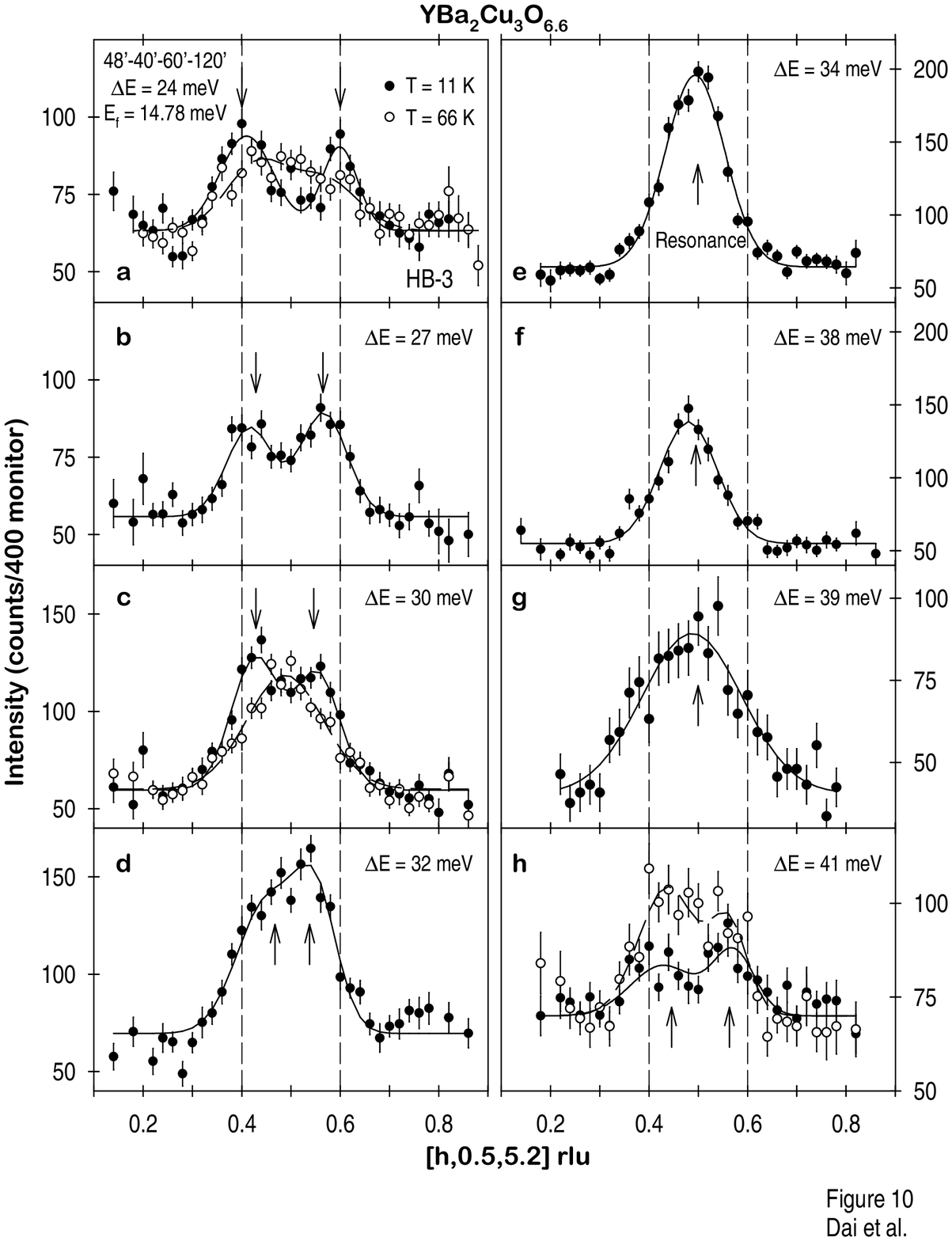}
\caption{
Constant-energy scans along the  
$[h,0.5,5.2]$ direction for YBa$_2$Cu$_3$O$_{6.6}$. 
The data were taken with a Be(002) monochromator
and a PG analyzer with a neutron final energy fixed at $E_f=14.78$ meV. 
The energy transfers are $\hbar\omega=24$ meV (a),
27 meV (b), 30 meV (c), 32 meV (d), 34 meV (e), 38 meV (f), 39 meV (g), and 41 meV (h).
The flattish top profiles indicative of the 
incommensurate spin fluctuations are clearly observed in the normal state 
at $\hbar\omega=24$ meV and 41 meV. However, at $\hbar\omega=30$ meV, 
the scattering is commensurate in the normal state. 
In the superconducting state, the incommensurability of spin fluctuations decreases 
with increasing energy as the resonance is approached and splits up again at energies 
above the resonance.
The solid lines are Gaussian fits to the data.
}
\end{figure}

Finally, we discuss measurements for 
YBa$_2$Cu$_3$O$_{6.7}$. Figure 11 shows the result of
constant-$q$ scans above and below $T_c$. As the energy of 
the resonance moves up to 37 meV, the phonon scattering at 31 meV becomes 
more prominent. Constant-energy scans at several energies below the resonance 
are shown in Fig. 12. In contrast to the well-defined double peaks for
YBa$_2$Cu$_3$O$_{6.6}$ [Fig. 10(a)], the scattering at 24 meV and 27 meV 
[Figs. 12(a) and (b)] shows a very 
weak structure above and below $T_c$. For 
YBa$_2$Cu$_3$O$_{6.6}$, the susceptibility increases on cooling from 
the normal to the superconducting state at the incommensurate positions, 
accompanied by a suppression of fluctuations at the commensurate point \cite{dai98}.
In an attempt to determine whether the same is true for YBa$_2$Cu$_3$O$_{6.7}$,
we plot the difference spectra between 11 K and 80 K. 
As emphasized
in an earlier study \cite{dai98}, these difference spectra can be simply  regarded as changes in the
dynamical susceptibility, i.e., 
$\chi^{\prime\prime}(11 K)-\chi^{\prime\prime}(80 K)$. Figures 12(d) 
and (e) show the outcome of this subtraction for 24 meV and 27 meV, respectively. 
For $\hbar\omega=24$ meV, the difference spectrum shows negative 
scattering centered around $(\pi,\pi)$ with no evidence
of susceptibility gain at the expected incommensurate positions. 
This is consistent with the opening of a 24 meV gap in the spin fluctuations
spectra below $T_c$. 
At $\hbar\omega=27 $ meV, both the raw data [Fig. 12(b)] and the  
difference spectrum show evidence of incommensurate spin fluctuations
at $\delta\approx 0.1$ rlu, consistent with previous measurements 
using an integrated technique that gives a 
incommensurability of $\delta=0.11\pm 0.01$ \cite{mook98}. 
However, the clarity of incommensurate 
spin fluctuations is weak as compared to that for YBa$_2$Cu$_3$O$_{6.6}$.
Experiments on a different crystal of 
YBa$_2$Cu$_3$O$_{6.7}$ by Arai {\it et al.} \cite{arai} 
show clear incommensurate spin fluctuations at $\delta=0.11\pm 0.01$ rlu.   
At energies close to the resonance ($\hbar\omega=34$ meV and 37 meV), 
the scattering 
[Figs. 12(c) and (f)] shows conventional Gaussian profiles.

\begin{figure}
\includegraphics[width = 3 in]{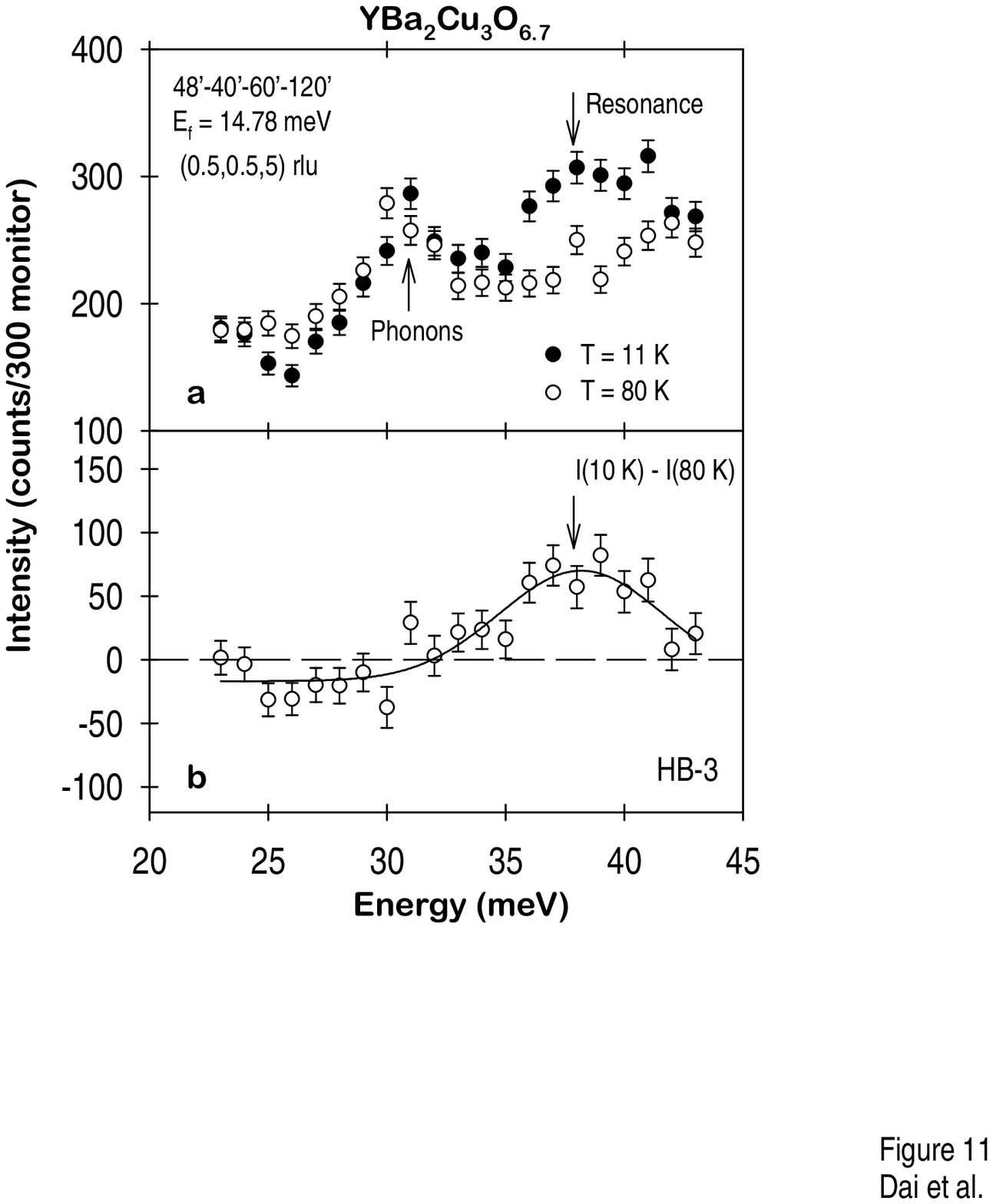}
\caption{
Constant-${\bf q}$ scans at (0.5,0.5,5) rlu above and below $T_c$ 
for YBa$_2$Cu$_3$O$_{6.7}$.  (a) Raw scattering at two different 
temperatures with phonons and resonance marked by arrows. 
(b) The temperature difference between 10 K ($T_c-64$ K) and 80 K. The positive scattering around 37 meV
shows the resonance. The solid line is a Gaussian fit to the 
data.
}
\end{figure}

\begin{figure}
\includegraphics[width = 3 in]{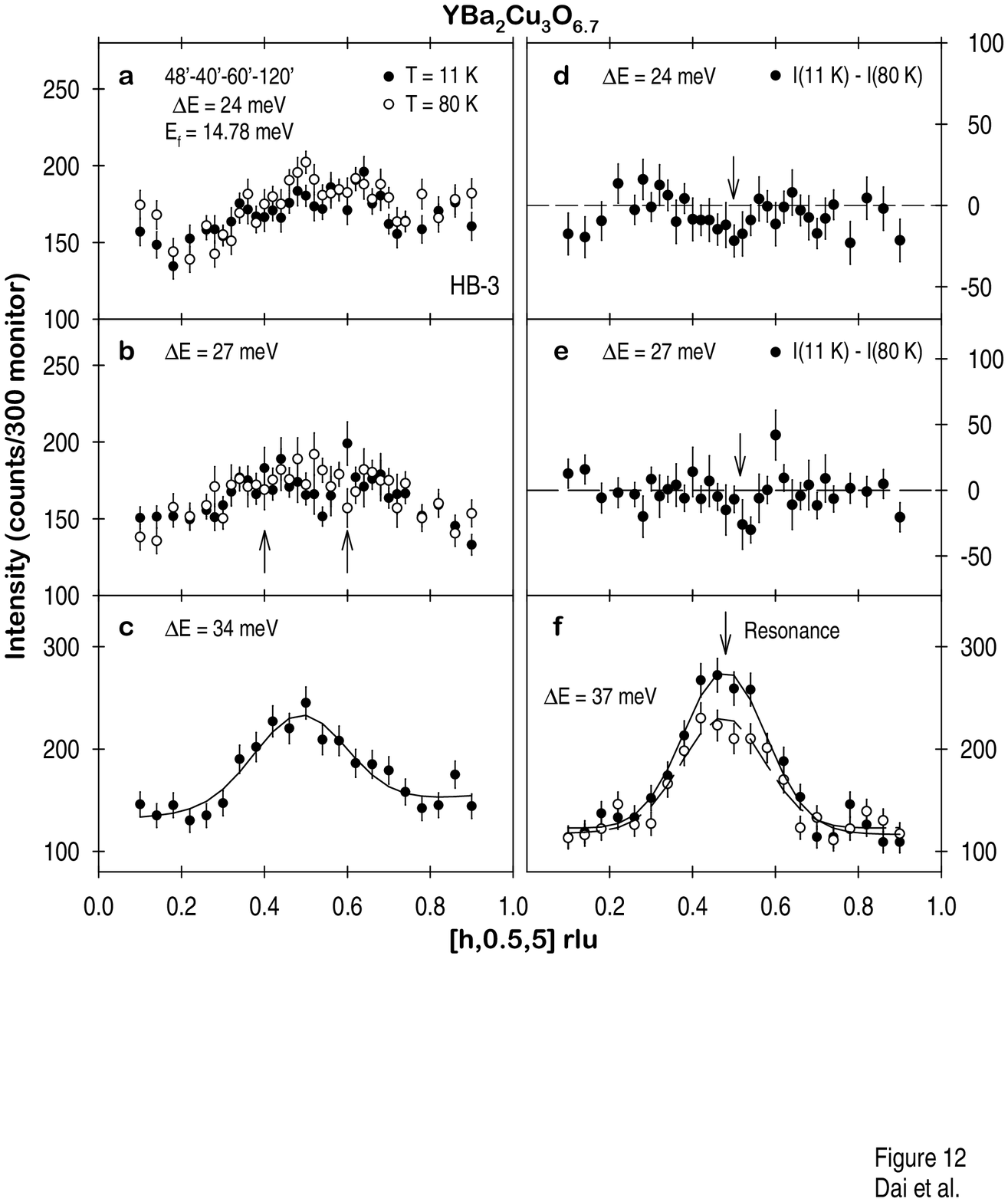}
\caption{
Constant-energy scans along the  
$[h,0.5,5.0]$ direction for YBa$_2$Cu$_3$O$_{6.7}$. The energy transfers are $\hbar\omega=24$ meV (a),
27 meV (b), 34 meV (c), and 37 meV (f). (d) and (e) show the difference spectra
between 11 K and 80 K. 
In the superconducting state, 
weak incommensurate peaks are observed
at $\hbar\omega =27$ meV (b), and a spin gap opens up at 24 meV (d). 
The solid lines are Gaussian fits to the data, and the 
arrows in (b) and (e) indicate the incommensurate positions. The integrated 
method described in Ref. [23] confirms the present measurements. 
}
\end{figure}

\subsubsection{YBa$_2$Cu$_3$O$_{6.8}$}

Much of the discussion in this section will be focused on our 
analysis of separating phonon and spurious events from magnetic scattering
for energies below the resonance. While the magnetic 
signal usually spreads throughout the energy and momentum space 
at high temperatures and becomes unobservable at any particular $\hbar\omega$-$Q$ point, 
the intensity of phonons should follow the Bose statistics with 
increasing temperature. Therefore, careful measurements as a 
function of temperature and wave vector should allow the distinction of 
magnetic scattering from phonons.

\begin{figure}
\includegraphics[width = 3 in]{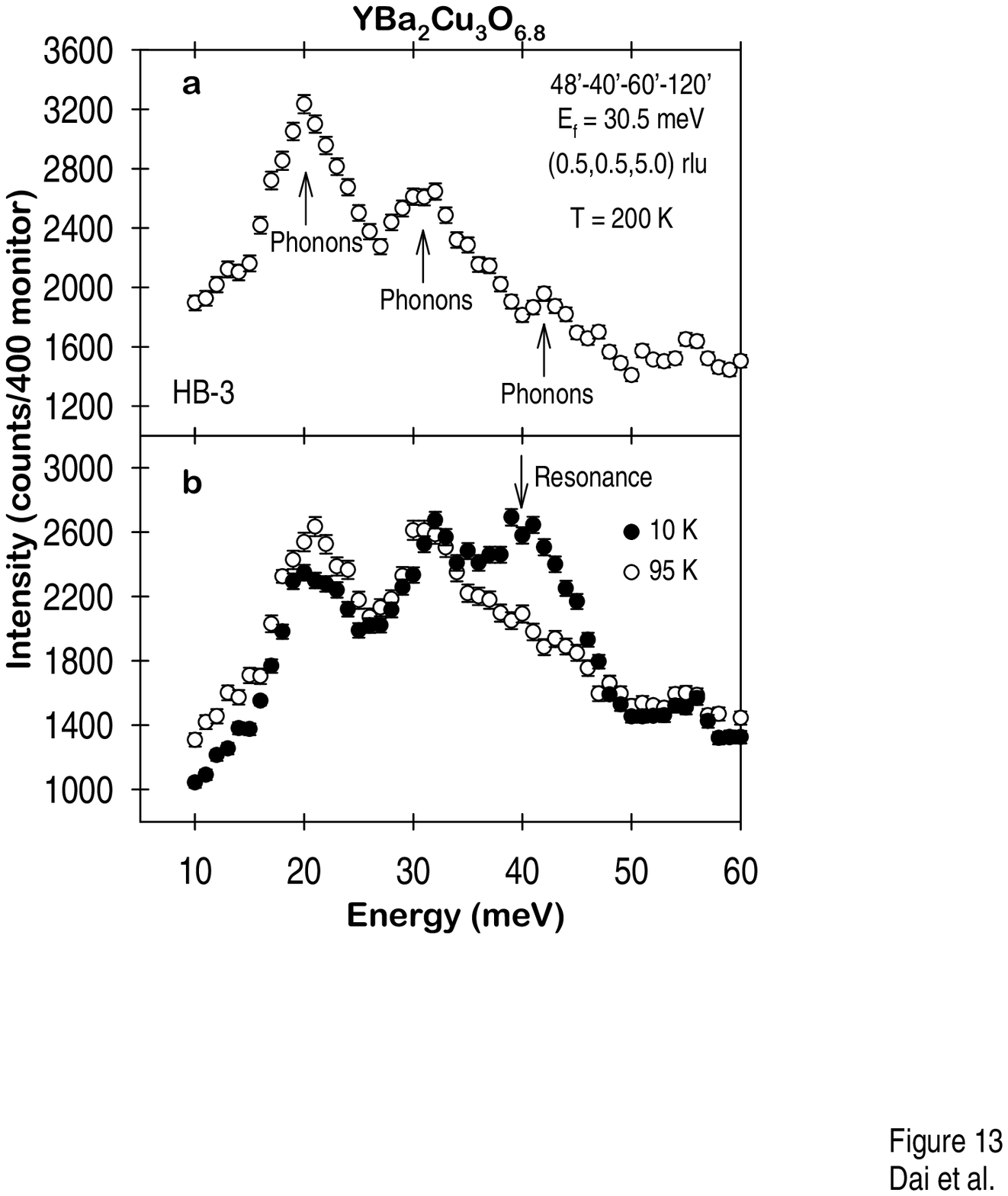}
\caption{
Constant-${\bf q}$ scans at (0.5,0.5,5) rlu  
for YBa$_2$Cu$_3$O$_{6.8}$.  (a) Raw scattering at 200 K
with phonons marked by arrows. 
(b) The same scans at 10 K and 95 K with the resonance marked by an arrow.
}
\end{figure}

\begin{figure}
\includegraphics[width = 3 in]{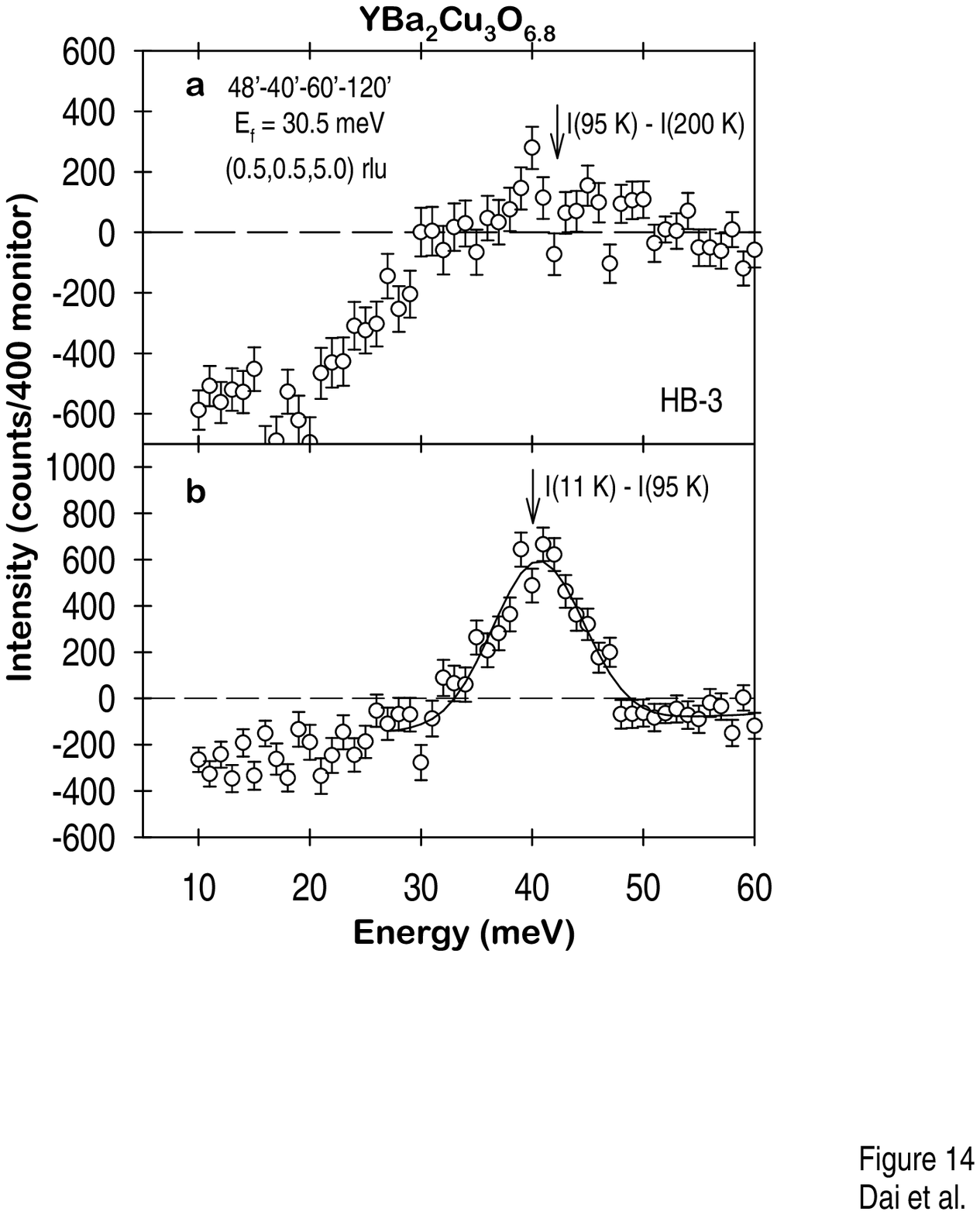}
\caption{
Difference spectra in constant-${\bf q}$ scans at (0.5,0.5,5) rlu 
above and below $T_c$  
for YBa$_2$Cu$_3$O$_{6.8}$. 
(a) The temperature difference between 95 K ($T_c+13$ K) and 200 K. The positive scattering around 40 meV
is consistent with the precursor of the resonance in the normal state. (b) The temperature 
difference between 11 K ($T_c-71$ K) and 95 K. The solid line is a Gaussian fit to the 
data.
}
\end{figure}

\begin{figure}
\includegraphics[width = 3 in]{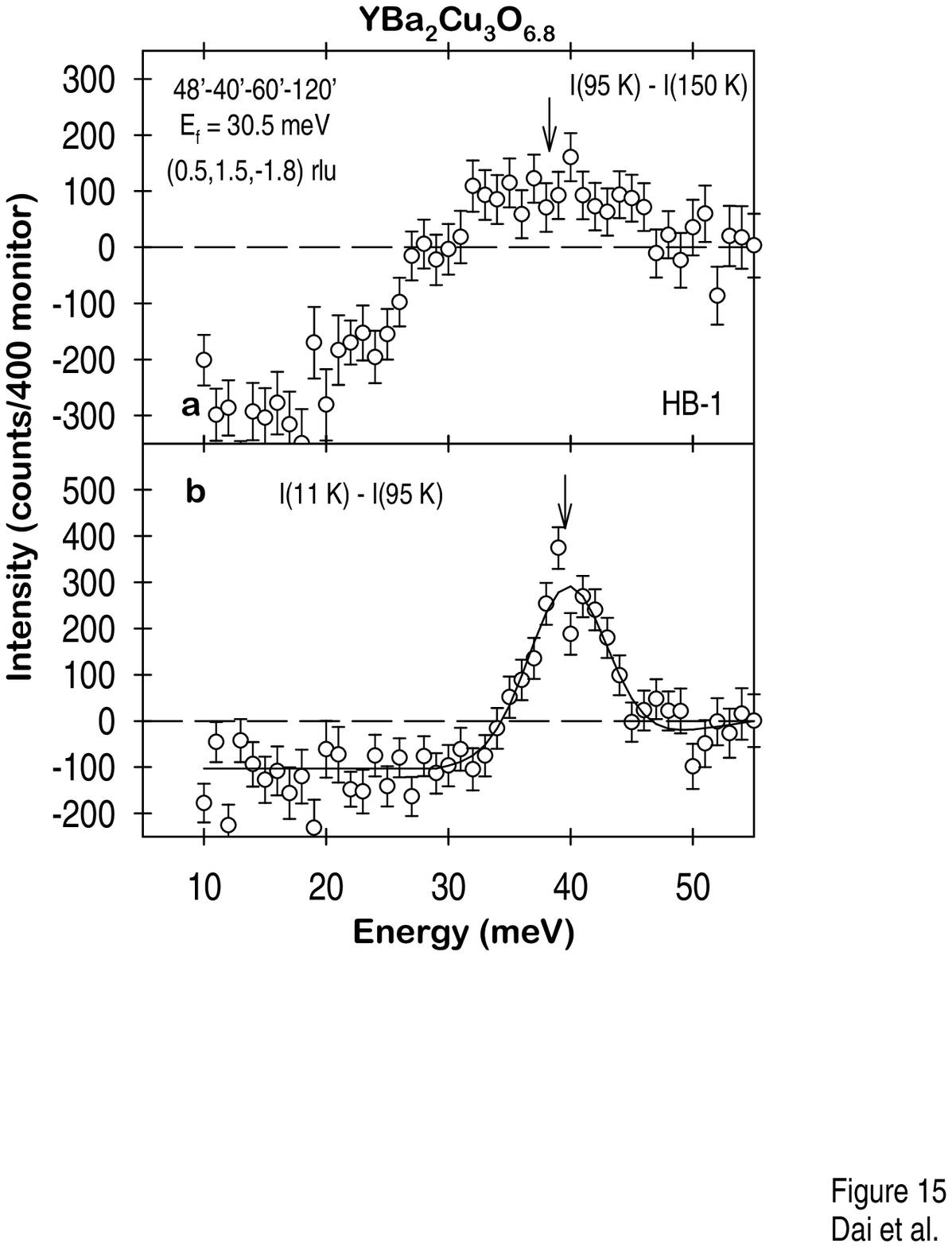}
\caption{
Difference spectra in constant-${\bf q}$ scans at (0.5,1.5,-1.8) rlu 
above and below $T_c$  
for YBa$_2$Cu$_3$O$_{6.8}$ with the sample in the $(h,3h,l)$ zone. 
(a) The temperature difference between 95 K ($T_c+13$ K) and 150 K. The positive scattering around 40 meV
again confirms the presence of the precursor for the resonance in the normal state. (b) 
The temperature difference between 11 K ($T_c-71$ K) and 95 K. The solid line is a Gaussian fit to
the  data.
}
\end{figure}

As an example, we show in Fig. 13 raw constant-$Q$ scans at the 
wave vector ${\bf q}=(0.5,0.5,5)$ rlu at various temperatures. The comparison of the 
results for temperatures above and below $T_c$ [Fig. 13(b)] 
indicates a clear enhancement of the scattering at $\hbar\omega\approx 40$ meV,
consistent with the presence of the resonance.
In addition to the enhancement of the resonance, there are general intensity 
reductions in the 
scattering below $\sim30$ meV from the normal to the superconducting state. 
Such intensity reductions at low energies are likely due to the drop in 
the single- and multi- phonon scattering although a simultaneous reduction in magnetic signal cannot
be ruled out. Nevertheless, the strong enhancement of the scattering at 
$\hbar\omega=$ $\sim20$,
$\sim31$,  and $\sim41$ meV at 200 K [Fig. 13(a)] indicates that the peaks in the constant-$Q$
scans  at these energies are mostly phonons in origin. 
To demonstrate the enhancement of the 
magnetic scattering on cooling above and below $T_c$, we show  
the difference spectra between low and high temperatures 
in two experimental geometries in 
Figs. 14 and 15. The data reveal two 
important features: first, the resonance in this underdoped compounds 
occurs at $\sim40$ meV, almost identical to the optimally and overdoped 
doped compound \cite{mignod,mook93,fong95}; second, there is an enhancement of the 
magnetic signal around the resonance energy above $T_c$, consistent with the 
presence of a precursor for the resonance in the underdoped
YBa$_2$Cu$_3$O$_{6+x}$ \cite{dai99}.

\begin{figure}
\includegraphics[width = 3 in]{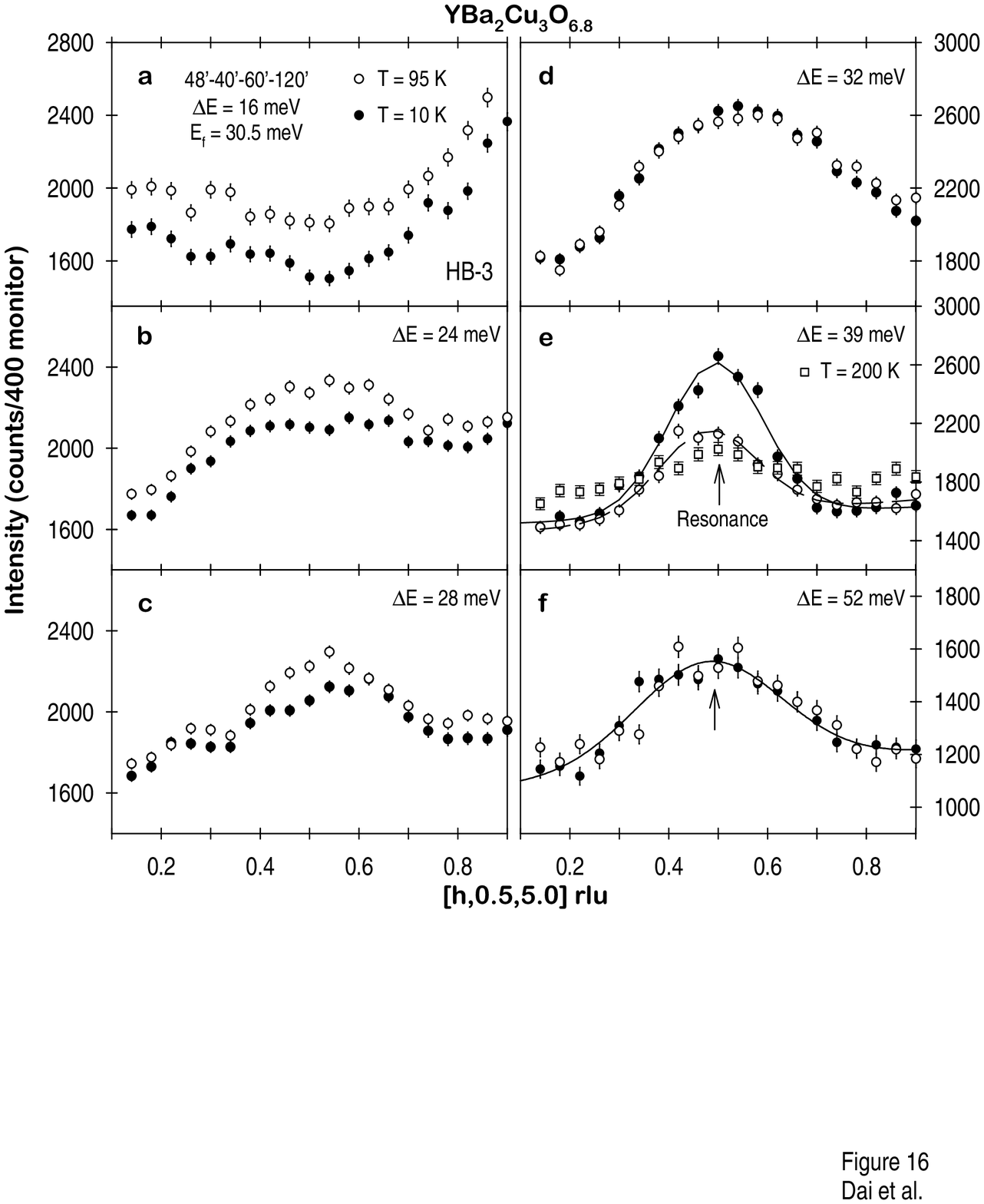}
\caption{
Raw constant-energy scans along the  
$[h,0.5,5.0]$ direction for YBa$_2$Cu$_3$O$_{6.8}$ in the low-resolution measurements with
$E_f=30.5$ meV. The energy transfers are $\hbar\omega=16$ meV (a),
24 meV (b), 28 meV (c), 32 meV (d), 39 meV (e), and 52 meV (f). 
The resonance is clearly seen at $\hbar\omega=39$ meV.
}
\end{figure}

\begin{figure}
\includegraphics[width = 3 in]{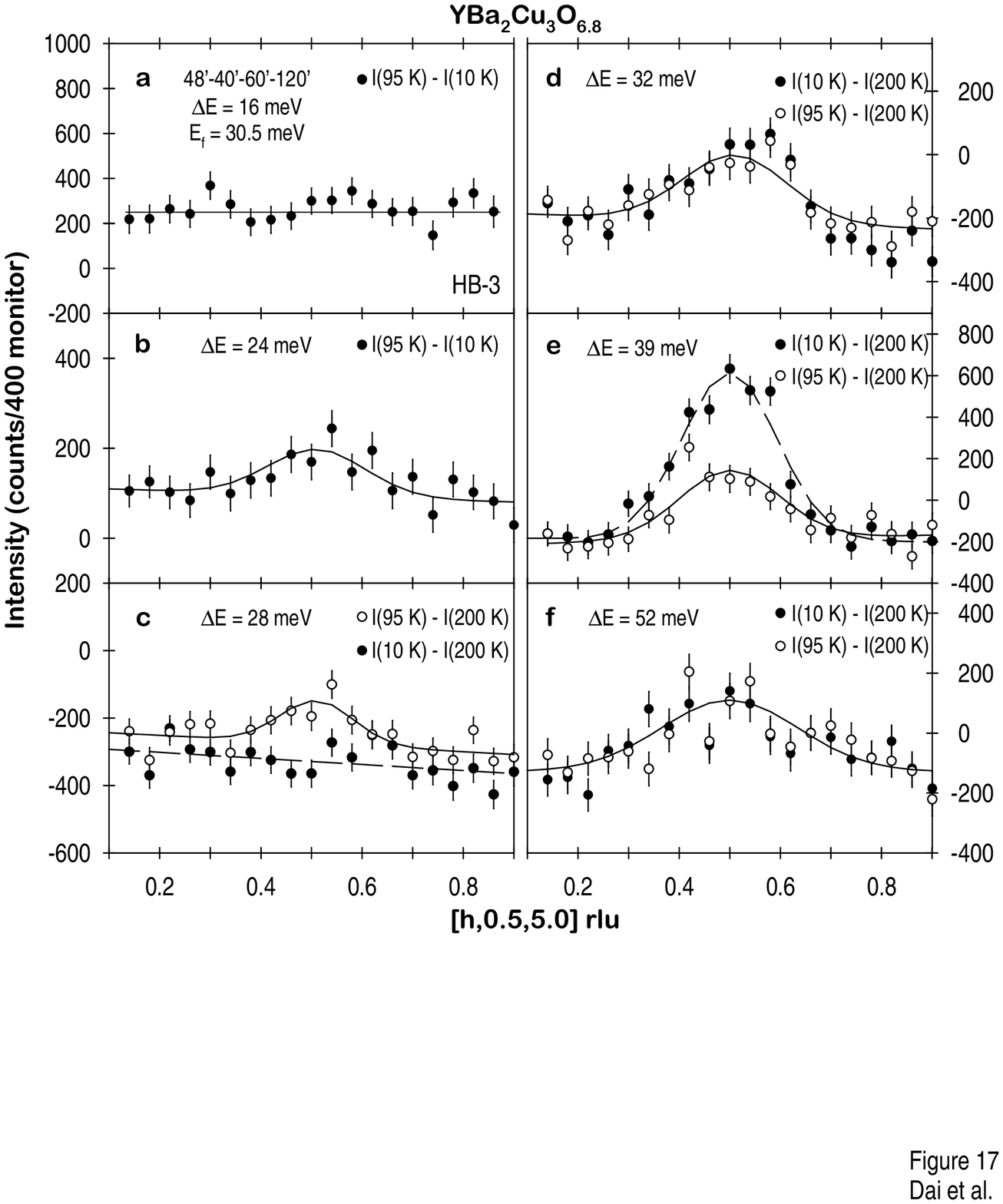}
\caption{
Difference spectra in constant-energy scans along the  
$[h,0.5,5.0]$ direction for YBa$_2$Cu$_3$O$_{6.8}$.  
The energy transfers are $\hbar\omega=16$ meV (a),
24 meV (b), 28 meV (c), 32 meV (d), 39 meV (e), 52 meV (f).
(a) suggests no change in susceptibility between the normal and superconducting states.
(b) and (c) indicate that Gaussian-like normal state scattering is suppressed 
in the superconducting state at $\hbar\omega=24$ and 28 meV. 
The situation for $\hbar\omega=32$ meV is not clear in this low-resolution 
measurement although there are clear magnetic intensities in the normal and the 
superconducting states. The intensity gain of the resonance is seen in (e), and the 
magnetic scattering appears to have weak temperature dependence at 52 meV (f). 
The solid lines are Gaussian fits to the data.
}
\end{figure}

To search for incommensurate spin fluctuations in YBa$_2$Cu$_3$O$_{6.8}$, we
scanned along the wave vector $[h,0.5,5]$ direction at energies 
below and above the resonance. Figure 16 summarizes the low-resolution raw data at temperatures 
just above and well below $T_c$. 
At $\hbar\omega=16$ meV
[Fig. 16(a)], the scattering  shows a sloped background with minimum intensity at around
$(\pi,\pi)$ in both the normal and superconducting states. This suggests the existence of 
a normal state spin gap of 16 meV. 
On increasing the energy to $\hbar\omega=24$ meV [Fig. 16(b)],
the intensity of the broad peak around $(\pi,\pi)$ in the normal state
is somewhat reduced below $T_c$ with no clear evidence of an incommensurate structure.
A similar behavior is also observed at $\hbar\omega=28$ meV [Figs. 16(c)]. 
The scattering profiles  
between the normal and superconducting states are almost indistinguishable at 
$\hbar\omega=32$ meV and 52 meV [Figs. 16(d) and (f)]. Near the resonance energy [Fig. 16(e)],
the scattering can be well described by Gaussians on linear backgrounds and 
shows a clear increase in intensity on entering the superconducting state from the normal state.
In addition, the scattering at 200 K (open squares) shows a broad  
peak around
$(\pi,\pi)$, which indicates weak magnetic scattering at this temperature. 

\begin{figure}
\includegraphics[width = 3 in]{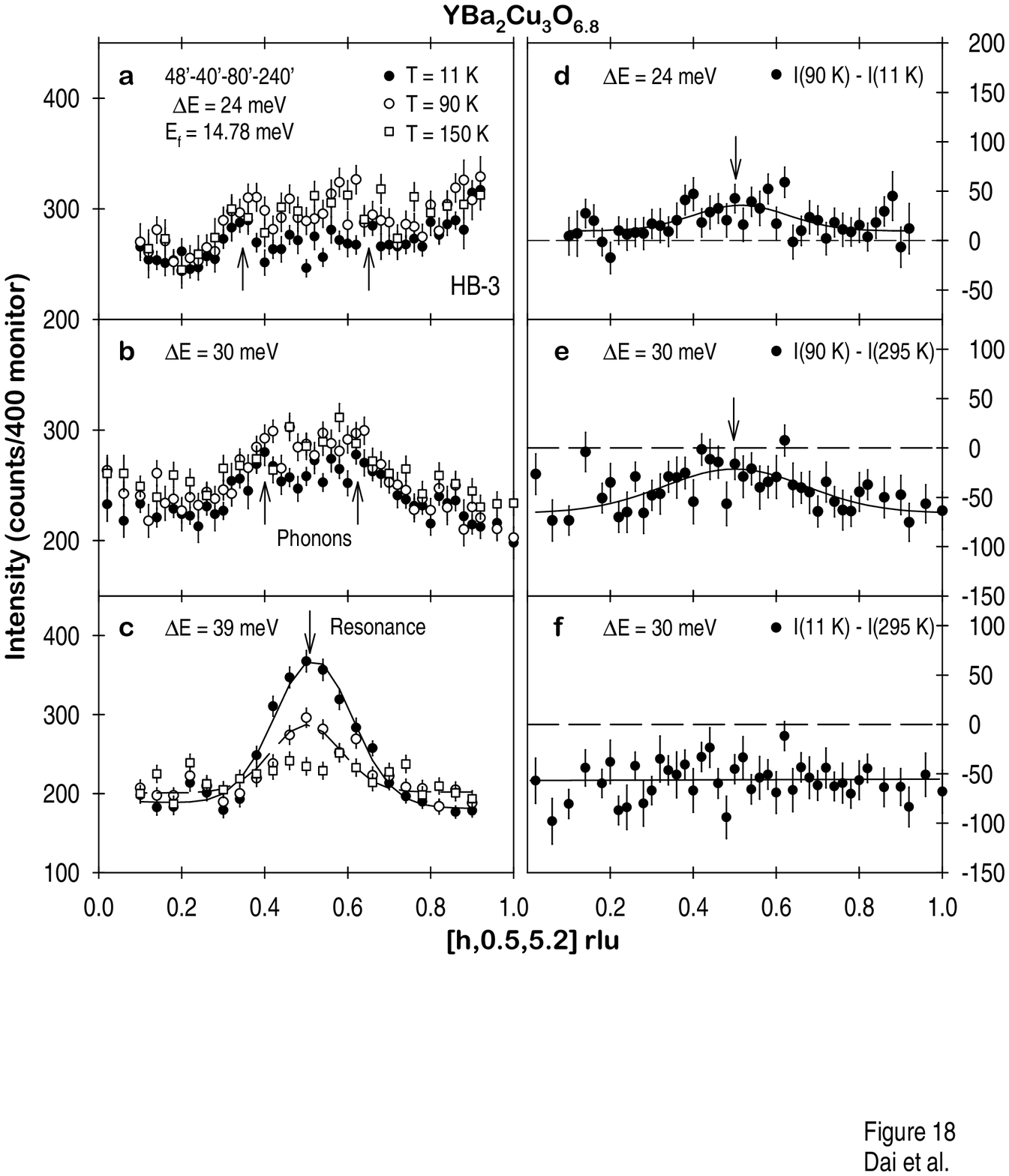}
\caption{
High-resolution constant-energy scans along the  
$[h,0.5,5.2]$ direction for YBa$_2$Cu$_3$O$_{6.8}$. 
The energy transfers are $\hbar\omega=24$ meV (a),
30 meV (b), and 39 meV (c). The weak structures in (a) and (b) at 11 K marked by the 
arrows are nonmagnetic scattering. (d) The difference spectrum for
$\hbar\omega=24$ meV between 90 K and 11 K 
shows a Gaussian profile in the normal state, consistent with Fig. 17(b). (e) and (f)
demonstrate the opening of the spin gap from the normal to the superconducting state 
at $\hbar\omega=30$ meV. No evidence for incommensurate spin fluctuations is  
found at $\hbar\omega=30$ meV.  
The solid lines are Gaussian fits to the data and guides to the eye.
}
\end{figure}

\begin{figure}
\includegraphics[width = 3 in]{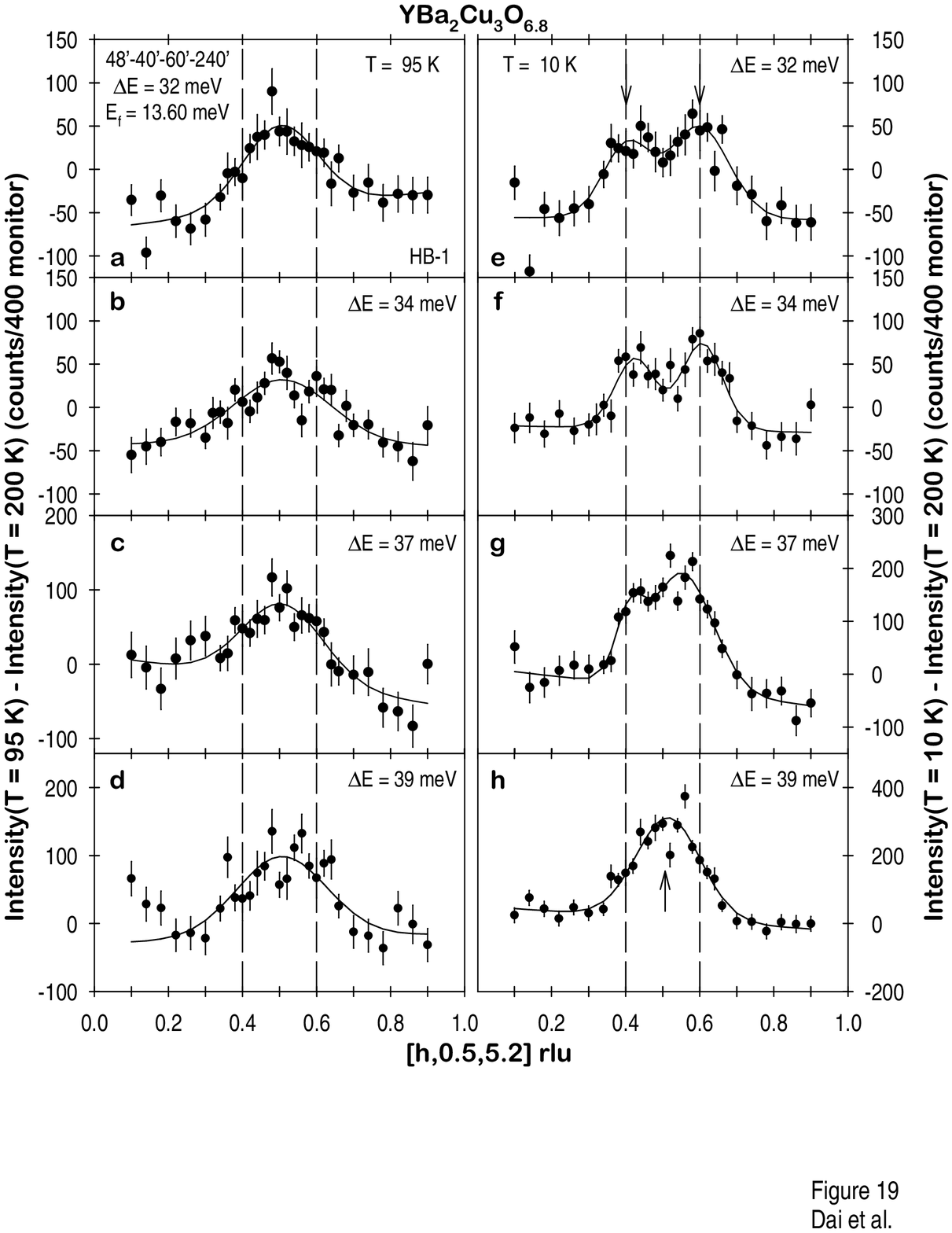}
\caption{
Difference spectra in high-resolution constant-energy scans along the  
$[h,0.5,5.2]$ direction above and below $T_c$ for YBa$_2$Cu$_3$O$_{6.8}$. 
The energy transfers are $\hbar\omega=32$ meV (a) and (e),
34 meV (b) and (f), 37 meV (c) and (g), and 39 meV (d) and (h). 
The scattering below the 39 meV resonance 
changes from commensurate in the normal state to incommensurate in the
superconducting state. Similar behavior have been reported by Bourges {\it et al.} 
for YBa$_2$Cu$_3$O$_{6.85}$ [51]. 
The solid lines are Gaussian fits to the data.
}
\end{figure}

Since the scattering at 200 K is mostly nonmagnetic for energies
below 39 meV [Fig. 16(e)], we systematically subtracted 200 K data from that obtained at
lower  temperatures. Figures 17(c-f) show the outcome of 
such procedure, and the net intensity gain around $(\pi,\pi)$ above the general reduction 
in the multi-phonon background should represent the enhancement of 
magnetic scattering at low temperatures. For $\hbar\omega = 28$ meV [Fig. 17(c)],
the broad peak in the normal state vanishes below $T_c$, which suggests the enlargement
of the normal state spin gap in the superconducting state. 
On increasing the energy to $\hbar\omega=32$ meV, 
the scattering shows little change from the normal state to the 
superconducting state [Figs. 17(d)]. In addition, the maximum scattering intensity  
no longer seems to peak at $(\pi,\pi)$, but shows weak evidence for
incommensurate scattering. Unfortunately, the poor resolution of the 
measurements does not allow a conclusive identification of incommensurate 
spin fluctuations.
At $\hbar\omega=39$ meV [Fig. 17(e)], the enhancement of resonance in Fig. 16(e) is confirmed.
Finally for an energy above the resonance ($\hbar\omega=52$ meV), 
the scattering shows no observable change above and below $T_c$ [Fig. 17(f)].

Because the intensity of low energy phonons 
is more sensitive to the increasing temperature and the Bose population factor, 
the subtraction procedure described above 
does not work for low energy excitations. 
However, we can compare the difference spectra at temperatures just above and below $T_c$. 
At $\hbar\omega=16$ meV [Fig. 17(a)],  
the difference spectra between 95 K ($T_c+13$ K) and 10 K ($<T_c$) show
no identifiable feature which is consistent with the presence of a
normal state spin gap. Similar subtraction at $\hbar\omega=24$ meV shows a clear peak
around $(\pi,\pi)$ [Fig. 17(b)] which indicates the suppression of 24 meV spin fluctuations in the 
superconducting state. Therefore, the 
normal state spin fluctuations in YBa$_2$Cu$_3$O$_{6.8}$ consist of a gap ($\geq 16$ meV) 
and broad excitations centered around $(\pi,\pi)$. In the 
superconducting state, the normal state spin gap increases in magnitude, and the intensity 
of the low energy fluctuations is suppressed. 

To further determine the magnitude of the spin gap in the superconducting state, we
carried out high-resolution measurements with the neutron final energy fixed at $E_f = 14.78$ meV.  
Figures 18(a-c) show constant-energy scans at $\hbar\omega=24$, 30, and 39 meV. The
difference  spectra at $\hbar\omega=24$ and 30 meV between low and high temperatures are
shown in Figs. 18(d-f).  Careful temperature dependent measurements suggest that 
the weak structure marked by arrows in the low temperature raw data of 
Figs. 18(a) and (b)
is not magnetic in origin. The opening of the spin gap at $\hbar\omega=30$ meV 
in the superconducting state is demonstrated by the difference spectra of Figs. 18(e) and
(f),  thus confirming the result of Fig. 17.

Finally, we present results on the incommensurate spin
fluctuations at energies just below the
resonance in YBa$_2$Cu$_3$O$_{6.8}$. These high-resolution measurements were
carried out on the HB-1 triple-axis spectrometer 
with $E_f = 13.6$ meV. Following the procedures established above, we
show in Fig. 19 the temperature difference spectra at $\hbar\omega=32$, 34, 37, and
39 meV above and below $T_c$. Incommensurate spin fluctuations 
 were observed below $T_c$
at $\hbar\omega=32$, 34, and 37 meV. 
As in the case of YBa$_2$Cu$_3$O$_{6.6}$, the incommensurability
of these fluctuations decreases with increasing energy and becomes commensurate at the 
resonance energy. In addition, the incommensurability of 
YBa$_2$Cu$_3$O$_{6.8}$ at $\hbar\omega\approx34$ meV is essentially the same as in the more 
underdoped YBa$_2$Cu$_3$O$_{6.6}$ at 24 meV [Fig. 10(a)]. However, the normal state 
spin fluctuations of YBa$_2$Cu$_3$O$_{6.8}$ are clearly commensurate and centered 
around $(\pi,\pi)$ whereas the fluctuations in YBa$_2$Cu$_3$O$_{6.6}$ show a flattish 
top indicative of the incommensurability at temperatures above $T_c$. 
Therefore, the 
spin fluctuation spectrum in YBa$_2$Cu$_3$O$_{6.8}$ has a gap 
in the normal state ($\hbar\omega\approx16$ meV),
a superconducting gap ($\hbar\omega\approx 30$ meV), a resonance at $\hbar\omega\approx39$ meV,  
and commensurate-to-incommensurate 
transition for spin fluctuations below the resonance on entering 
the superconducting state.  

\subsubsection{YBa$_2$Cu$_3$O$_{6.93}$ and YBa$_2$Cu$_3$O$_{6.95}$}
In the studies of ideally doped and overdoped YBa$_2$Cu$_3$O$_{6+x}$,
the important issues are the nature of the normal state magnetic scattering 
and the evolution of the incommensurate spin fluctuations. To determine 
whether there is detectable magnetic scattering around $\hbar\omega\approx40$ meV in
the normal state of YBa$_2$Cu$_3$O$_{6.93}$ and YBa$_2$Cu$_3$O$_{6.95}$, we performed 
a series of constant-$Q$ scans at $(\pi,\pi)$ at various temperatures 
below and above $T_c$. Figures 20
and 21 display the low temperature minus high temperature 
difference spectra for YBa$_2$Cu$_3$O$_{6.93}$ and YBa$_2$Cu$_3$O$_{6.95}$,
respectively. In both compounds, we find the resonance peak 
at $\hbar\omega\approx 40$ meV below $T_c$ and no 
trace or reminiscence of 
the resonance above $T_c$ as seen in all of the underdoped 
materials. Careful constant-energy scans for YBa$_2$Cu$_3$O$_{6.95}$ 
at $\hbar\omega=28$, 30, 32 and 40 meV are shown in Figs. 22(a-c). 
The temperature 
difference spectra in Figs. 22(d-f) yield a spin gap value of $\hbar\omega=32$ meV in the
superconducting state and show no discernible normal state magnetic scattering
\cite{fong95}. 

\begin{figure}
\includegraphics[width = 3 in]{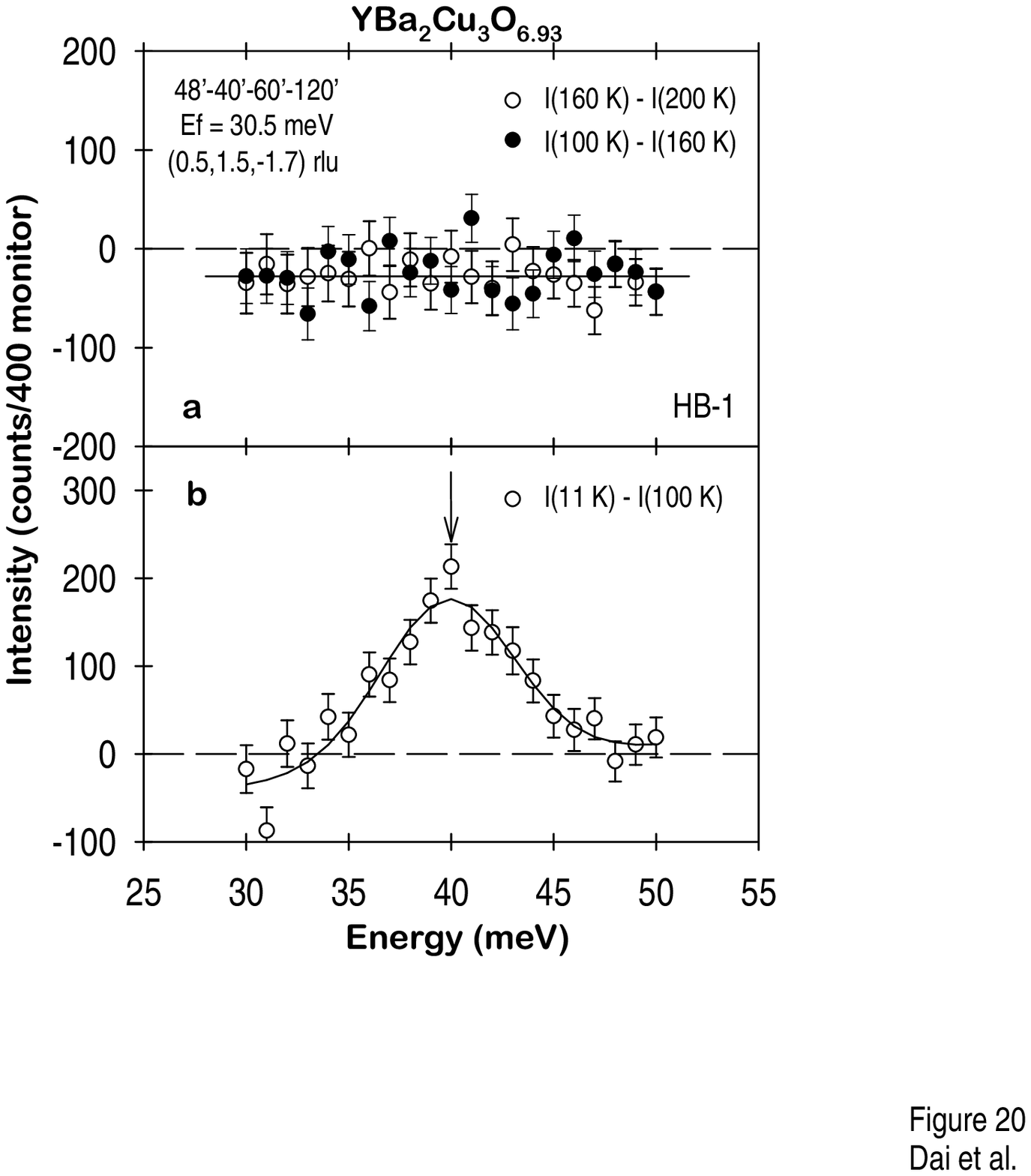}
\caption{
Difference spectra in constant-${\bf q}$ scans at (0.5,1.5,-1.7) rlu 
above and below $T_c$  
for YBa$_2$Cu$_3$O$_{6.93}$ with the crystal in the $(h,3h,l)$ zone. (a)
The filled circles 
show the difference between 100 K ($T_c+7.5$ K) and 160 K. 
The open circles are the difference between 160 K and 200 K.
Note that there is no positive scattering around 40 meV in the normal state, which 
suggests no precursor of the resonance. (b)
The temperature difference between 11 K ($T_c-81.5$ K) and 100 K. The 
solid line in (b) is a Gaussian fit to the 
data.
}
\end{figure}

\begin{figure}
\includegraphics[width = 3 in]{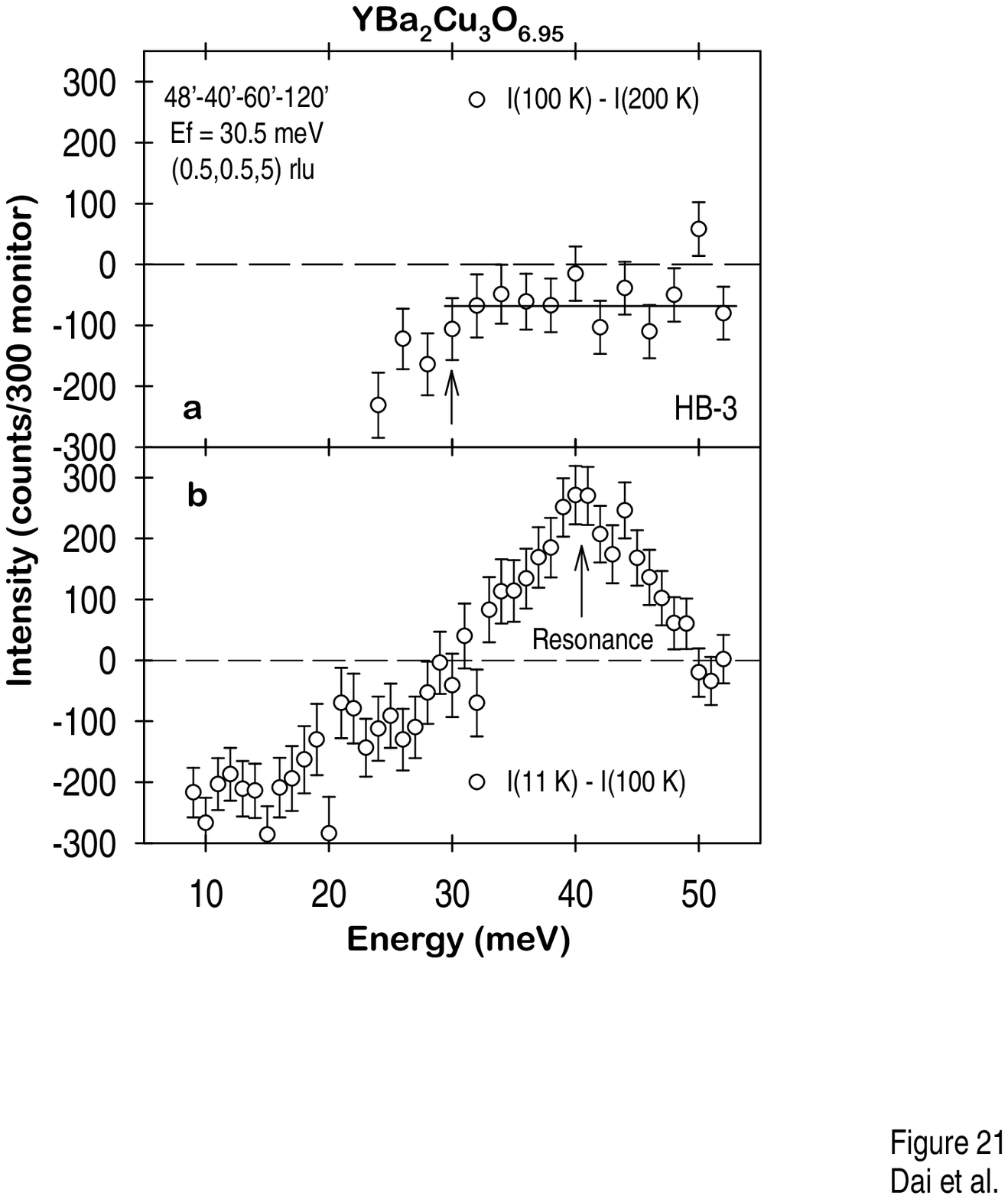}
\caption{
Difference spectra in constant-${\bf q}$ scans at (0.5,0.5,5) rlu 
above and below $T_c$  
for YBa$_2$Cu$_3$O$_{6.95}$. (a) The difference spectrum between 100 K ($T_c+8$ K) and 200 K. 
Note that there are no observable features above the background around 40 meV in the normal state,
again confirming no precursor of the resonance. (b) 
The temperature difference between 11 K ($T_c-81$ K) and 100 K.
}
\end{figure}

To further elucidate the nature of the incommensurate spin fluctuations, 
we performed a series of constant-energy scans at 
energies just below the 40 meV resonance for highly doped 
YBa$_2$Cu$_3$O$_{6.95}$. Figure 23 summarizes the outcome below and above $T_c$ at 
$\hbar\omega=34$, 35.5, 37 and 40 meV. Clear incommensurate spin fluctuations 
were found at $\hbar\omega=34$, 35.5, and 37 meV in the low temperature 
superconducting state. 
 Surprisingly, the incommensurability of the fluctuations at 
$\hbar\omega=34$ meV is 
$\delta=0.11\pm0.0128$ rlu, which is 
indistinguishable from all the other underdoped compounds with $x\geq 0.6$.
The incommensurability changes 
to $\delta=0.089\pm0.008$ and $0.084\pm0.02$ rlu for $\hbar\omega=35.5$ meV and 37 meV, 
respectively. 
Although there are 
weak features in the constant-energy scans just above $T_c$
[Figs. 23(a-d)], 
careful temperature dependent studies of the 
constant-energy profiles indicate that these features 
are nonmagnetic in origin [Fig. 23(e), open circles].
Thus, incommensurate fluctuations in highly doped YBa$_2$Cu$_3$O$_{6+x}$ 
appear only in the low temperature superconducting state
at energies close to the resonance. Furthermore, the 
incommensurabilities of the low energy spin fluctuations  
are insensitive to the doping for YBa$_2$Cu$_3$O$_{6+x}$ with 
$x\geq 0.6$ and saturate at $\delta=1/10$. In contrast, 
incommensurate spin fluctuations in the single-layer La$_{2-x}$Sr$_x$CuO$_4$ 
near optimal doping  
survive to temperatures well above $T_c$ \cite{aeppli}, and 
the incommensurabilities  
of the low energy spin fluctuations increase smoothly to the saturation value of  
$\delta = 1/8$ for materials with optimal doping \cite{yamada}. 

\begin{figure}
\includegraphics[width = 3 in]{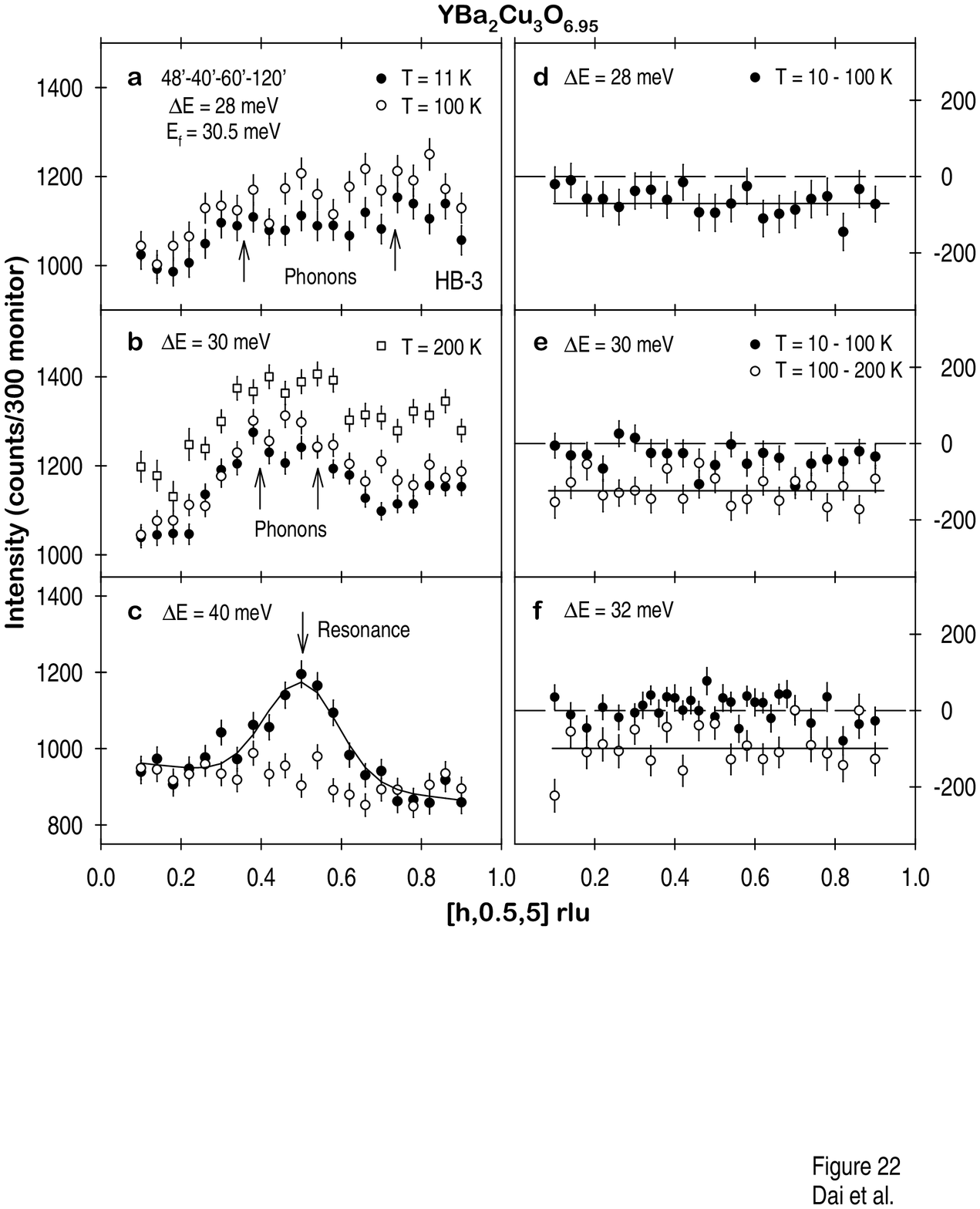}
\caption{
Constant-energy scans along the  
$[h,0.5,5]$ direction at various temperatures for YBa$_2$Cu$_3$O$_{6.95}$. 
The energy transfers are $\hbar\omega=28$ meV (a),
30 meV (b), and 40 meV (c). The weak structures in (a) and (b) at 11 K marked by  
arrows are nonmagnetic scattering. (d) The temperature difference spectrum for
$\hbar\omega=28$ meV between 10 K and 100 K shows no observable feature around $(\pi,\pi)$.
(e) and (f) suggest no discernible magnetic scattering above and below $T_c$
at $\hbar\omega=30$ and 32 meV. In addition, no evidence for incommensurate spin fluctuations is  
found below $\hbar\omega=32$ meV.  
The solid lines are Gaussian fits to the data and guides to the eye.
}
\end{figure}

\begin{figure}
\includegraphics[width = 3 in]{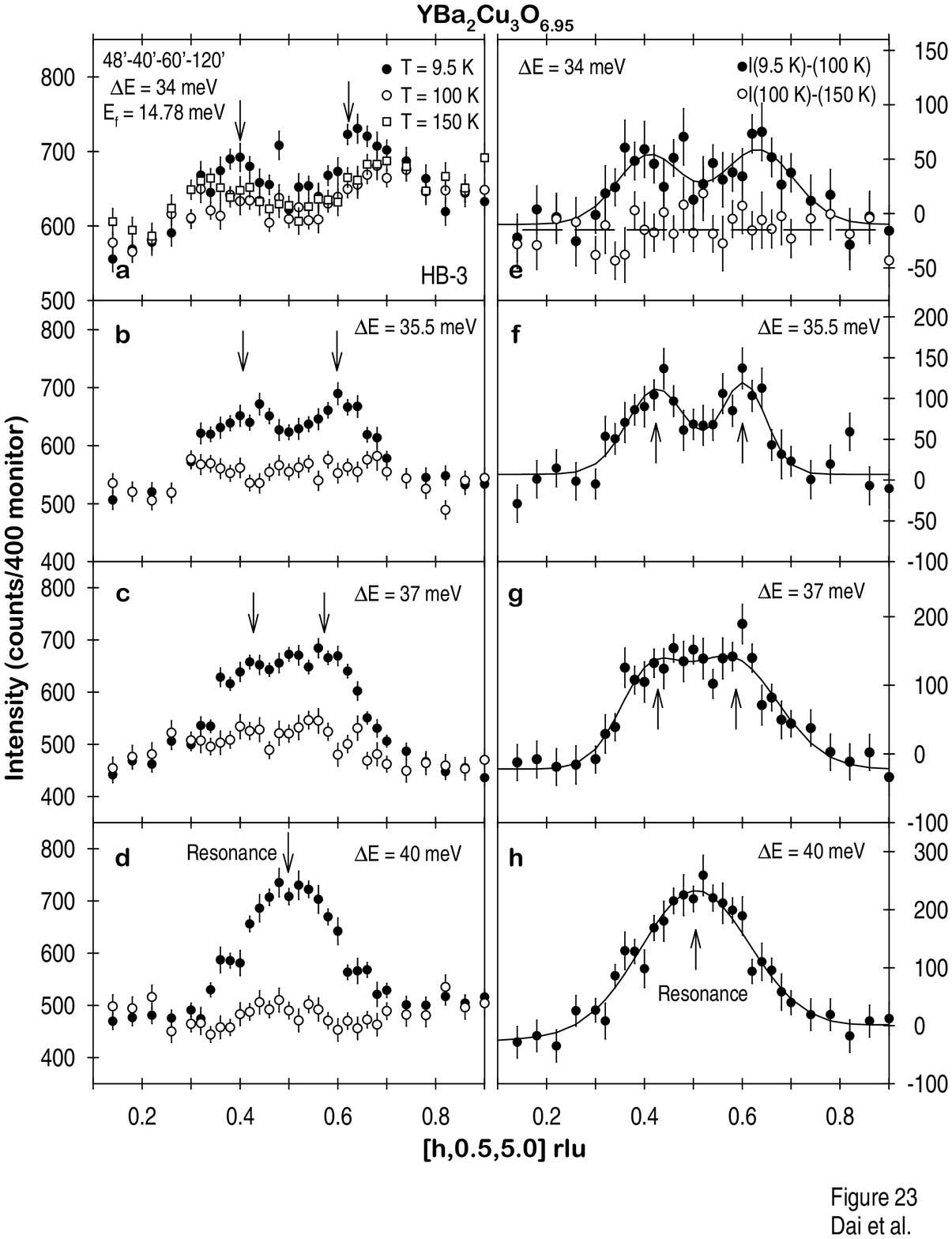}
\caption{
High-resolution constant-energy scans along the  
$[h,0.5,5]$ direction at various temperatures for YBa$_2$Cu$_3$O$_{6.95}$. 
The energy transfers are $\hbar\omega=34$ meV (a),
35.5 meV (b), 37 meV (c), and 40 meV (d). 
The enhancement of the scattering from 100 K to 9.5 K is marked by  
arrows. (e) The filled circles 
show the difference between 9.5 K ($T_c-82.5$ K) and 100 K. 
The open circles are the difference between 100 K and 150 K. Therefore,
while there is no discernible magnetic intensity in the normal state,
clear incommensurate spin fluctuations are observed in the superconducting state.  
(f) and (g) show the presence of incommensurate magnetic scattering 
at $\hbar\omega=35.5$ and 37 meV with the 
incommensurability marked by arrows. 
(h) shows the commensurate resonance. 
The solid and dashed lines are Gaussian fits to the data and guides to the eye, respectively.
}
\end{figure}

\subsection{Discussion}
\subsubsection{Doping dependence of the incommensurability}
The existence of incommensurate 
spin fluctuations now appears to be a common feature of the YBa$_2$Cu$_3$O$_{6+x}$ 
family of superconductors \cite{noteincomm}. Such fluctuations have also been observed
in Sr$^{2+}$- and oxygen- doped samples of La$_{2}$CuO$_4$ \cite{kastner}. Thus, 
it becomes clear that incommensurability is the common feature between these  
two classes of the most studied superconductors. However, before attempting 
any quantitative comparison of the similarities between YBa$_2$Cu$_3$O$_{6+x}$ 
and La$_{2-x}$Sr$_{x}$CuO$_4$, we must first consider the number of holes ($p$)
in the CuO$_2$ plane in these two systems. For La$_{2-x}$Sr$_{x}$CuO$_4$, 
Yamada and coworkers \cite{yamada} found an excellent linear relationship between
$\delta$ and the effective hole concentration up to around $p\approx 0.12$. 
Moreover, the incommensurability $\delta$ is energy independent for $\hbar\omega < 15$ meV 
\cite{mason}.
The 
situation for YBa$_2$Cu$_3$O$_{6+x}$ is more subtle. Using the established 
approximate parabolic relationship $T_c/T_{c,max}=1-82.6(p-0.16)^2$, where 
$T_{c,max}$ is the maximum transition temperature of the system, and $p$ is the 
hole-doping \cite{tallon}, we can calculate the effective hole doping $p$ 
of YBa$_2$Cu$_3$O$_{6+x}$ from their $T_c$ values. However, 
as shown in Figs. 5, 10, 19, and 23, the incommensurability 
in YBa$_2$Cu$_3$O$_{6+x}$ decreases with increasing energy close to the resonance.
As a consequence, it is difficult to compare directly the doping dependence of the 
incommensurability in these two families of materials as $\delta$ is not a well defined 
quantity for YBa$_2$Cu$_3$O$_{6+x}$. In order to make such comparison, we chose to  
systematically use the $\delta$ 
for the incommensurate low frequency spin fluctuations furthest away in energy from the 
resonace. 

In Figure 24, we plot the incommensurability $\delta$ obtained 
in such procedure as a function of $p$. 
For completeness, we also 
included the data of Arai {\it et al.} (open square) for YBa$_2$Cu$_3$O$_{6.7}$ \cite{arai}  
and very recent data of 
Bourges {\it et al.} (filled square) for YBa$_2$Cu$_3$O$_{6.85}$ \cite{bourges2000}. As we can see
from the figure, the incommensurability increases initially with doping, but it   
saturates quickly at $\delta\approx 0.1$ for hole doping $p\geq 0.1$. This result is 
different from La$_{2-x}$Sr$_{x}$CuO$_4$ where the incommensurability
is found to be linear with $\delta$ until reaching $0.125$ at large doping \cite{yamada}. 
For La$_{2-x}$Sr$_{x}$CuO$_4$, the saturation of incommensurability above $x\approx 0.12$   
appears to be related to the saturation of the density of O($2p$)-type holes in the CuO$_2$ plane as 
suggested from optical measurements \cite{uchida}. In the case of 
YBa$_2$Cu$_3$O$_{6+x}$, the incommensurability saturates for $x\geq 0.6$ whereas 
the hole concentration $p$ in the CuO$_2$ plane continues to increase with 
increasing oxygen doping and $T_c$ \cite{tallon}.
Therefore, the saturation of the incommensurability in YBa$_2$Cu$_3$O$_{6+x}$ may have 
a different microscopic origin from that of La$_{2-x}$Sr$_{x}$CuO$_4$.    

\begin{figure}
\includegraphics[width = 3 in]{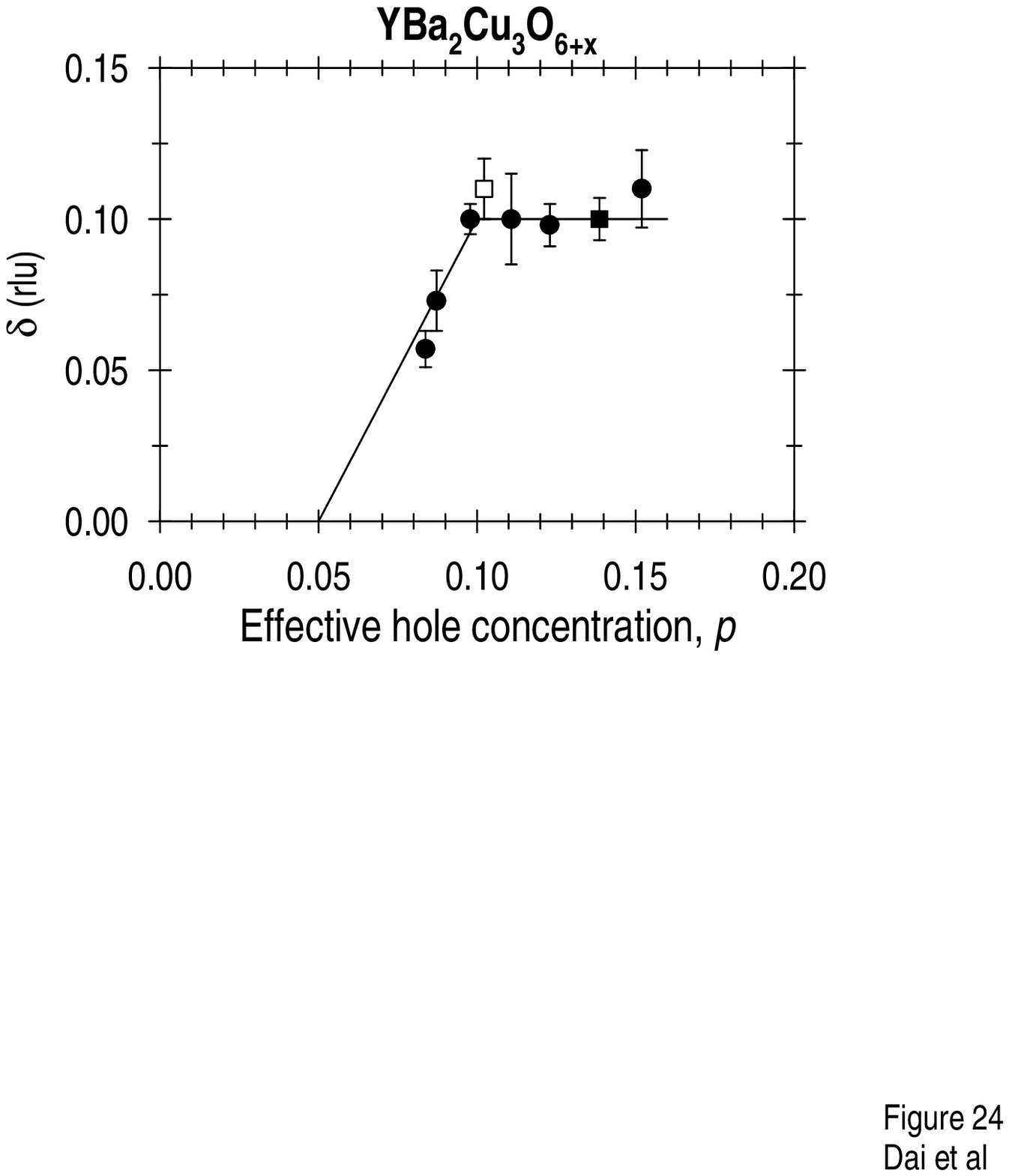}
\caption{
The incommensurability $\delta$ as a function of effective hole concentration, $p$, 
calculated from the parabolic relationship discussed in the text. $\delta$ is seen to saturate
for $p\geq 0.1$. The open and filled squares are from Ref. [25] and [51], respectively. 
The filled circles are from the present work. The solid lines are guides to the eye. Note that
superconductivity in cuprates occurs for $p\geq 0.05$ [44].
}
\end{figure}

\begin{figure}
\includegraphics[width = 3 in]{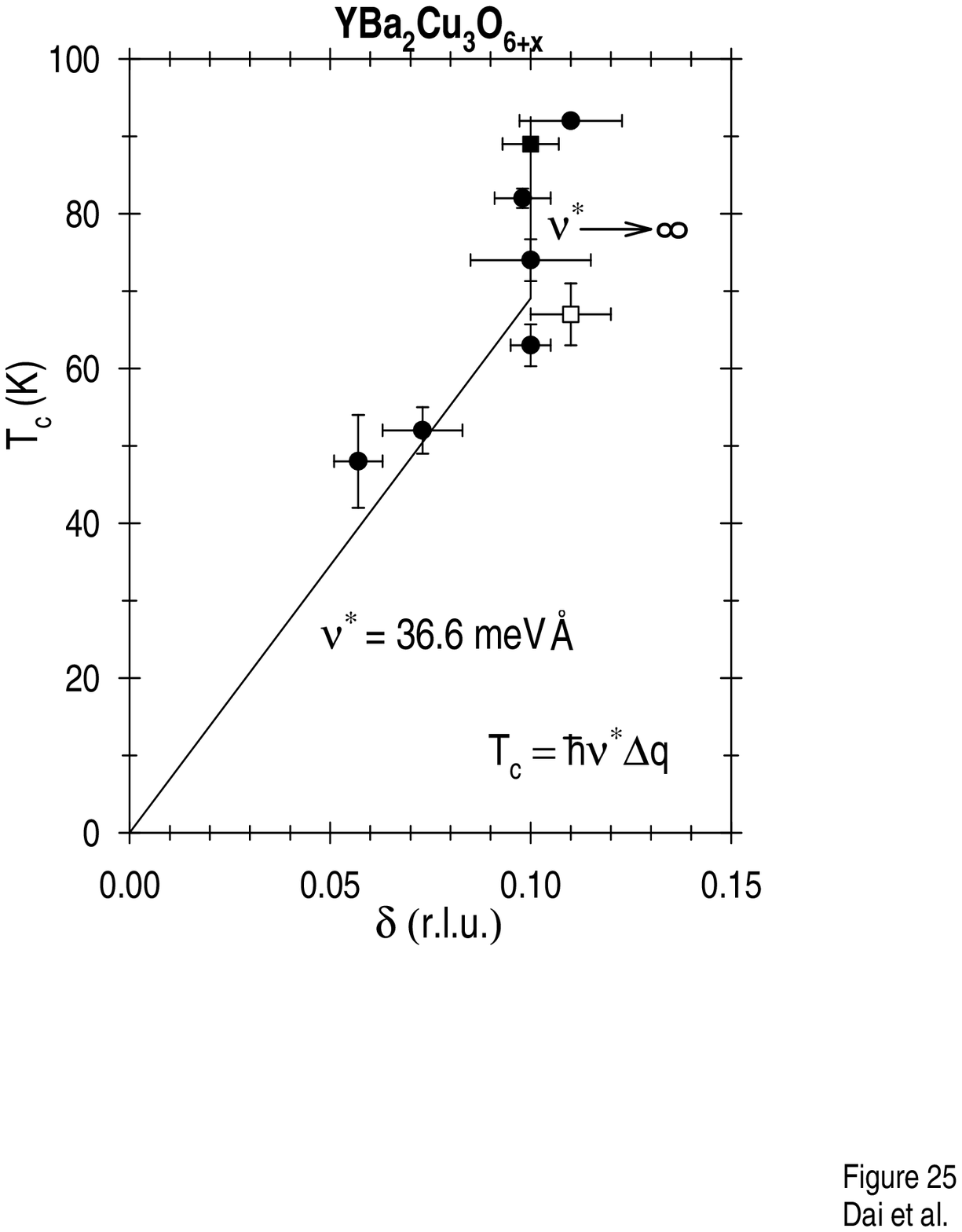}
\caption{
The incommensurability $\delta$ plotted as a function of $T_c$. 
The solid line for $T_c\leq 63$ K is a fit using 
$T_c=\hbar\nu^\ast\delta$ with $\hbar\nu^\ast\approx 36.6$ meV\AA. For $T_c\geq 63$ K, $\delta$ saturates to 0.1 rlu thus 
giving $\nu^\ast\longrightarrow\infty$.
}
\end{figure}

Although it is interesting to compare the incommensurability as a function of effective hole concentration in 
the CuO$_2$ plane of YBa$_2$Cu$_3$O$_{6+x}$ and La$_{2-x}$Sr$_{x}$CuO$_4$, a more robust test of 
the similarities in these two families of materials can be made by  
comparing the relation between $T_c$ and $\delta$ because 
these two quantities are free from any uncertainties associated with
the doping level or oxygen stoichiometry. For La$_{2-x}$Sr$_{x}$CuO$_4$, 
Yamada {\it et al.} found a linear relationship between $T_c$ and $\delta$ up to the 
optimal doping regime \cite{yamada}. Using $T_c=\hbar\nu^\ast_{214}\delta$, Balatsky
and Bourges \cite{balatsky} extracted a velocity $\hbar\nu^\ast_{214}=20$ meV\AA. 
For YBa$_2$Cu$_3$O$_{6+x}$, these authors suggested a similar linear relation
between $T_c$ and the $q$ width in momentum space, $\Delta q$ (HWHM), for the whole
doping range: $T_c=\hbar\nu^\ast\Delta q$, with $\hbar\nu^\ast=35$ meV\AA.
They then claimed strong similarities in the magnetic states of YBa$_2$Cu$_3$O$_{6+x}$
and La$_{2-x}$Sr$_{x}$CuO$_4$ \cite{balatsky}. In Fig. 25, we plot the 
relation between $T_c$ and the measured incommensurability $\delta$ for
YBa$_2$Cu$_3$O$_{6+x}$ in the whole doping range. For 
superconducting transition temperatures below $\sim$60 K, 
our data are consistent with 
 the proposed 
linear relation between $T_c$ and $\delta$ with  $\hbar\nu^\ast=36.6$ meV\AA\ \cite{balatsky}. 
However, the incommensurability is essentially unchanged for all samples with $T_c$ 
larger than 60 K. Therefore, if there were excitations associated with such a 
velocity, its magnitude would approach infinity for samples with transition temperatures
above 60 K.

\subsubsection{Doping dependence of the spin gap in the superconducting state}

One of the salient features of the spin excitations spectra in YBa$_2$Cu$_3$O$_{6+x}$ is the 
presence 
of spin gaps in the normal and superconducting states. For example, 
the low-energy spin fluctuations of YBa$_2$Cu$_3$O$_{6.8}$ show well defined 
Gaussian peaks centered around 
$(\pi,\pi)$ at energies above 24 meV in the normal state. On cooling the 
system to below $T_c$, the normal state scattering is 
suppressed which indicates the opening of a gap in the spin fluctuation 
spectrum (Figs. 17 and 18). In Figure 26, we plot the magnitude of the spin gap in the 
low temperature superconducting state as a function of $T_c$ for the whole
doping range. The solid line shows  
 the linear relation between $T_c$
and  the magnitude of the spin gap ($E_{sg}$) up to the optimal doping with 
$E_{sg}/k_BT_c=3.8$. 
This empirical result connects directly the spin gap and the 
superconductivity. The value of $E_{sg}/k_BT_c=3.8$ is not far away from 
the $2\Delta/k_BT_c=3.52$ for the weak-coupling limit of the BCS theory of 
superconductivity, where $\Delta$ is the superconducting gap. 
For samples with low oxygen doping and low $T_c$, the linear relation between $T_c$ and 
$E_{sg}$ breaks down, possibly due to the oxygen doping inhomogeneity and
broad  superconducting transitions in 
weakly-doped materials.

\begin{figure}
\includegraphics[width = 3 in]{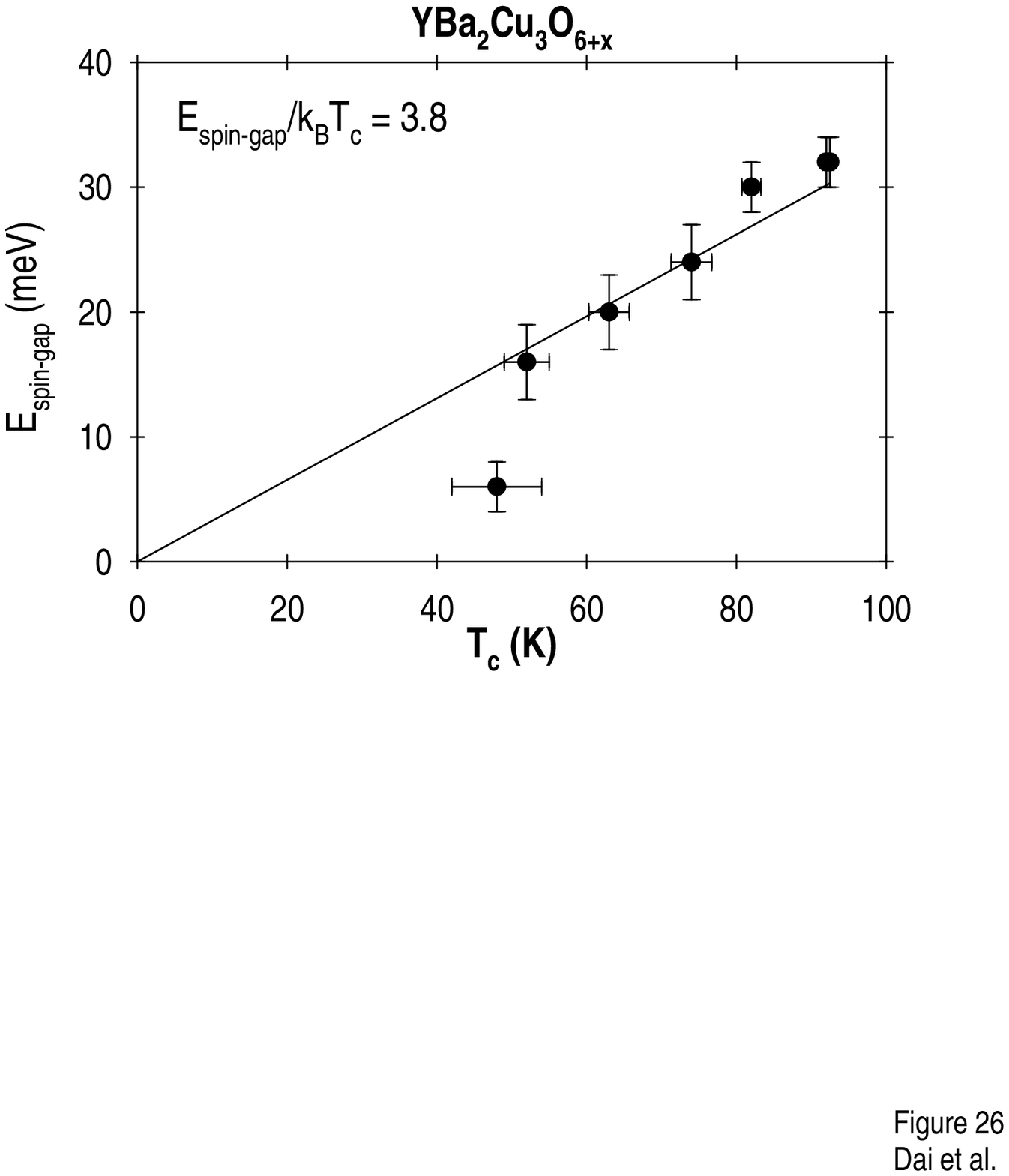}
\caption{
The magnitude of the spin-gap in the superconducting state 
as a function of $T_c$. 
The solid line gives $E_{sg}/k_BT_c=3.8$. For oxygen concentrations $x$ below 0.45,
the linear relationship between $E_{sg}$ and $k_BT_c$ breaks down.
}
\end{figure}

\subsubsection{Doping dependence of the resonance and superconducting condensation energy}

Among the many interesting features observed in the spin excitations spectra of 
YBa$_2$Cu$_3$O$_{6+x}$, perhaps the most-discussed is the collective acoustic 
spin excitation named the ``resonance''. 
First reported by Rossat-Mignod {\it et al.} \cite{mignod} for YBa$_2$Cu$_3$O$_{6.92}$, 
the resonance at $\sim$40 meV in the highly doped YBa$_2$Cu$_3$O$_{6+x}$ 
was shown to be magnetic in origin, narrow in energy \cite{mook93}, and to appear exclusively 
below $T_c$ \cite{fong95}. In subsequent experiments \cite{dai96,fong97,bourges97}, 
the resonance was also found in underdoped YBa$_2$Cu$_3$O$_{6+x}$ with the 
resonance energy ($E_r$) approximately scaling with $T_c$: $E_r\propto k_BT_c$.
Consistent with previous results \cite{balatsky}, 
the coherence length of 
the resonance ($\xi$) is weakly doping dependent except for 
 YBa$_2$Cu$_3$O$_{6.6}$ which has considerably larger $\xi$ than for all the 
other dopings (Fig. 27). The coherence length of the incommensurate spin fluctuations (Fig. 28)
also appears to be weakly doping dependent. In addition, we note 
that on approaching the optimal oxygen doping for
YBa$_2$Cu$_3$O$_{6+x}$, the resonance energy ($E_r$) is insensitive to the changing
$T_c$ (Fig. 29). Therefore, the linear relationship between $E_r$ and
$T_c$ appears to break down for materials with oxygen concentrations close to its optimal doping. 
Finally, we compared 
the peak intensity of 
the lowest-energy observable incommensurate fluctuations with that of the resonance (Fig. 30). While the 
ratio clearly decreases with increasing doping for $p\leq 0.1$, it saturates to a value 
of $\sim$ 0.25 for $p\geq 0.1$. 

\begin{figure}
\includegraphics[width = 3 in]{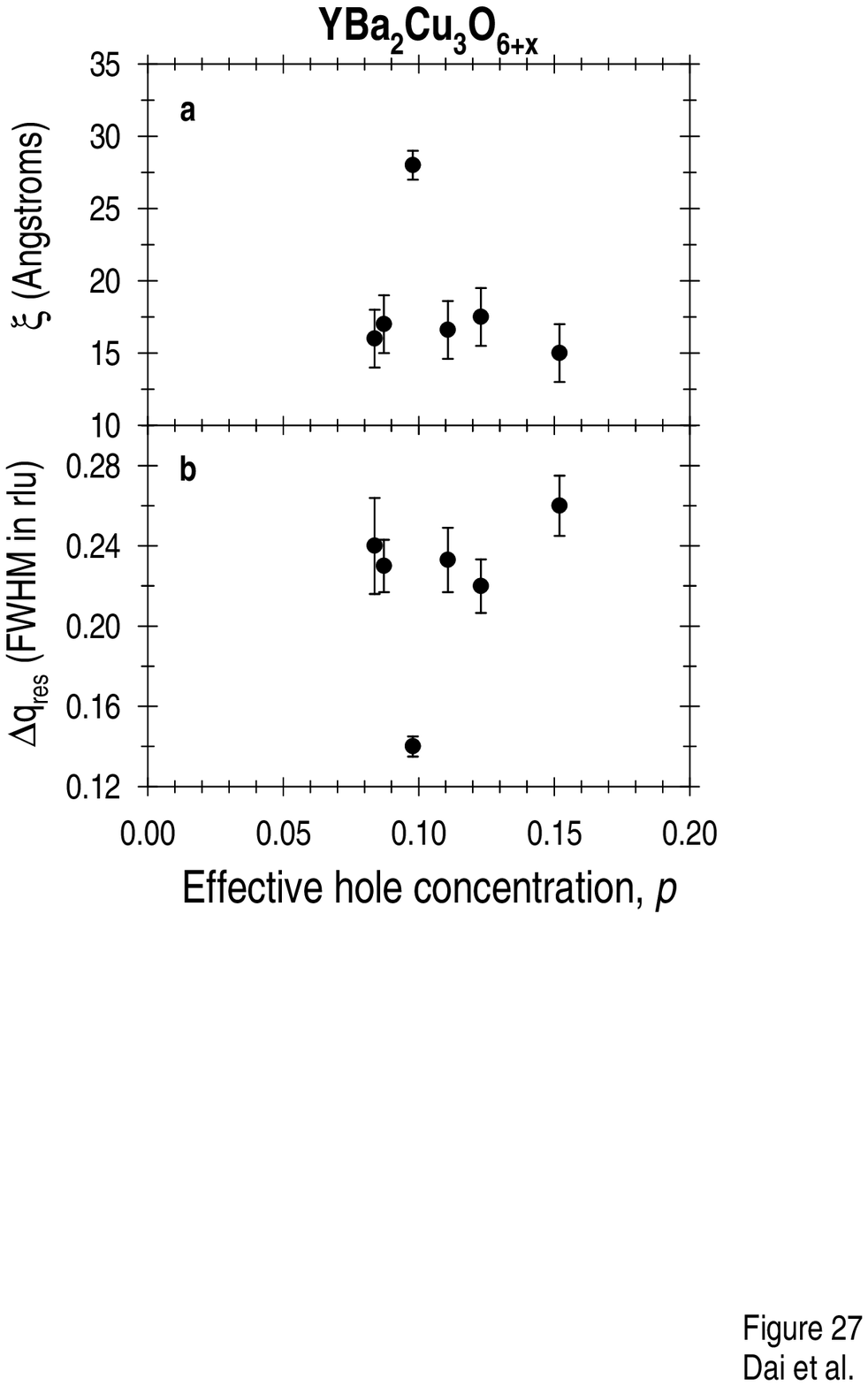}
\caption{
The minimum coherence length and dynamical $q$ width (FWHM) of the resonance 
as a function of effective hole concentration, $p$. At present, the origin of 
the anomalous long coherence length of the resonance 
for YBa$_2$Cu$_3$O$_{6.6}$ is unclear. 
The instrumental resolutions along the scan direction are $\sim$0.12 rlu
for the PG(002) monochromator and 0.07 rlu for the Be(002) monochromator at $\hbar\omega=34$ meV.
Slightly different values are obtained at other energies. Since the instrumental 
resolution is much smaller than the observed $q$ width of the resonance shown in (b),
the resonance coherence lengths in (a) are mostly intrinsic and not resolution limited.
}
\end{figure}

\begin{figure}
\includegraphics[width = 3 in]{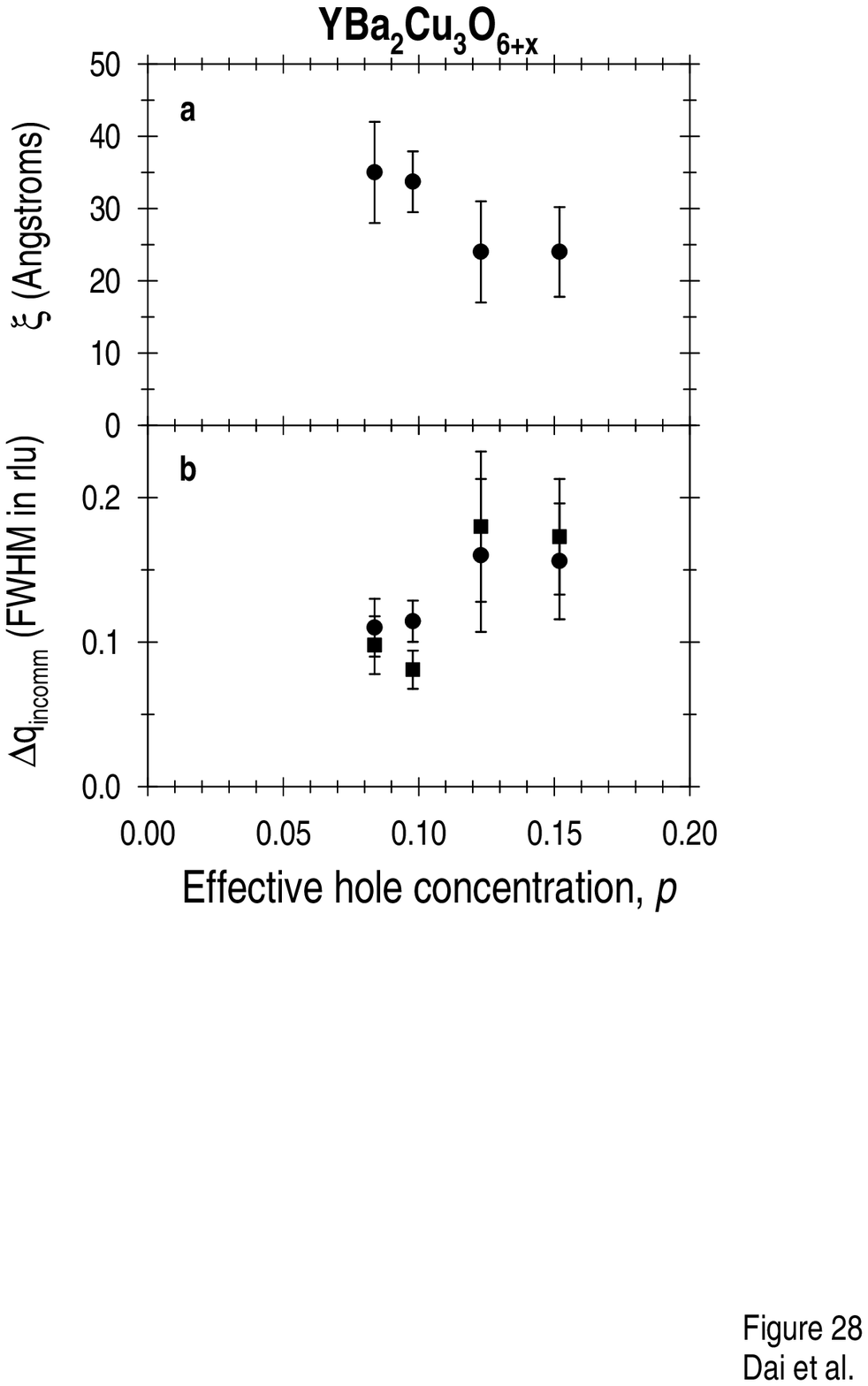}
\caption{
The minimum coherence length and dynamical $q$ width (FWHM) of the incommensurate
spin fluctuations 
as a function of effective hole concentration, $p$, for YBa$_2$Cu$_3$O$_{6+x}$
with $x=0.45$ at $\hbar\omega=24$ meV, 0.6 at 24 meV, 0.8 at 32 meV, 0.93 at 34 meV. 
The filled circles and squares in (b) are the measured 
widths for the left and right incommensurate peaks, respectively.  
The instrumental resolutions along the scan direction are $\sim$0.09 rlu
for PG(002) monochromator and 0.049 rlu for Be(002) monochromator at $\hbar\omega=24$ meV.
They change to $\sim$0.12 rlu for $\hbar\omega= 32$ and 34 meV with PG(002) monochromator.
Therefore, the slightly larger observed widths at $\hbar\omega= 32$ and 34 meV in (b) are 
due to the broadened instrumental resolution at these energies. 
}
\end{figure}

One interesting aspect of the resonance peak is its unusual temperature dependence.  
For optimal ($x=0.93$) and over doped ($x=1$) samples, the onset of the resonance occurs at a temperature
$T^\ast$ which almost coincides with $T_c$ \cite{mook93}. 
For underdoped $x=0.8$ and 0.6, $T^\ast$ increases to temperatures well above $T_c$ while both
$T_c$ and the resonance energy itself are reduced \cite{dai99}. By empirically comparing 
the cross-over temperature $T^\ast$ and the pseudogap temperature obtained by 
other techniques \cite{ito,wuyts,takigawa}, we associate the initial occurrence of the resonance
with the pseudogap temperature \cite{dai99}. In recent published \cite{fong2000a} 
and unpublished \cite{bourges2000a} papers,
Bourges {\it et al.} \cite{bourges2000a} claimed that 
there is no justification for 
a separation of the normal state spin excitations spectrum into resonant and nonresonant parts. 
Although these authors agreed that the broad maximum of the spin susceptibility in the normal state 
occurs at the same energy as the resonance peak in some underdoped materials \cite{dai99,bourges2000b}, 
they argued that ``the apparent equivalence of the normal state energy and the resonance peak 
energy breaks down in underdoped samples closer to optimal doping'' \cite{bourges2000a}.

\begin{figure}
\includegraphics[width = 3 in]{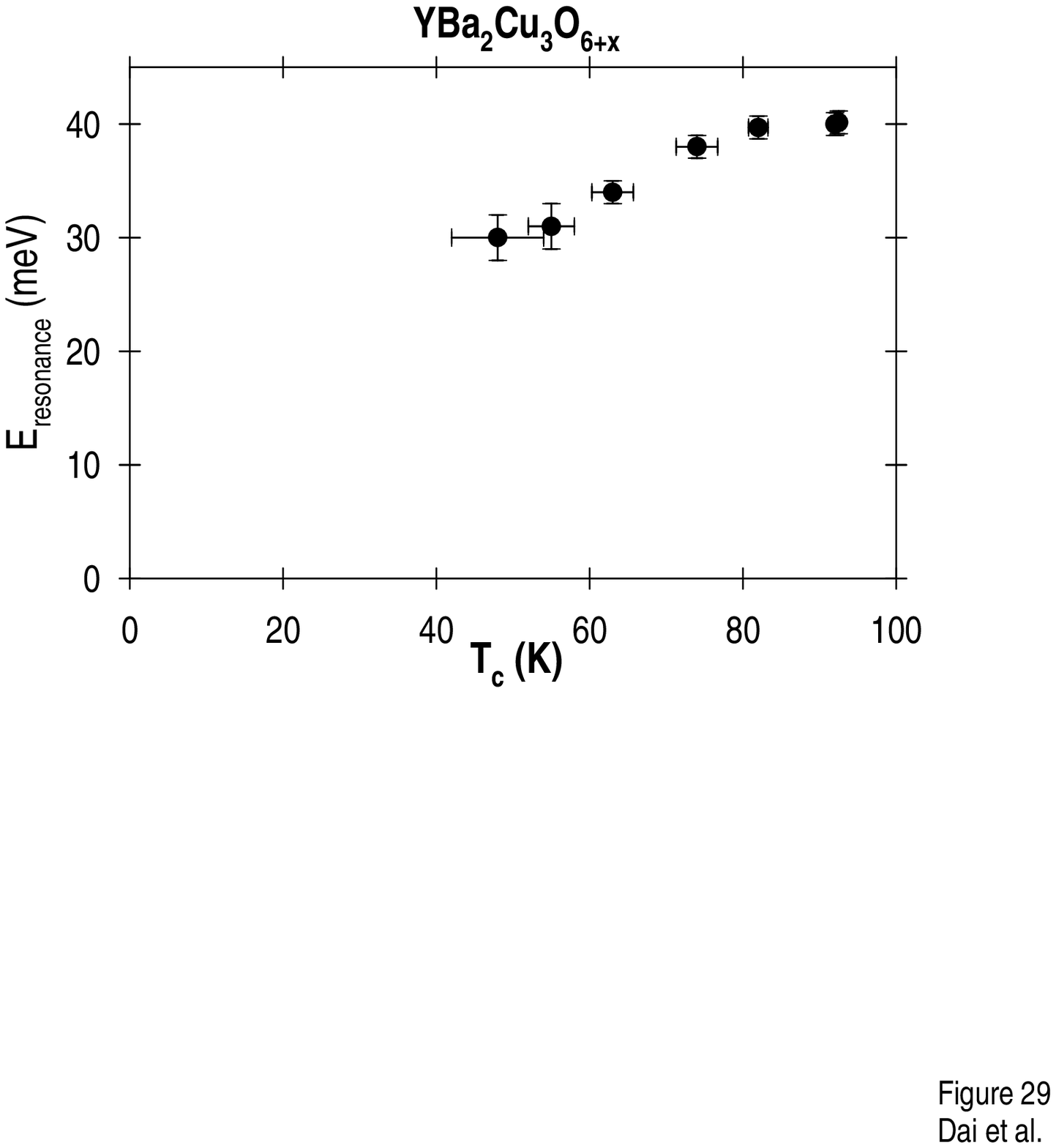}
\caption{
The resonance energy as a function of $T_c$ for samples used in
this study. The horizontal error bars are the superconducting transition widths.
}
\end{figure}

\begin{figure}
\includegraphics[width = 3 in]{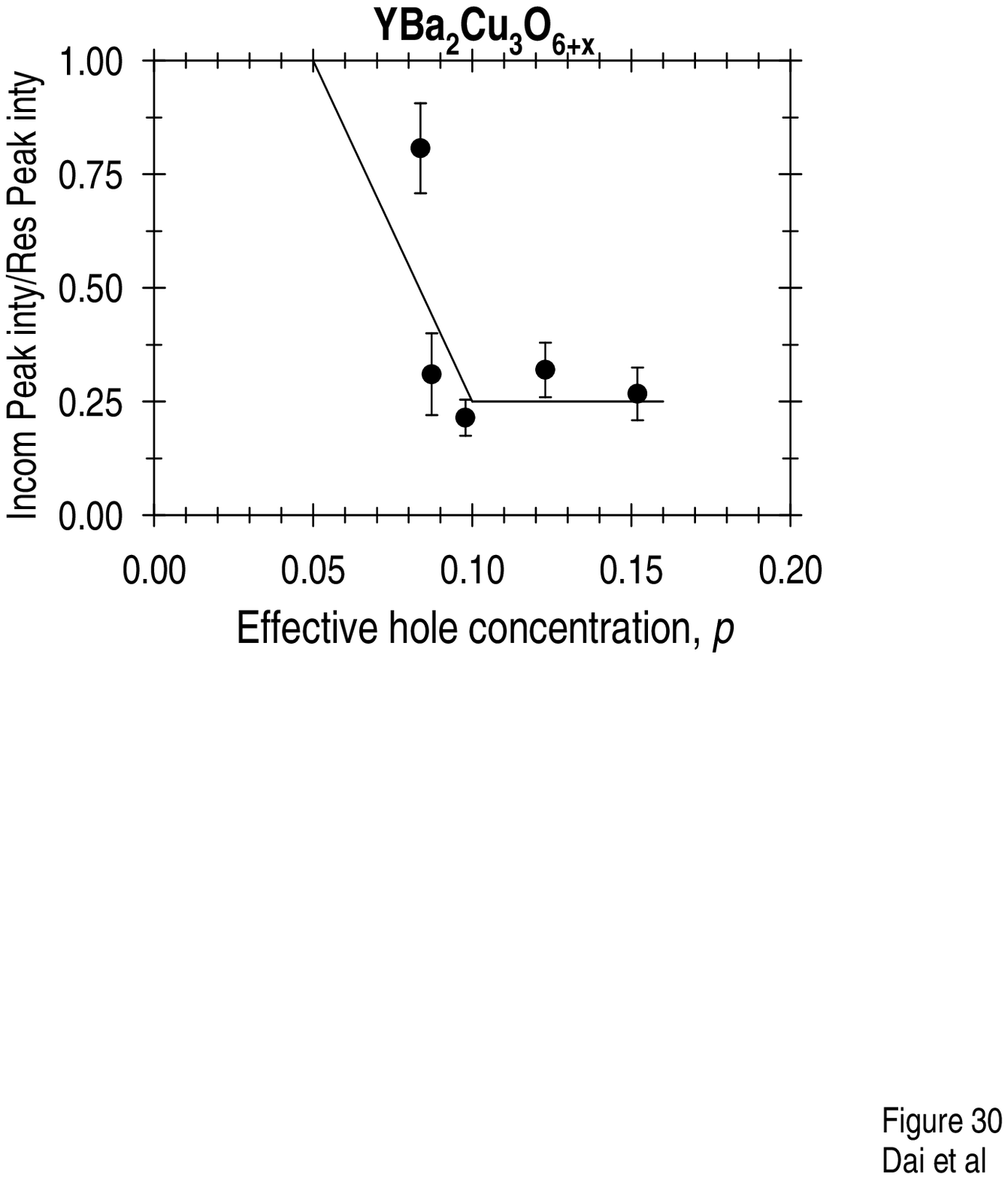}
\caption{
The ratio between the peak intensity of incommensurate spin fluctuations 
and the resonance as a function of effective hole concentration, $p$, 
for YBa$_2$Cu$_3$O$_{6+x}$. The intensity of incommensurate spin fluctuations is
at $\hbar\omega=24$ meV for $x=0.45$, 0.5, and 0.6, at 32 meV for $x=0.8$,
and at 34 meV for $x=0.95$. The solid lines are guides to the eye.
}
\end{figure}

Here we point out that the claims by Bourges {\it et al.} are incorrect. 
As can be seen from our results in Sec. IIIA1--3, there are clear enhancements in the 
spin susceptibility around the resonance energy at temperatures above $T_c$ for all the 
underdoped compounds including YBa$_2$Cu$_3$O$_{6.8}$.
These observations are in sharp contrast
to the claims of
Bourges  {\it et al.} \cite{bourges2000a}. Moreover,  
from the temperature dependence of the 
resonance intensity as a function of doping \cite{dai99}, we find that the 
intensity of the resonance changes smoothly and shows progressively
larger pretransitional regime above $T_c$ as a function of decreasing doping. As a consequence,
it is natural to associate the enhancement of the spin susceptibility above $T_c$ around 
the resonance energy in underdoped materials as the precursor of the resonance.

We also emphasize that the temperature dependence of the resonance \cite{dai99} is closely related 
to the thermodynamics of YBa$_2$Cu$_3$O$_{6+x}$ near $T_c$ \cite{loram} 
and hence the 
superconducting condensation energy \cite{scalapinowhite,demlerzhang,sudip1,norman1}. 
If the electronic entropy changes around $T_c$ are mostly due to spin excitations \cite{dai99}, 
they must 
respond to an external magnetic field in qualitatively the  
same way \cite{janko}. 
In a conventional type-II BCS superconductor \cite{schrieffer},
the effect of a field is to simply shift the superconducting specific heat anomaly ($C^{el}(x,T)$) 
to lower
temperatures and to reduce its amplitude until the upper critical field, 
$B_{c2}(0)$,
is reached \cite{junod1}. 
For high-$T_c$ superconductors such as YBa$_2$Cu$_3$O$_{6+x}$, 
the electronic specific heat anomaly near $T_c$ is drastically     
suppressed by a moderate field along the $c$-axis  
while it is little affected by the same field parallel to the $ab$-plane \cite{junod2}. 

In a recent experiment on YBa$_2$Cu$_3$O$_{6.6}$ \cite{dai2000}, we discovered that 
a moderate magnetic field ($B=6.8$
T) along the direction approximately perpendicular to the CuO$_2$ planes ($c$-axis) can drastically
suppress the intensity of the resonance at 34 meV 
while leaving other scattering relatively unaltered. 
Similar, but much smaller effect also   
occurs for the same field parallel to
the CuO$_2$ planes. Our results thus provide 
further support for the intimate connection between
the magnetic excitation spectrum and thermodynamics of high-$T_c$ superconductors \cite{dai99}.
The persistence of a field effect above $T_c$ 
is consistent with the precursor of the resonance in the normal state of 
 the underdoped YBa$_2$Cu$_3$O$_{6+x}$.

\subsubsection{Stripes versus Fermi-liquid}
Fermi-liquid theory, based on the premise that electron correlation effects in
the metallic state of matter correspond to the noninteracting ``quasiparticles'' on
a Fermi surface, has been successful in solving many of the important problems in 
condensed-matter physics, including conventional superconductivity. In the case of 
copper-oxides, where high-$T_c$ superconductivity is obtained  
by hole-doping of the
AF insulating parent compounds, an important class of models \cite{zaanen,schulz,poilblanc,kivelson} 
argues that electrons in these materials cannot be described 
by quasiparticles and Fermi-liquid theory.
Rather, the electrons have been incorporated in complex, inhomogeneous (fluctuating) patterns of the
charge  and spin confined to separate linear regions in the crystal resembling ``stripes''
\cite{orenstein}.   Since their discovery
\cite{yoshizawa,birgeneau,cheong,mason,thurston,yamada,yamadaprl,lake,kastner}, the incommensurate
spin fluctuations in the superconducting  La$_{2-x}$Sr$_x$CuO$_4$ have been interpreted 
as due to the presence of dynamic spin stripes \cite{zaanen1,emery,whitescalapino,martins} or
as the consequence of spin-flip scattering
across a nested Fermi surface \cite{bulut1,si,littlewood,brinckmann,morr,kao,norman2}. In 1995, Tranquada and 
coworkers \cite{tranquada95} found evidence supporting the former interpretation 
in a closely related material La$_{1.88-y}$Nd$_y$Sr$_{0.12}$CuO$_4$. 
For Nd substitution of $y=0.4$,
the distortion of the crystal structure causes the condensation of the incommensurate 
spin fluctuations in Nd-free La$_{2-x}$Sr$_x$CuO$_4$ into a spin density wave (SDW) at the 
same wave vector \cite{tranquada95}. The analysis of the elastic magnetic SDW peaks and 
their corresponding charge-ordering peaks associated with the modulation of the underlying lattice 
indicate the presence of a static striped phase in La$_{1.88-y}$Nd$_y$Sr$_{0.12}$CuO$_4$. 
This strongly suggests that the incommensurate fluctuations 
in superconducting La$_{2-x}$Sr$_x$CuO$_4$ 
are a fluctuating version (dynamic) of the stripes. 

However, if the stripe picture is universal for all high-$T_c$ superconductors, it should also be 
able to explain the incommensurate spin fluctuations of
YBa$_2$Cu$_3$O$_{6+x}$. In the case of La$_{2-x}$Sr$_x$CuO$_4$, the incommensurability saturates at
``1/8''. Whereas the incommensurability of the low-energy spin 
fluctuations 
in YBa$_2$Cu$_3$O$_{6+x}$ 
follows the expected linear behavior as a function of $T_c$ for low doping 
materials \cite{yamada} but abruptly saturates 
at ``1/10'' for $x\geq 0.6$. This could mean that the minimum stripe spacing in 
YBa$_2$Cu$_3$O$_{6+x}$ is five unit cells compared to four in La$_{2-x}$Sr$_x$CuO$_4$ \cite{tranquada95}. 
However, the reduction of the 
incommensurability with increasing energy close to the resonance in YBa$_2$Cu$_3$O$_{6+x}$, 
first
reported by  Arai {\it et al.} \cite{arai} and subsequently confirmed by Bourges {\it et al.}
\cite{bourges2000} in underdoped compounds, is different from the energy independent incommensurability of the
low-frequency spin fluctuations in La$_{2-x}$Sr$_x$CuO$_4$ \cite{mason}. In addition, the
incommensurate fluctuations  in La$_{2-x}$Sr$_x$CuO$_4$ are observed at temperatures well above
$T_c$ \cite{aeppli}. Such fluctuations in 
YBa$_2$Cu$_3$O$_{6+x}$ 
are present above $T_c$ for underdoped compounds, but
for highly doped materials they are only 
found close to the resonance in the superconducting state.

At present, it remains unclear how to explain the observed 
behavior differences in these two families of materials from a stripe model.
In particular, the saturation of the incommensurability at 
$\delta=1/10$ for effective hole concentration $p\geq 0.1$ (Fig. 24) 
is puzzling. Assuming that the incommensurate spin fluctuations 
in YBa$_2$Cu$_3$O$_{6+x}$ at all doping levels are due to 
dynamic stripes composed of the hole-rich metallic regions separated by the antiphase
insulating domains \cite{tranquada95}, one would expect the growth of 
the metallic stripes with increasing doping at 
the expense of the insulating domains \cite{zaanen,schulz,poilblanc,kivelson}.
As a consequence, the incommensurability should increase with increasing 
hole doping which is in contrast to the observation of a 
saturating $\delta$ for YBa$_2$Cu$_3$O$_{6+x}$ with $x\geq0.6$.

Although we have emphasized the differences,
the incommensurate spin fluctuations for these two classes of materials also have  
remarkable similarities in their structure and doping dependence for
oxygen concentrations $x\leq 0.6$ ($p\leq 0.1$). 
These similarities and  
the one-dimensional nature of the incommensurate spin fluctuations in underdoped 
YBa$_2$Cu$_3$O$_{6.6}$ \cite{mook00} and the associated phonon anomaly \cite{mookn99a} 
are consistent with the stripe formation in 
YBa$_2$Cu$_3$O$_{6+x}$ for effective hole doping below 0.1. 
To truly understand the microscopic origin 
of the incommensurate spin fluctuations 
in different cuprates using a stripe model, perhaps one needs to first explain 
the presence of resonance in the bilayer compounds,   
which takes much of the weight of the total magnetic scattering, and its absence in the 
single-layer materials \cite{mook00}. For highly doped materials, 
the incommensurate spin fluctuations in YBa$_2$Cu$_3$O$_{6+x}$
appears to be intimately related to the resonance as they occur only at 
energies very close to the resonance. From the magnetic field dependence of
the spin excitations \cite{dai2000}, it is clear that the resonance measures the pairing 
and long-range phase coherence. Thus, if a stripe model can explain the resonance,
it may also be able to account for the differences in the incommensurability in 
La$_{2-x}$Sr$_x$CuO$_4$ and YBa$_2$Cu$_3$O$_{6+x}$. 

Alternatively, assuming that 
the observed incommensurate fluctuations in YBa$_2$Cu$_3$O$_{6+x}$
come from a nested Fermi surface \cite{brinckmann,morr,kao,norman2},
one may ask whether such models can predict the observed intensity and doping 
dependent behavior. 
Unfortunately, there are no explicit predictions about 
the doping dependence of the incommensurability 
and its relationship with the resonance from a nested Fermi surface model 
that can be directly compared with 
our experiments. Therefore, 
it is still not clear whether the observed incommensurate fluctuations 
in YBa$_2$Cu$_3$O$_{6+x}$ are consistent with the Fermi-liquid description at all doping levels. 
In particular, the abrupt saturation of the incommensurability for
YBa$_2$Cu$_3$O$_{6+x}$ with $x\geq 0.6$ 
suggests that the physics accounting for this behavior may have a different microscopic  
origin from that of the lower doping compounds. 

\subsubsection{Models for the resonance}

Since the discovery of the resonance in highly doped and underdoped YBa$_2$Cu$_3$O$_{6+x}$ 
\cite{mignod,mook93,fong95,bourges96,fong96,dai96,fong97,bourges97}, 
many fundamentally different microscopic mechanisms have been proposed to explain its origin. 
In the earlier theoretical studies \cite{maki,lercher,lavagna,onufrieva}, the resonance is 
interpreted as the consequence of the $d$-wave gap symmetry of the cuprate superconductors. 
Subsequent workers have noticed that $d$-wave superconductivity alone is insufficient to explain
the resonance, and that very strong Coulomb correlations among electrons 
must also play an important role. In some models, the band structure 
anomaly effects alone can produce a peak in spin susceptibility
that grows with the opening of the superconducting gap \cite{blumberg,abrikosov,yin}, 
while other workers have applied a random-phase approximation (RPA) to treat the effects 
of strong Coulomb correlations with the Lindhard 
function appropriate for either a Fermi liquid \cite{bulut,liu,mazin,millis} or 
a spin-charge separated metal \cite{brinckmann,tanamoto,normand}. In particular, 
the resonance has been treated as a disordered magnon-like collective mode in the 
particle-hole channel whose energy 
is bound by the superconducting energy gap \cite{morr,abanov,sachdev}. Using a 
completely different approach \cite{demler}, the resonance is viewed as 
a pseudo-Goldstone boson mode in the particle-particle channel instead of the more conventional
particle-hole channel. In the framework of this theory, antiferromagnetism and 
$d$-wave superconductivity in copper oxides are treated in equal 
footing \cite{sczhang}, and
resonance is directly  responsible 
for the superconducting condensation energy \cite{demlerzhang}.

While the purpose of this article is not to review the similarities and differences 
among various models of the resonance, we wish to make several general remarks in 
light of the results described here and in recent field-dependent studies of 
the resonance \cite{dai2000}. First, it is now clear that the incommensurability and
the resonance 
are inseparable parts of the general features of the spin dynamics in YBa$_2$Cu$_3$O$_{6+x}$
at all dopings. Thus any microscopic model used to describe the resonance must also
be able to explain the doping and energy dependence of the incommensurate spin fluctuations. 
While current theoretical models do not provide explicit predictions that can be directly 
compared our experiments \cite{voo}, it is hoped that the work described here will stimulate future 
theoretical efforts in this direction. The remarkable field dependence of the 
resonance \cite{dai2000} indicates that it measures the superconducting pairing and phase 
coherence. In particular, the larger $c$-axis field effect is 
difficult to understand within the framework of the 
interlayer tunneling theory, according to which the resonance should be strongly affected by the 
in-plane field that disrupts the coherent Josephson coupling along $c$-axis \cite{yin}. 
On the other hand, the field data \cite{dai2000} are consistent with mechanism where
the dominant loss of entropy on entering the superconducting state 
is due to the growth of magnetic correlations in the CuO$_2$ planes 
\cite{scalapinowhite,demlerzhang,sudip1}. Finally, 
if the resonance is a magnon-like collective mode \cite{morr,abanov,sachdev}, it must not be   
a spin-wave in the conventional sense as the intensity of spin-wave excitations
in ferromagnets and antiferromagnets  
should be magnetic field independent for the relatively small fields used \cite{lovesey,dai2000}.

\section{Summary and Conclusions}
In this article, we have 
described in detail the extensive polarized and unpolarized neutron 
measurements of the dynamical magnetic susceptibility 
$\chi^{\prime\prime}({\bf q},\omega)$ of superconducting 
YBa$_2$Cu$_3$O$_{6+x}$
above and below the transition temperature $T_c$.
Using a careful polarization and temperature dependence analysis, we find that the resonance and
incommensurate spin fluctuations are general features of the spin dynamical behavior
in YBa$_2$Cu$_3$O$_{6+x}$ for all oxygen doping levels. 
We establish the doping dependence of the incommensurability,
the resonance, and the superconducting spin-gap energy.
In the superconducting state, we find that the spin-gap energy is 
proportional to $k_BT_c$, but the resonance energy itself deviates 
from the linear behavior with $k_BT_c$ for oxygen doping close 
to the optimal values. Similarly, we show that the doping dependence 
of the incommensurability saturates to $\delta=0.1$ for oxygen doping $x\geq 0.6$.  
We compare the observed behaviors 
in the two most studied superconducting copper-oxides 
La$_{2-x}$Sr$_x$CuO$_4$ and YBa$_2$Cu$_3$O$_{6+x}$ 
and discuss their similarities and differences.
We stress that any comprehensive theoretical understanding of the spin dynamical behavior in these
two families of materials must take into account the fact that the resonance so clearly seen in the
bilayer  YBa$_2$Cu$_3$O$_{6+x}$ is absent in the single-layer La$_{2-x}$Sr$_x$CuO$_4$.
For underdoped YBa$_2$Cu$_3$O$_{6+x}$, the evolution of the resonance as a function of doping 
is illustrated. In particular, we find the enhancement of the susceptibility at the resonance
energy above $T_c$ for all underdoped compounds including YBa$_2$Cu$_3$O$_{6.8}$.
If we define the ``resonance'' as the enhancement of the susceptibility 
in a small energy range of the magnetic excitations spectra as a function of 
decreasing temperature, it is natural to associate the susceptibility gain 
around the resonance energy above $T_c$ 
in the underdoped compounds as the precursor of the resonance. 
This assertion is supported by the systematic studies of the evolution of the temperature
dependence of the resonance that show the progressively larger pretransitional regime above 
$T_c$ for more underdoped YBa$_2$Cu$_3$O$_{6+x}$ \cite{dai99}. 
Furthermore, the resonance and incommensurate fluctuations appear to be 
intimately connected. Thus, any microscopic model for  
the magnetic fluctuations in YBa$_2$Cu$_3$O$_{6+x}$ must be able to account for both the 
incommensurate spin fluctuations and the resonance.


%
%

%

\begin{acknowledgments}
We are grateful to G. Aeppli, S. Chakravarty, 
T. Egami, V. J. Emery, S. M. Hayden, S. A. Kivelson, D. K. Morr, D. Pines, D. J. Scalapino, 
J. Zaanen, and S. C. Zhang
for many enlightening
discussions. We also wish to thank J. A. Fernandez-Baca, Jiandi Zhang, and 
S. C. Zhang for a critical reading of the manuscript, to thank J. R. Thompson and K. J. Song for 
measuring the superconducting transition temperatures for all the samples discussed in this article,
and to thank B. C. Chakoumakos and T. B. Lindemer for allowing the use of their unpublished data. 
Finally, we acknowledge much technical support over the years by R. G. Maple, S. A. Moore, G. B. Taylor, and
C. M. Redmon. 
This work was supported by U.S. DOE under contract 
DE-AC05-00OR22725 with 
UT-Battelle, LLC.
\end{acknowledgments}


\begin{references}
\bibitem{noteincommb} To facilitate the comparison of the results between La$_{2-x}$Sr$_x$CuO$_4$ 
and YBa$_2$Cu$_3$O$_{6+x}$, we adopted the definition of incommensurability $\delta$ used in [9]. 
Note that this definition of incommensurability $\delta$ is different (by a factor of 2)  
from that in [24].
\bibitem{note} Strictly speaking, this equation is only true if the efficiency of the
Mezei coil is 100\%.
\bibitem{pseudogap} Although
there is also a spin pseudogap in the normal state of YBa$_2$Cu$_3$O$_{6+x}$, we 
did not systematically 
investigate its temperature and doping dependence because the 
difficulty in separating the low-energy magnetic scattering 
from phonons and other spurious processes.
\bibitem{notemosaic} Although one expects incommensurate spin fluctuations 
at this doping level, detection of explicit incommensurability  
may require samples to have single grain with narrow mosaic spread.
The mosaic of YBa$_2$Cu$_3$O$_{6.5}$ is considerably larger than other
crystals in this study. As a consequence, we only an observe broad flattish top peak 
around $(\pi,\pi)$ at $\hbar\omega=24$ meV [Fig. 8(c)]. The incommensurability 
was determined by fitting the 24 meV profile with a pair of Gaussian peaks separated by
$\delta$ rlu from $(\pi,\pi)$.
\bibitem{noteincomm} Although we have demonstrated the clear presence incommensurate spin
fluctuations for YBa$_2$Cu$_3$O$_{6+x}$ with $x=0.45$, 0.6, 0.7, 0.8, and 0.95, no explicit
incommensurability is observed for our YBa$_2$Cu$_3$O$_{6.5}$ sample and the clarity of 
incommensurate spin fluctuations appears to be sample dependent. At present, we do not understand
the microscopic origin for such behavior. Perhaps the detailed incommensurate structure is
very sensitive to the oxygen uniformity of the material.
\end{references}

\end{document}